\documentclass[12pt,letterpaper]{article}
\pdfoutput=1
\usepackage{jheppub}
\usepackage{amsmath}
\usepackage{bm}
\usepackage{comment}
\usepackage[utf8]{inputenc}

\newcommand{\tr}{\mathrm{tr}\,}

\newlength{\dummysp}
\usepackage[font=scriptsize]{caption}
\usepackage{mwe}

\def\R{{\mathbb R}}
\def\S{{\mathbb S}}
\def\Z{{\mathbb Z}}

\def\T{{\mathbb T}}

\def\tr{\,{\rm tr}\,}

\def\beq{\begin{equation}}
\def\eeq{\end{equation}}

\title{The gaugino condensate from asymmetric four-torus with twists}
 \author[a]{Mohamed M. Anber,}\author[b]{Erich Poppitz} 
  
\affiliation[a]{Centre for Particle Theory, Department of Mathematical Sciences, Durham University, South Road, Durham DH1 3LE, UK}
\affiliation[b]{Department of Physics,   University of Toronto, 60 St George St., 
Toronto, ON M5S 1A7, Canada}
\emailAdd{mohamed.anber@durham.ac.uk}\emailAdd{poppitz@physics.utoronto.ca}

 \abstract{We calculate the gaugino condensate in $SU(2)$ super Yang-Mills theory on an asymmetric four-torus $\mathbb T^4$ with  't Hooft's twisted boundary conditions. The $\mathbb T^4$ asymmetry is controlled by a dimensionless detuning parameter $\Delta$, proportional to $L_3 L_4 - L_1 L_2$, with $L_i$ denoting the  $\mathbb T^4$ periods. We perform our calculations via  a path integral on a  $\mathbb T^4$. Its size is  taken much smaller than the inverse strong scale $\Lambda$ and the theory is  well inside the semi-classical weak-coupling regime. The instanton background, constructed for $\Delta\ll 1$ in \cite{GarciaPerez:2000aiw}, has fractional topological charge $Q=\frac{1}{2}$ and supports two gaugino zero modes, yielding a non-vanishing bilinear condensate, which we find to be $\Delta$-independent. Further, the theory has a mixed discrete chiral/$1$-form center anomaly leading to double degeneracy of  the energy eigenstates on any size torus with 't Hooft twists. In particular, there are two vacua, $|0\rangle$ and $|1\rangle$, that are exchanged under chiral transformation. Using this information, the $\Delta$-independence of the condensate, and assuming further that the semi-classical theory is continuously connected to the strongly-coupled  large-$\mathbb T^4$  regime,  we determine the numerical coefficient of the gaugino condensate: $\langle 0| \mbox{tr}\lambda\lambda|0\rangle=|\langle 1| \mbox{tr}\lambda\lambda|1\rangle|=32\pi^2 \Lambda^3$, a result equal to twice the known $\mathbb R^4$ value. We discuss possible loopholes in the continuity approach that may lead to this discrepancy.  }

\begin{document}

\maketitle

\section{Introduction}

Strongly coupled gauge theories have been under intense study over the past few years, thanks to the recent developments of generalized global symmetries \cite{Gaiotto:2014kfa}. These are operations that implement the group multiplication laws via topological constructions such that the symmetry operations are supported on topological surfaces that are robust under small deformations. An ordinary $0$-form symmetry $G$, which acts on point-like particles, is implemented by operators supported on co-dimension $1$ surfaces. These surfaces obey the group-multiplication laws via fusion rules and give rise to phases valued in $G$ when they cross the charged objects.   Likewise, a $1$-form symmetry acts on $1$-dimensional objects and is implemented via co-dimension $2$ surfaces. For example, Wilson lines in $4$-D $SU(N)$ pure or super Yang-Mills theories are charged under $\mathbb Z_N^{(1)}$ 1-form symmetry. The $1$-form symmetry is implemented via topological $2$-dimensional surfaces that obey the $\mathbb Z_N$ group multiplication laws and give rise to $\mathbb Z_N$ phases when they link with Wilson lines. Like ordinary $0$-form symmetries, $1$-form symmetries organize the spectrum of a theory into representations, satisfy Ward identities, and may become anomalous if one tries to gauge them. In particular, 't Hooft anomaly matching conditions (or 't Hooft anomalies for short), which impose stringent constraints on quantum field theory (QFT), can be generalized to include anomalies of $1$-form  symmetries. Detecting the anomaly in $4$-D requires defining a given QFT on manifolds with nontrivial $2$-cycles, the typical example being the $4$-torus $\mathbb T^4$.   Generalized 't Hooft anomalies provide a framework for classifying QFT and its phases. Many recent works entertained this generalized framework to shed light on a plethora of asymptotically free gauge theories, including vector-like \cite{Gaiotto:2017yup, Tanizaki:2018wtg, Anber:2019nze, Bolognesi:2019fej, Anber:2020gig, Sulejmanpasic:2020zfs,Anber:2019nfu,Lohitsiri:2022jyz} and chiral theories \cite{Bolognesi:2020mpe, Anber:2021iip}. 

Assuming a gauge theory generates a mass gap in the infrared (IR), one way to match its 't Hooft anomalies is via the formation of condensates. A famous example, and the subject of this paper, is the formation of gaugino condensates in ${\cal N}=1$ $SU(N)$ super Yang-Mills theory (SYM), which are needed to match a generalized 't Hooft anomaly. This anomaly can be understood as follows. In addition to the $\mathbb Z_N^{(1)}$ $1$-form symmetry that acts on the Wilson lines, SYM enjoys a $\mathbb Z_{2N}^{d\chi}$ $0$-form global discrete chiral, or $R$, symmetry. We may gauge the $1$-form  symmetry by turning on a background gauge field of $\mathbb Z_N^{(1)}$, which is done either by coupling SYM to a $\mathbb Z_N$ TQFT \cite{Kapustin:2014gua} or by activating a 't Hooft flux \cite{tHooft:1979rtg,tHooft:1981nnx}. The latter is a field configuration on  $\mathbb T^4$ that carries a fractional flux $B \in H^2(\mathbb T^4, \mathbb Z_N)$ (i.e. the flux piercing $2$-cycles obeys the quantization rule $\int_{\mathbb T^2 \subset\mathbb T^4} B\in \frac{2\pi \mathbb Z}{N}$) and  fractional topological charge $Q=\frac{1}{8\pi^2}\int_{\mathbb T^4}B \wedge B\in \frac{\mathbb Z}{N}$.  In mathematical language, we consider the $PSU(N)\equiv SU(N)/\mathbb Z_N$ bundle instead of the $SU(N)$ bundle. The former has a non-trivial Brauer class $w \in H^2(BPSU(N), \mathbb Z_N)$ that obstructs the lifting of $PSU(N)$ to $SU(N)$, and we set $B=w$. In practice, the $PSU(N)$ flux is turned on by imposing twisted boundary conditions on the fields that live on $\mathbb T^4$. 
 Next, we examine the partition function ${\cal Z}[PSU(N)]$ of SYM in this background by performing a $\mathbb Z_{2N}^{d\chi}$ rotation. We find ${\cal Z}[PSU(N)]\rightarrow e^{i\frac{2\pi}{N}}{\cal Z}[PSU(N)]$, and thus, the theory stops being invariant under the action of $\mathbb Z_{2N}^{d\chi}$ once we gauge the $1$-form symmetry. This is the sought generalized 't Hooft anomaly---a mixed anomaly between the $\mathbb Z_N^{(1)}$ $1$-form symmetry and the $\mathbb Z_{2N}^{d\chi}$ $0$-form discrete chiral symmetry. The anomaly indicates that the IR theory cannot be trivially gapped. Assuming confinement, the theory breaks its $\mathbb Z_{2N}^{d\chi}$ symmetry and forms $N$ vacua separated by domain walls. The order parameter of $\mathbb Z_{2N}^{d\chi}$ symmetry is the bilinear gaugino condensate $\langle\mbox{tr}\lambda^2 \rangle$ or higher-order condensates $\langle(\mbox{tr}\lambda^{2} )^n \rangle$, $n>1$ and $ n\, \mbox{mod}\, N \neq 0$. The existence of $N$ vacua is also in accordance with the Witten index \cite{Witten:1982df}. 
 
 \subsection{Gaugino condensate: a historical prelude}

Generally speaking, the matching conditions can only provide kinematical constraints and do not, by  themselves yield insights into the details of the IR dynamics of a gauge theory. One needs an extra guide if, at all, there is a hope to understand the dynamics, e.g., the condensates. Thanks to supersymmetry, such studies are possible in SYM. These date back to the early eighties \cite{Novikov:1983ee,Rossi:1983bu, Amati:1984uz,Affleck:1983rr,Novikov:1985ic} (see \cite{Amati:1988ft, Shifman:1999mv,Shifman:1999kf,Dorey:2002ik,Vandoren:2008xg,Terning:2006bq} for reviews). The condensate calculations are based on the Belavin-Polyakov-Schwartz-Tyupkin (BPST) instanton calculus \cite{Belavin:1975fg}:  BPST instantons are (anti) self-dual Yang-Mills configurations that violate the non-renormalization theorems of SYM and hence,  give a non-zero vacuum expectation value to the condensates. On dimensional-analysis grounds one can write  $\langle\frac{\mbox{tr}\lambda^2}{16\pi^2} \rangle=c\Lambda^3$, where $\Lambda$ is the strong-coupling scale and $c$ is a numerical factor. The exact value of $c$ was a controversial issue that caused many debates in the past. Generally, there are two methods to compute the bilinear gaugino condensate in $4$-D SYM: the strong-coupling and the weak-coupling instanton methods. 

In the first method, we start directly from the $4$-D SYM in its strong-coupling regime and do instanton calculus, as in \cite{Novikov:1983ee, Rossi:1983bu}. A single $SU(N)$ BPST instanton carries integer topological charge $Q\in \mathbb Z$, and the configuration with the lowest topological charge $Q=1$ admits $2N$ gaugino zero modes. The saturation of the zero modes in the $Q=1$ instanton background gives a nonzero value to the $2N$-point function $\langle(\mbox{tr}\lambda^{2} )^N \rangle$, from which one can naively extract the value of the $2$-point function $\langle(\mbox{tr}\lambda^{2} ) \rangle=\left[\langle(\mbox{tr}\lambda^{2} )^N \rangle\right]^{1/N}$. A detailed calculation, keeping track of all numerical coefficients, gives $\langle(\mbox{tr}\lambda^{2} ) \rangle= 2((N-1)!(3N-1))^{-1/N}(16\pi^2\Lambda^3)e^{i\frac{2\pi k}{N}}$, with $k=0,1,...,N-1$. The complex phase results from taking the $N$th root of unity, in accordance with the expectation that the theory admits $N$ distinct vacua needed to match the generalized 't Hooft anomaly.  

In the weak-coupling instanton method, we consider super QCD with $N-1$ fundamental flavors $\Phi_i$, $i=1,.., N-1$, where $\Phi_i$ is a chiral superfield, and give all the flavors small masses $m$. We work in the limit $|\Phi_i|\gg \Lambda_Q$, where $\Lambda_{Q}$ is the strong scale in the presence of quarks. Since there are $N-1$ flavors, the gauge group fully abelianizes and we are well inside the weak-coupling regime. The total superpotential of this theory takes the from ${\cal W}=m_{j}^i\bar \Phi^j\Phi_i+\frac{\Lambda_Q^{2N+1}}{\mbox{Det}(\bar\Phi\Phi)}$, where the second term is the Affleck-Dine-Seiberg (ADS) superpotential \cite{Affleck:1983rr}. The ADS term is nonperturbative in nature and is based on holomorphy and the symmetry structure of super QCD. It also results from saturating the quarks' zero modes in the BPST instanton background (the numerical coefficient was obtained\footnote{The comparison between the weak-coupling and strong-coupling instanton methods in $SU(2)$ was first performed in \cite{Novikov:1985ic}, where the correct ratio between the two methods was given.} in \cite{Cordes:1985um}, and corrected in \cite{Finnell:1995dr}). Since we are in a weak-coupling limit,  the instanton calculations are reliable. Minimizing the energy, we obtain the supersymmetric vacuum $\bar\Phi^j\Phi_i=\left(m^{-1}\right)_i^j \left[\Lambda_Q^{(2N+1)} \mbox{Det}m\right]^{1/N}$. Finally, we substitute this result back into ${\cal W}$ to find ${\cal W}=N\left[\Lambda_Q^{2N+1}\mbox{Det}m\right]^{1/N}$. We then decouple the quarks by taking $m\gg \Lambda_Q$, thus, leaving the weak-coupling regime.  Using holomorphy, we can write ${\cal W}_{\scriptsize\mbox{eff}}=N\Lambda^{3}$, where $\Lambda$ is the strong scale at the mass threshold, and it exactly coincides with SYM strong scale at the decoupling limit.  Recalling that one can write the holomorphic strong scale as $\Lambda=\mu e^{2\pi i\tau/3N}$, with $\tau=\frac{4\pi i}{g^2_h(\mu)}$ ($g_h$ is the holomorphic gauge coupling, running at one loop only and $\mu$ is some arbitrary energy scale) and that $\langle \mbox{tr}\lambda^2 \rangle=-8\pi i \frac{\partial {\cal W}_{\scriptsize\mbox{eff}}}{\partial \tau}$, one obtains $\langle\mbox{tr}\lambda^2 \rangle=16\pi^2 \Lambda^3e^{i\frac{2\pi k}{N}}$ in the $k$-th vacuum.\footnote{The definition of the strong coupling scale we follow in this paper is given by $\Lambda^3=\mu^3\frac{e^{-8\pi^2/N g^2}}{g^2}$, the one used in \cite{Davies:2000nw,Dorey:2002ik,Vandoren:2008xg}.}

Having two different methods that yield two different answers resulted in many debates in the literature about the validity of both methods. It was earlier understood that the strong-coupling instanton method is in tension with the cluster decomposition principle (CDP). Consider the correlator $\langle \mbox{tr}\lambda^2(x) \mbox{tr}\lambda^2(x') \rangle$. In the limit $|x-x'|\rightarrow \infty$ we expect $\langle \mbox{tr}\lambda^2(x) \mbox{tr}\lambda^2(x') \rangle= \langle \mbox{tr}\lambda^2\rangle^2$. However, since a BPST instanton cannot saturate $2$ gaugino zero modes, one finds $\langle \mbox{tr}\lambda^2 \rangle=0$, contradicting CDP. A possible resolution of this puzzle was proposed in \cite{Kovner:1997im}. It was hypothesized that SYM admits an extra phase with vanishing bilinear condensate $\langle \mbox{tr}\lambda^2\rangle=0$ and that averaging over the chirally symmetric and non-symmetric phases gives the result of the strong-coupling instanton method. This hypothesis was carefully examined in  \cite{Ritz:1999mq} by considering a softly broken ${\cal N}=2$ Seiberg-Witten theory down to SYM. It was shown that the chirally-symmetric phase is absent, casting doubt on the averaging hypothesis. Further considerations in \cite{Cachazo:2002ry} excluded the symmetric vacuum. We also note that the anomaly-matching argument of \cite{Csaki:1997aw} also precludes such a phase. It was further shown in \cite{Hollowood:1999qn} that strong-coupling multi-instanton calculations are inconsistent with CDP. 

Extra support to the weak-coupling calculations came from studying SYM on $\mathbb R^3\times \mathbb S^1_L$, where $\mathbb S^1_L$ is a small spatial circle with circumference $L\ll (N\Lambda)^{-1}$, and both gauge fields and gauginos are given periodic boundary conditions on $\mathbb S^1_L$ \cite{Davies:1999uw, Davies:2000nw}. Compactification over a small circle causes this theory to fully abelianize and enter its weakly-coupled regime.\footnote{The $4$-D coupling constant on $\R^3 \times \S^1_L$ ceases to run at scale $\sim 1/NL$, roughly the $W$-boson mass. Taking $L\ll (N\Lambda)^{-1}$, we stay in the weakly-coupled regime.}  The theory on $\mathbb R^3\times \mathbb S^1_L$ admits monopole-instantons, the microscopic constituents of calorons.\footnote{\label{caloron note}A $SU(N)$ caloron with a unit topological charge is composed of $N$ monopole instantons. Calorons are Yang-Mills (anti) self-dual configurations defined on $\mathbb R^3\times \mathbb S^1$  with integral topological charges and non-trivial holonomy (the expectation value of the Polyakov's loop) along $\mathbb S^1$. A BPST instanton, in contrast, (or more precisely, the Harrington-Shepard solution \cite{Harrington:1978ve} defined on $\mathbb R^3\times \mathbb S^1$) has a trivial holonomy.   Calorons were discovered by Kraan and van Baal \cite{Kraan:1998pm} using the Atiyah-Drinfeld-Hitchin-Manin (ADHM) construction \cite{Atiyah:1978ri} and independently by  Lee and Yi  \cite{Lee:1997vp} and Lee and Lu \cite{Lee:1998bb}  in the context of D-branes. Let $L$ and $P$ be the size of $\mathbb S^1$ and the holonomy (in units of $1/L$). If $LP\gtrsim 1$, then the monopole constituents are well-separated in space, and one can make sense of them in a semi-classical treatment, as in the case of SYM on $\mathbb R^3\times \mathbb S_L^1$. In particular, in the supersymmetric vacuum, which preserves the $0$-form $\mathbb Z_N$ center symmetry of the theory, the constituent monopoles are of equal action $S=\frac{8\pi^2}{N g^2}$. In the opposite limit, $LP\ll 1$, the monopoles hide inside the caloron core.} At the supersymmetric vacuum, the monopoles have topological charges of $1/N$. Using the index theorem, we deduce that a single monopole can saturate $2$ gaugino zero modes giving rise to the bilinear condensate.  Detailed calculations give  $\langle\frac{\mbox{tr}\lambda^2}{16\pi^2} \rangle=\Lambda^3$, the exact same result from $4$-D weak-coupling instanton calculations. The advantage of the compactified theory over the $4$-D theory at weak coupling is that in the former, one can understand the dynamics responsible for the condensate formation without relying on the miracle of holomorphy. As a bonus, the proliferation of the monopoles causes the compactified theory to generate a mass gap and confine, a result that is prohibitively difficult to understand in $4$-D.

The continuity of confinement and condensate between the small and large $L$ limits may imply that the fractional instantons are responsible for the dynamics on $\mathbb R^4$, even though one may not make analytical sense of them in a strongly-coupled setup. The continuity conjecture was taken seriously over the past decade in supersymmetric and nonsupersymmetric theories; see \cite{Dunne:2016nmc, Poppitz:2021cxe} for reviews. The conjecture withstood many tests, but we are still far from a firm conclusion about the role of fractional microscopic objects in strongly-coupled phenomena. This continuity was mainly tested on $\mathbb R^3\times \mathbb S^1_L$, and one wishes to examine whether it holds in other geometries. 

One such geometry is $\mathbb T^4$, very natural from the point of view of lattice practicalities. Right after 't Hooft presented his twisted solutions (solutions with twisted boundary conditions, i.e.~$PSU(N)$ bundle solutions, with fractional topological charges) \cite{tHooft:1981nnx}, van Baal studied their mathematical properties \cite{vanBaal:1982ag,vanBaal:1984ar}.  Later, it was argued in a series of works \cite{Gonzalez-Arroyo:1995ynx, Gonzalez-Arroyo:1995isl, Gonzalez-Arroyo:1996eos}, see also the review \cite{GarciaPerez:2009mmu}, that (anti) self-dual 't Hooft fractional instantons can be seen in realistic simulations of $4$-D pure Yang-Mills theory and that such configurations could be utilized to explain confinement.\footnote{Even before these studies, a program known as the ``femtouniverse'' utilized the Hamiltonian formalism on  $\R \times \T^3$, to study Yang-Mills theories at small volumes \cite{Luscher:1982ma}; see \cite{vanBaal:2000zc} for a review.}   Further, it was argued in \cite{GarciaPerez:1999hs}, via extended lattice simulations, that 't Hooft fractional instantons on $\mathbb T^4$  are ultimately connected to monopole instantons in the infinite $3$-volume limit and finite time\footnote{Notice that these are simulations in pure Yang-Mills theory, and thus, unlike SYM, there is no distinction between thermal and spatial circles.} direction.  In particular, it was shown that an exact $SU(2)$ caloron solution with a unit topological charge and equal-action constituent monopoles (see Footnote \ref{caloron note}) can be constructed on the twisted $\mathbb T^4$ by gluing two twisted solutions, each carrying $Q=\frac{1}{2}$ charge, along the space directions.\footnote{In fact, there is an obstruction to the existence of $Q=1$ (anti) self-dual caloron on $\mathbb T^4$ with untwisted boundary conditions \cite{Braam:1988qk}. Yet, in practice, one can find a very good approximate self-dual solution even in the absence of twists.}

\subsection{Gaugino condensate on asymmetric $\mathbb T^4$: a summary of the procedure and results}

The possible connection between 't Hooft's solutions and monopole-instantons \cite{GarciaPerez:1999hs} calls for a serious examination of this finding. Since both the $\mathbb R^3\times\mathbb S_L^1$ monopole-instanton and $4$-D weak-coupling instanton methods give the same gaugino condensate, one wonders whether the same result can be obtained in the 't Hooft flux background.\footnote{We interchangeably use the term ``'t Hooft flux background'' and ``'t Hooft twists." The latter is more precise, since it refers to the twist of the boundary conditions, which does not always lead to nonzero ``flux,'' i.e. nonzero gauge field strength. Whether such field strength is  present or not is a dynamical issue, see e.g.~\cite{Witten:1982df}. We hope that our abuse of terminology does not lead to confusion. } This is especially timely after the advent of the new generalized anomalies, which can be easily detected when a QFT is put on $\mathbb T^4$ with a $PSU(N)$ bundle (or 't Hooft twisted boundary conditions). Recently, \"Unsal advocated that a refinement of the instanton sum in the partition functions of $SU(N)$ theories has to be considered: while fractional instantons of $PSU(N)$ bundles contribute to observables like the gaugino condensate and vacuum energy, the sum over the fractional objects has to be constrained to yield the integer topological charges of the $SU(N)$ bundle \cite{Unsal:2020yeh} (see also \cite{vanBaal:2000zc} for an earlier assertion). 

In this paper, we put this proposal under scrutiny and perform detailed calculations of the gaugino condensate on $\mathbb T^4$ and  a $PSU(2)$ bundle.\footnote{The gaugino condensate on a symmetric $\mathbb T^4$ in the background of 't Hooft flux was first considered long ago in \cite{Cohen:1983fd}. These calculations were solely based on supersymmetry transformations along with dimensional analysis, and no attempt to determine the numerical coefficient was made.} Our conclusion is that the condensate in any of the two vacua is given by $\langle0|\mbox{tr}\lambda^2 |0\rangle=|\langle1|\mbox{tr}\lambda^2 |1\rangle|=32\pi^2 \Lambda^3$, with a coefficient that is twice that obtained via the weak-coupling instanton method on $\mathbb R^4$ and the semi-classical calculations on $\mathbb R^3\times \mathbb S^1_L$. The extra factor of $2$ is unexpected and calls for further serious examinations of the role of the $PSU(N)$ bundles in $SU(N)$ gauge theories and of the continuity conjecture.

The calculations that lead to this puzzling result are surprisingly rich. The simplest $SU(2)$ pure Yang-Mills (i.e. Yang-Mills equations with zero source) fractional instanton with topological charge $Q=\frac{1}{2}$ and action $S_0=\frac{4\pi^2}{g^2}$ was constructed by 't Hooft in his seminal work \cite{tHooft:1981nnx}. This solution is abelian in nature, i.e.~the gauge field components are along the Cartan direction. The solution must be (anti) self-dual, otherwise, the fluctuations in this background will have negative modes signaling instability. Let $L_1, L_2,L_3,L_4$ be the lengths of the periods in $\mathbb T^4$. Then, the self-duality of the abelian solution is guaranteed if and only if $L_1L_2=L_3L_4$. We dub the $\mathbb T^4$ that obeys this relation as the self-dual torus,  the simplest one being the symmetric $\mathbb T^4$ with $L_1$=$L_2$=$L_3$=$L_4$. Given the simplicity of the abelian solution on the symmetric $\mathbb T^4$, one is tempted to use it as a source for the gaugino condensate. According to the index theorem, this background must saturate $2$ gaugino zero modes.  Nevertheless, the direct solution of the Dirac equation yields $6$ zero modes (four ``undotted'' and two ``dotted" ones). To make things worse, the extra zero modes source the super Yang-Mills equations of motion, hinting that such an abelian solution might not be a consistent background. The resolution of both puzzles could have been achieved had we been able to show that higher-order corrections (including loops) lift the extra gaugino zero modes. Although we believe that this should be the case, we were not able to show it in a satisfactory way.

In order to circumvent this difficulty, we chose to depart from the symmetric self-dual to the non-self-dual torus. This is both a  blessing and a curse. The curse is the technical difficulty of the problem, while the blessing is that the asymmetric torus enables us to take various interesting limits, e.g.~$\R \times \T^3$ and $\R^2 \times \T^2$. If the relation $L_1L_2=L_3L_4$ is violated, the abelian solution must be modified to include non-abelian pieces, ensuring that the solution persists to be (anti) self-dual. To date, there exists no analytical solution with $Q=\frac{1}{2}$ on $\mathbb T^4$ with arbitrary shape. However, a systematic procedure to deal with non-self-dual tori was developed in \cite{GarciaPerez:2000aiw}. This method gives an approximate analytical self-dual solution, with charge $|Q|=\frac{1}{2}$ and action $S_0=\frac{4\pi^2}{g^2}$, as an expansion in a ``detuning" parameter $\Delta\equiv (L_3L_4-L_1L_2)/\sqrt{L_1L_2L_3L_4}$ measuring the deviation from a self-dual $\mathbb T^4$.

Solving the Dirac equation on a small asymmetric $\mathbb T^4$ with $\Delta\ll1$ and limiting our explicit treatment to the lowest order in $\Delta$, we find exactly $2$ gaugino zero modes as per the index theorem. We also verified that the explicit solutions of the fermion zero modes are consistent with supersymmetry transformations.   These modes do not source the Yang-Mills equations of motion, implying that our approximate solution is a consistent background.  The gauge field fluctuations admit $4$ bosonic zero modes (also in accordance with the index theorem),  interpreted as translational moduli. Supersymmetry guarantees the cancelation between the bosonic and fermionic excited states, leaving only the bosonic and fermion zero mode integrals to deal with. 

The contributions of zero modes to the path integral are taken into account using the method of collective coordinates, and therefore, we need to integrate over the moduli space ${\cal M}$.  To correctly identify the shape and size of ${\cal M}$, we carefully examine all gauge invariant observables in the background of the fractional instanton on $\mathbb T^4$. This includes both local gauge-invariant densities and Wilson loops. While gauge-invariant densities are invariant under translations over a period on $\mathbb T^4$, a Wilson loop acquires a $\mathbb Z_2$ phase.\footnote{Effectively, a Wilson loop measures the $\mathbb Z_2$ flux of the twisted solution.}  Therefore, we find ${\cal M}\sim\mathbb T^4$ with double the periods of the physical torus to account for the gauge inequivalent classes. This result is further supported by investigating the Hamiltonian formalism of the theory.    Interestingly, the metric on ${\cal M}$ is found to be proportional to the background action and is, hence,  independent of $\Delta$.  Putting the pieces together, we finally obtain $\langle\mbox{tr}\lambda^2 \rangle=64\pi^2 \Lambda^3$, with a coefficient four times the expected number on $\R^4$. 

To understand the significance of this result, we need to interpret the expectation value $\langle\mbox{tr}\lambda^2 \rangle$  using the Hamiltonian formalism. Due to the mixed discrete chiral/one-form center anomaly,  the energy eigenstates are doubly degenerate in the background of a 't Hooft flux on any torus size: there are $2$ vacua that are exchanged under the operation of chiral transformation, which explains the extra factor of $2$, while the relative phase in the condensate in the two vacua is compensated by the 't Hooft twist in the Euclidean time direction  \cite{Cox:2021vsa}. Thus, restricting the condensate to one vacuum, upon taking the limit of large $\T^4$, we obtain $\langle0|\mbox{tr}\lambda^2 |0\rangle=|\langle1|\mbox{tr}\lambda^2 |1\rangle|=32\pi^2 \Lambda^3$. This coefficient is twice the known value on $\mathbb R^4$ (using the weak-coupling instanton method)  or on $\mathbb R^3\times \mathbb S_L^1$.      

We emphasize that our calculations are performed on a small $\mathbb T^4$, compared to the strong scale of the theory, and thus, we are well inside the semi-classical regime. The calculations in this regime are under analytic control, thanks to the smallness of the coupling constant. However, to make a connection to the result on $\mathbb R^4$, we made a few assumptions.  The invalidation of any of these assumptions can explain the discrepancy between our result and the expected one. We now list the  assumptions that led to our result:
\begin{enumerate}

\item We assume that there is a unique fractional instanton on the asymmetric $\mathbb T^4$ with topological charge $Q=\frac{1}{2}$. This solution is nonabelian in nature and is obtained from 't Hooft's abelian solution on a symmetric $\mathbb T^4$ as an expansion in the detuning parameter $\Delta$ \cite{GarciaPerez:2000aiw}.  However, we were not able to prove the uniqueness\footnote{To avoid possible confusion, we note that the series in $\Delta$ determines a unique self-dual configuration for fixed values of the moduli (see the proof in Appendix \ref{constructing}, near eqn.~(\ref{cequations})). The possibility of non-uniqueness mentioned here refers to existence of a genuinely nonabelian $Q = {1 \over 2}$ solution not connected to 't Hooft's abelian solution. The numerical study of \cite{GarciaPerez:2000aiw} appears to support uniqueness in this sense, but does not  constitute  a proof.} of the solution. Instead, we rely on the numerical evidence in \cite{GarciaPerez:2000aiw}.

\item Another lingering issue is the radius of convergence of the expansion, which is yet to be understood. Although our final result does not depend on $\Delta$, one needs to be cautious when interpreting the result.   In order to interpret the result and compare with the condensate on $\mathbb R^4$, we had to make use of the Hamiltonian formalism. Here, first, one puts the theory on a small spatial torus $\mathbb T^3$, with a 
't Hooft twist, and with periods of length $L\ll \Lambda^{-1}$, and then takes the limit of a large Euclidean time direction, $L_4\gg \Lambda^{-1}$. This is the limit $\Delta\sim \sqrt{\frac{L_4}{L}}\gg1$, well outside the small-$\Delta$ regime used to compute the condensate.

\item After taking $\Delta$ large, one needs to take the size of the remaining $
\T^3$ periods beyond the inverse strong-coupling scale. Thus, we necessarily leave the semi-classical regime, yet we assume no extra instantons contribute to the condensate in this limit. (However, we can not help but note that this is similar, at least in spirit, to the  extrapolation of the $\R^3 \times \S^1_L$ semiclassical calculation of the gaugino condensate to $\R^4$.)

\item In carrying out our calculations, we assume that quantum corrections due to bosonic, fermionic, and ghost fluctuations cancel exactly in our supersymmetric background, to all loop orders, and for arbitrary $\mathbb T^4$ size.  Although the cancellation of the determinants to any loop order is well-established in $\mathbb R^4$, one may need to examine this assumption more carefully in the fractional instanton background of $\mathbb T^4$.
\end{enumerate}
In view of the above discussion,  in the bulk of the paper, we present our calculations in sufficient detail to help the interested readers follow all numerical factors and enable them to dwell on the procedures used---and on improving the interpretation of our result. 

\subsection{Future directions}
\label{sec:future}
Here, in lieu of a conclusion, we point out that our study warrants a few expansion directions:
\begin{enumerate}
\item An immediate step would be generalizing the $SU(2)$ result to $SU(N\geq 3)$. Self-dual instantons with topological charge $\frac{1}{N}$, $N\geq 3$, are necessarily non-abelian solutions of Yang-Mills equations \cite{tHooft:1981nnx}. This makes the treatment more involved, especially as we deviate from the self-dual torus.\footnote{The definition of a self-dual torus in $SU(N)$ is different from the $SU(2)$ case \cite{tHooft:1981nnx}.}  Fortunately, a systematic analysis to deal with this problem, a generalization of the method in \cite{GarciaPerez:2000aiw}, appeared in \cite{Gonzalez-Arroyo:2019wpu}.  One can follow the same line of thought in our work to compute the condensate in $SU(N)$.

\item It is tempting to study the condensates in other gauge groups, e.g., $Sp(N)$ and $Spin(2N)$ groups. These groups have small center groups: $Sp(N)$ has a $\mathbb Z_2$  and $Spin(2N)$ has a $\mathbb Z_4$ or a $\mathbb Z_2 \times \mathbb Z_2$ center group depending on whether $N$ is odd or even. Owing to the small centers, 't Hooft fluxes (or fractional instantons) of these groups will, in general, support more than $2$ gaugino zero modes. This situation is different from the calculations on $\mathbb R^3\times \mathbb S^1_L$, where a monopole-instanton always supports $2$ zero modes for all gauge groups, even those with no center symmetry. Therefore, at first, it is not clear how the twists and monopoles are connected. Investigating this problem is important for the continuity program.  
 
\item It is also interesting to carry out our procedure to higher orders in $\Delta$. The convergence of the series in $\Delta$ may shed more light on the problem of continuity.  In this regard, we note that the closely related work \cite{Gonzalez-Arroyo:2004fmv} in the two-dimensional abelian Higgs model was able to carry the $\Delta$-expansion  to  51-st order, with the results indicating convergence to the infinite volume limit.

\item Another interesting geometry is to consider SYM on $\mathbb T^2\times \mathbb R^2$ with a 't Hooft flux turned on $\mathbb T^2$.  This setup was considered previously in \cite{Tanizaki:2022ngt}. It was shown via dimensional reduction from $\mathbb T^2\times \mathbb R^2$ to $\mathbb R^2$ that the theory admits $\mathbb Z_N$ vortices and that the gaugino condensate forms in the $2$-D effective theory $\langle\mbox{tr}\lambda^2\rangle\sim \Lambda^3$, as expected. However, the authors did not attempt to compute the numerical coefficient of the condensate. Such a calculation would mandate a more careful treatment of the dimensionally reduced $2$-D theory, presumably using the power of supersymmetry. We also note that $\mathbb T^2\times \mathbb R^2$ is the limiting geometry of our asymmetric $\mathbb T^4$ after taking $\Delta $ to infinity.  
\end{enumerate}

\subsection{Outline}
\label{sec:outline} 

Our paper is organized as follows:

 In Section \ref{Fractional instantons on the symmetric torus}, we formulate the theory and spell out all the necessary ingredients to define the partition function and the condensate on $\mathbb T^4$ with twisted boundary conditions. Further, we explain that a self-dual torus gives rise to extra, unexpected, zero modes. Then, we present the solution on the asymmetric $\mathbb T^4$ in Section \ref{Fractional instantons and gaugino zero modes on the asymmetric torus} and discuss both the fermion and bosonic zero modes. In Section \ref{The path integral: bosonic and fermionic measures}, we calculate the measures of the bosonic and fermion zero modes. In Section \ref{The Hamiltonian formalism, Wilson loops, and the moduli space}, we deviate to the Hamiltonian formalism to discuss two important aspects. First, we argue that the moduli space of the bosonic zero modes is isomorphic to $\mathbb T^4$ with a period size twice the size of the physical period. Next, we recall that the $0$-form/$1$-form mixed anomaly implies that the energy eigenstates on $\T^3$ are doubly degenerate and the theory admits $2$ degenerate vacua. These features are important for the interpretation of our calculation of the gaugino bilinear. Finally, in Section \ref{The gaugino condensate}, we put all the pieces together  to obtain our result of the gaugino condensate. 

Owing to the mismatch between our and both $\mathbb R^4$ and $\mathbb R^3\times \mathbb S^1_L$ results for the gaugino condensate, and in order to offer the reader the opportunity to catch mistakes, if any,  we present our rather detailed calculations of various quantities in two appendices.

 In Appendix \ref{Constructing instantons on the asymmetric T4 with twists}, we review in great detail  the construction of \cite{GarciaPerez:2000aiw} of  the fractional instanton to order $\Delta$, with emphasis on the dependence on the collective coordinates. We also provide expressions of the field strengths and Wilson loops in the background of the leading-order solution.
 
  In Appendix \ref{zeromodesection}, we construct the fermion zero modes by directly solving the Weyl equation as well as by using supersymmetry transformation, and we prove that both methods yield the same result. Then, we construct the bosonic zero modes to any order in $\Delta$, first by employing the fermionic zero modes and then by taking derivatives of the classical background solution w.r.t.~the collective coordinates modulo gauge transformation (we also determine these gauge transformations).  Finally, we determine the Jacobian of the bosonic zero modes moduli space needed to complete the calculations.

\section{Fractional instantons on the symmetric torus}
\label {Fractional instantons on the symmetric torus}

We consider the $SU(2)$ SYM theory on $\mathbb T^4$ with periods of lengths $L_1,L_2,L_3,L_4$. The Euclidean action of the theory is given by\footnote{The Euclidean action,  supersymmetry transformations, and the matrices $\sigma_n$, $\bar\sigma_n$, $\sigma_{mn}$, $\bar\sigma_{mn}$, are  as in \cite{Dorey:2002ik}, except   that we use hermitean gauge fields, necessitating the replacement $A^{\text{that ref.}} = i A^{\text{this paper}}$.  See also Appendices \ref{constructing} and \ref{fermionappx1}.}
\begin{eqnarray}
S_{SYM}=\frac{1}{g^2}\int_{\mathbb T^4}\mbox{tr}\left[  {1\over 2}  F_{mn} F_{mn} +{2}(\partial_n \bar\lambda_{\dot\alpha} + i [A_n,\bar\lambda_{\dot\alpha}])\bar\sigma_n^{\dot\alpha \alpha} \lambda_\alpha\right]\,,
\label{total action}
\end{eqnarray}
and $\lambda$ is a left-handed adjoint Weyl fermion, the gaugino. $D_n=\partial_n+i [A_n,\,]$ is the covariant derivative, $\sigma_n\equiv(i\vec\sigma,1)$, $\bar\sigma_n\equiv(-i\vec\sigma,1)$, $\vec \sigma$ are the Pauli matrices, and the Latin letters $n,m$ run over $1,2,3,4$. The field strength is given by $F_{mn}=\partial_m A_n-\partial_n A_m+i[A_m,A_n]$. This action is invariant under the supersymmetry transformations
 \begin{eqnarray}
 \nonumber
\delta A_n &=& \zeta^\alpha \; \sigma_{n \; \alpha \dot\alpha} \; \bar\lambda^{\dot\alpha} + \bar\zeta_{\dot\alpha} \; \bar\sigma_n^{\dot\alpha \alpha} \; \lambda_\alpha\,,\quad
\delta \lambda_\alpha = - \sigma_{mn \; \alpha}^{~~~~~\beta} \;\zeta_\beta \; F_{mn}\,,\quad
\delta \bar\lambda^{\dot\alpha} = - \bar\sigma^{~~~~~\dot\alpha}_{mn \;\;~~ \dot\beta} \;\bar\zeta^{\dot\beta} \; F_{mn} \,,\\
\label{susyhermitean1bulk}
\end{eqnarray}
where the spinors obey $\xi^1 = \xi_2, \xi^2 = - \xi_1$ and likewise for the dotted ones. The equations of motion that result from the variation of $S_{SYM}$ are 
\begin{eqnarray}
(D_m F_{mn})^A= - i\; \mbox{tr} \; \bar\lambda\bar \sigma_n[ T^A, \lambda]\,,\quad  \bar \sigma_{n }^{\dot\alpha \alpha}   D_n\lambda^A_\alpha =0\,, \quad \sigma_{n\; \alpha \dot\alpha}  D_n \bar \lambda^{A \dot\alpha} =0,
\label{classical EOM}
\end{eqnarray} where $A=1,2,3$ labels the color group generators $T^A = \tau^A/2$ with $\tau^A$ the Pauli matrices.
We shall consider SYM with twisted boundary conditions on $\mathbb T^4$. Without loss of generality, we can use the following transition functions:
\begin{equation}\label{omega1bulk}
\Omega_2(x) = e^{- i 2 \pi {x_1\over L_1} {\tau^3 \over 2}}, \; \; \Omega_4(x) = e^{- i 2 \pi {x_3 \over L_3} {\tau^3 \over 2}}, ~\text{while} ~ \Omega_1 = \Omega_3 = 1.
\end{equation}
$\Omega_2$ and $\Omega_4$ implement the twists along the $x_2$ and $x_4$ directions, while the transition functions along the $x_1$ and $x_3$ directions are trivial. The transition functions obey the cocycle conditions
\begin{equation}\label{cocyclebulk}
\Omega_i (x+ L_j \hat e_j) \; \Omega_j(x) = e^{i\pi n_{ij}\tau_3}\;  \Omega_j (x+ L_i\hat e_i)\; \Omega_i  (x),~ i,j = 1,2,3,4, \; \forall x \in \R^4,
\end{equation}
where $\hat{e}_n$ is a unit vector in the $x_n$ direction, $n_{12}=n_{34}=-n_{21}=-n_{43}=1$, and the rest of $n_{ij}$ are zeros.
 The  periodicity  condition on the gauge fields and gaugino in $\R^4$ defines the $\T^4$ fields:
\begin{eqnarray}
\nonumber
A(x + \hat{e}_n L_n) &=& \Omega_n(x) (A(x) - i d) \Omega_n^{-1}(x), ~ n=1,2,3,4, ~
\forall x \in \R^4,\\
\lambda (x + \hat{e}_n L_n)&=& \Omega_n(x) \lambda(x) \Omega_n^{-1}(x)\,,
\end{eqnarray}
where we denoted $A(x) =\sum\limits_{n=1}^4 A_n(x) d x_n$.  Let the $SU(2)$ non-abelian field $\Phi$ denote either the gauge field or the gaugino and expand $\Phi$ using the Cartan-Weyl basis:
\begin{eqnarray}
\Phi(x)=\Phi^3\frac{\tau^3}{2}+\Phi^+\tau^++\Phi^-\tau^-\,.
\end{eqnarray}
The Pauli matrices $\tau^3, \tau^{\pm}$ are the generators of $SU(2)$ in this basis. Using the commutation relations $[\tau^3,\tau^{\pm}]=\pm\tau^3$, $[\tau^+, \tau^-]=\tau^3$ along with the Baker-Campbell-Hausdorff formula, one finds that the boundary conditions on $\Phi$ are satisfied if and only if:
\begin{eqnarray}
\nonumber
\Phi^3(x_i+L_i)&=&\Phi^3(x)+{\cal I}\,,\\
\nonumber
\Phi^\pm(x_1+L_1,x_2,x_3,x_4)&=&\Phi^\pm(x_1,x_2,x_3,x_4)\,,\\
\Phi^\pm(x_1,x_2+L_2,x_3,x_4)&=&e^{\mp i\frac{2\pi x_1}{L_1}}\Phi^\pm(x_1,x_2,x_3,x_4)\,,
\label{basic BCS}
\end{eqnarray}
where ${\cal I}$ is a nonhomogeneous term that contributes to the gauge field and appears upon shifting the $x_2$ coordinate: $A_n^3(x+\hat e_2 L_2)=A_n^3(x)+\frac{2\pi}{L_1}\delta_{n,1}$. Similar boundary conditions hold in the 3-4 plane. As we shall show,  there is a classical solution to the equations of motion (\ref{classical EOM}), which satisfies the boundary conditions (\ref{basic BCS}) and has a topological charge $Q=\frac{1}{2}$. 

The Euclidean partition function of our system is given by the path integral:
\begin{eqnarray}
{\cal Z}_{SYM}[n_{12}=1, n_{34}=1]=\sum_{\nu \in\mathbb Z}\int [D A_\mu][D\lambda][D\bar\lambda]e^{-S_{SYM}-i(\nu + {1 \over 2})\theta}|_{n_{12}=1, n_{34}=1}\,,
\label{main partition function}
\end{eqnarray}
where we have emphasized that this path integral is to be computed with the twists in the 1-2 and 3-4 plane, as per the transition functions (\ref{omega1bulk}), and thus, all the fields need to satisfy the boundary conditions (\ref{basic BCS}). Notice that one also needs to sum over the integer topological sectors $\nu \in \Z$ in order for ${\cal Z}_{SYM}$ to respect locality and unitarity.\footnote{If $\nu \ne 0$, the transition functions (\ref{omega1bulk}) (but not the cocycle conditions (\ref{cocyclebulk})) have to be appropriately modified.} The vacuum angle $\theta$ can be transformed away by applying an axial rotation on the gauginos. Our main aim is to find the bilinear gaugino condensate starting from (\ref{main partition function}). 

The solution of ${\cal Z}_{SYM}$, as well as the condensate, will proceed by using the semi-classical techniques, which amount to computing the path integral as the sum of paths of small fluctuations in the background of instantons. The semiclassical approach is justified in the limit of small volume. 

Notice, however, that this partition function vanishes identically, which can be understood in two different ways. First, as we shall show, the theory exhibits $2$ fermion zero modes. These are the zero modes that are saturated in the twisted background and give rise to the fermion condensate. We can also argue that ${\cal Z}_{SYM}=0$ because of the mixed anomaly between the $0$-form discrete chiral and $1$-form center symmetries, as was previously shown in \cite{Cox:2021vsa}. 

The gaugino condensate calculations can proceed by inserting the gaugino bilinear $\tr\lambda\lambda$ in the path integral (\ref{main partition function}), where the trace is taken in the color space. We define the expectation value of the gaugino condensate as:
\begin{eqnarray}
\langle \tr \lambda\lambda \rangle={\cal N}^{-1}\sum_{\nu \in\mathbb Z}\int [D A_\mu][D\lambda][D\bar\lambda]\; \tr \lambda\lambda \;e^{-S_{SYM}-i(\nu + {1 \over 2}) \theta}|_{n_{12}=1, n_{34}=1}\,,
\label{the main gaugino condensate}
\end{eqnarray}
with normalization constant ${\cal N}$. 
Only the $\nu=0$ sector (in the presence of the twists) can contribute to the bilinear condensate, and thus, we restrict our calculations to this sector. The standard way of normalizing the expectation value of an operator is to divide by the partition function. Since ${\cal Z}_{SYM}=0$, we need to search for another appropriate way to normalize our physical observables. We choose to divide by the partition function with twists only in the 1-2 plane. 
Explicitly, 
\begin{eqnarray}
 {\cal N} = \sum_{\nu \in\mathbb Z}\int [D A_\mu][D\lambda][D\bar\lambda] \;e^{-S_{SYM}-i \nu  \theta}|_{n_{12}=1}\,.
\label{the main gaugino condensate normalization}
\end{eqnarray}
This partition function is saturated by the $\nu=0$ term. In the small-$\T^4$ limit, it  can be evaluated using a  semiclassical expansion around  two zero-action classical saddle point
configurations, related by the $\Z_2$ center symmetry in the $x_3$ direction, generated by the ``improper gauge transformation'' $T_3$. In  a gauge with constant transition functions\footnote{These zero-action configurations can also be exhibited in a gauge where $\Omega_{2}$ is as in (\ref{omega1bulk}), with $\Omega_{1,3,4}=1$, but details are to be given elsewhere.}  in the $x_1$-$x_2$ plane, these saddle points are $A = 0$ and $A= - i T_3 d T_3^{-1}$ \cite{Witten:1982df}. Each of these  saddle points gives an identical contribution to ${\cal N}$. There are no zero modes (see \cite{GonzalezArroyo:1987ycm} for calculations of the spectrum) and all determinants, fermionic and bosonic, are the ones computed in the $A=0$ background (we note that this is the normalization of instanton transition amplitudes already taken in \cite{tHooft:1976snw}). 
The partition function  (\ref{the main gaugino condensate normalization}), up to a normalization factor that is expected to cancel the numerical factor in the numerator of  (\ref{the main gaugino condensate}), is equal to the Witten index $\mbox{Tr}[(-1)^F]$, which counts the number of the ground states. Thus, we take ${\cal N}=2$. See Section \ref{The Hamiltonian formalism, Wilson loops, and the moduli space} for a Hamiltonian treatment. 
 
 As a first trial, let us find a self-consistent $Q=1/2$ fractional-instanton solution to the equations of motion (\ref{classical EOM}) with the boundary conditions (\ref{basic BCS}). It is easy to see that the abelian gauge configuration:
\begin{eqnarray}
\nonumber
\bar A_1^{3} &=& {2 \pi x_2 \over L_1 L_2} + {z_1 \over L_1}\,,\quad
\bar A_2^{3} = {z_2 \over L_2}\,,\\
\bar A_3^{3} &=& {2 \pi x_4 \over L_3 L_4} + {z_3 \over L_3}\,,\quad
\bar A_4^{3} = {z_4 \over L_4}\,,
\label{abelian background}
\end{eqnarray}
obeys (\ref{basic BCS}).
The bar is introduced here to serve a later convenience, and the constants $z_m$ are the collective coordinates, which will be set to zero in this section without loss of generality.  This field configuration was first found by 't Hooft, and it solves the Pure Yang-Mills equations $D_m F_{mn}=0$, owing to its abelian nature and constant field strength. It carries a topological charge $Q=\frac{1}{2}$ and its action is half the action of a BPST instanton: $S= S_0 \equiv \frac{4\pi^2}{g^2}$.  This solution must be self-dual or anti-self-dual, $F_{mn}=\pm\frac{1}{2}\epsilon_{mnpq}F_{pq}$, which guarantees that the bosonic and fermionic determinants in this background do not yield negative zero modes, a sign of instability of the solution on $\mathbb T^4$. Self-duality of the $Q=1/2$ solution  (\ref{abelian background}) implies the condition
\begin{eqnarray}
L_1L_2=L_3L_4\,.
\label{self-dual torus}
\end{eqnarray}

A torus that satisfies the relation (\ref{self-dual torus}) is said to be self-dual, and a simple choice that satisfies the condition (\ref{self-dual torus}) is the symmetric torus with $L_1=L_2=L_3=L_4$. The classical solution (\ref{abelian background}) will also hold in SYM provided that the right-hand side of the first equation in (\ref{classical EOM}) vanishes. To check that, we first need to solve the Weyl equations (second and third equations in (\ref{classical EOM})) in the background (\ref{abelian background}), along with the boundary conditions (\ref{basic BCS}).  We find $6$-independent zero modes: $4$ modes for $\lambda$ and $2$ modes for $\bar\lambda$: 
\begin{eqnarray}
\label{selfdualzeromodes}
\lambda&=&\frac{\tau^3}{2}\left[\begin{array}{c} \xi_1\\\xi_2\end{array}\right]+  \sum\limits_{n_1,n_3= -\infty}^\infty  e^{ -i 2\pi (n_1 {x_1 \over L_1} + n_3 {x_3 \over L_3})}  h_0^{12}(x_2 -   n_1L_2)  \; h_0^{34}(x_4 -   n_3 L_4)\tau^+\left[\begin{array}{c} 0\\\xi_3\end{array}\right] \nonumber
\\
\nonumber
&&\quad\quad\quad\quad+  \sum\limits_{n_1,n_3= -\infty}^\infty  e^{ i 2\pi (n_1 {x_1 \over L_1} + n_3 {x_3 \over L_3})}  h_0^{12}(x_2 -   n_1L_2)  \; h_0^{34}(x_4 -   n_3 L_4)\tau^-\left[\begin{array}{c} \xi_4\\0\end{array}\right]\,,\\
\bar\lambda&=&\frac{\tau^3}{2}\left[\begin{array}{c} \xi_5\\\xi_6\end{array}\right]\,,\end{eqnarray}
where $\xi_i$ are Grassmann numbers.
The functions $h_0^{12}$ and $h_0^{34}$ are the ground eigenstates of the simple harmonic oscillator, of frequencies $\omega_{12} = {2 \pi \over L_1 L_2}$ and  $\omega_{34} = {2 \pi \over L_3 L_4}$, respectively;  see Appendices \ref{constructing} and \ref{symmzeromodes}.   Owing to the abelian nature of the classical background, the Weyl equation yields the solutions in the $\tau^3$ directions. Substituting the fermion zero mode solutions into the right-hand side of the bosonic equation of motion in (\ref{classical EOM}), one easily finds  that the source  term is nonzero. This simple exercise shows that the abelian background (\ref{abelian background}) is either inconsistent or that some gaugino zero modes must be lifted by higher-order corrections in order to have $\mbox{tr}\bar\lambda \bar\sigma^n [T^A,\lambda] =0$. Presumably, higher-order corrections will gap $4$ of the fermion zero modes, and in the end, we will have a consistent story. 
We were not able to show that this is the case in a satisfactory way.\footnote{For completeness, we note that an analogous problem occurs with the bosonic zero modes around the solution (\ref{abelian background}) on the self-dual torus: there are 8 (four real and two complex), rather than 4, bosonic zero modes, as was shown long ago by  van Baal \cite{vanBaal:1984ar}. We also stress that the existence of the four undotted and two dotted fermionic zero modes (\ref{selfdualzeromodes}) is consistent with the index theorem, which only determines their difference. }

Instead, we chose to detune the self-dual torus, i.e., to relax the condition $L_1L_2=L_3L_4$, and modify the abelian background to include non-abelian pieces that are needed to guarantee the self-duality of the fractional instanton on an asymmetric $\T^4$. The detuned solution does not suffer from any of the above-mentioned problems. A further motivation for considering the asymmetric $\T^4$ connects to interesting semiclassical limits that have been considered.

\section{Fractional instantons and gaugino zero modes on the asymmetric torus}
\label{Fractional instantons and gaugino zero modes on the asymmetric torus}

In order to circumvent the problems of the self-dual $\mathbb T^4$, we instead search for a self-dual instanton on an asymmetric torus using a perturbation technique introduced in \cite{GarciaPerez:2000aiw}. The gauge field $A_n$ can be written in the general form
\begin{eqnarray}
\label{generalbkgrdbulk}
A_n^{cl}(x,z) = (\bar A_n^{3}(x,z) + S_n(x,z)) {\tau^3 \over 2} + W_n(x,z) \tau^+ + W_n^*(x,z) \tau^-\,, 
\end{eqnarray}
where we have split the $\tau^3$ component into two parts. The first part $\bar A_n^{3}(x,z)$ is the abelian background (\ref{abelian background}) that solves the sourceless Yang-Mills equations on the symmetric $\mathbb T^4$. The functions $S_n(x,z)$ and  $W_n(x,z)$ will be determined perturbatively.  We also introduce the dimensionless detuning parameter $\Delta\equiv\left(L_3L_4-L_1L_2\right)/\sqrt{L_1L_2L_3L_4}$, which measures the deviation from the self-dual torus, and take $\Delta>0$. The solution of the sourceless Yang-Mills equations on the asymmetric $\mathbb T^4$ is obtained by imposing the self-duality condition $F_{mn}=\frac{1}{2}\epsilon_{mnpq}F_{pq}$, where $\epsilon_{1234}=1$, etc. In order to reduce the gauge redundancy, one also imposes the background gauge condition $\partial_n A_n^{cl}+i[\bar A_n, A_n^{cl}]=0$.  

The details of the construction of the self-dual solution in an expansion in powers of $\Delta$ is given in Appendix \ref{constructing}. The presentation there follows \cite{GarciaPerez:2000aiw} but  for completeness we present it in detail using our notation, with an emphasis on the dependence of the solution of the $z_n$ variables.
Here, we only give an idea of the construction of the fractional instanton and present its main features. 
To simplify the equations, it is convenient to use a quaternion notation. Thus, we introduce the matrices:
 \begin{eqnarray} 
w = \sigma^n W_n\,,\quad w_c = C w^* C\,,\quad s = \sigma^n S_n\,,
\end{eqnarray}
where $\sigma^n = ( i\vec\sigma, 1_{2 \times 2}), ~\bar\sigma^n = (\sigma^n)^\dagger$ and $C = \left(\begin{array}{cc}0 & -i \cr i & 0 \end{array} \right)$.
Self-duality and the background gauge conditions (see (\ref{gaugecondition})) then yield the following equations:
\begin{eqnarray}
 \bar\sigma^n \partial_n s = {2 \pi  {\Delta} \over \sqrt{V}} i \tau^3- i (w_c^\dagger w_c - w^\dagger w)\,,\quad
\bar\sigma^n(\partial_n + i \bar{A}_n^{3}) w = - {i \over 2} (s^\dagger w - w_c^\dagger s)\,.
\label{selfdualitybulk}
\end{eqnarray}
These equations are subject to the boundary conditions (\ref{basic BCS}). The solutions to (\ref{selfdualitybulk}) is found as series expansions in $\sqrt{\Delta}$:
\begin{eqnarray}\label{expansion}
w=\Delta^{1/2}\sum_{j=0}^\infty w_j\Delta^j\,,\quad  s=\sum_{j=1}^\infty s_j\Delta^j\,.
\end{eqnarray}

The symmetry structure of the self-duality equations (\ref{selfdualitybulk}), as well as their solution to the leading order in $\Delta$, is discussed in Appendix \ref{constructing}. The final answer for the fractional instanton solution, giving the order $\sqrt{\Delta}$ terms in (\ref{generalbkgrdbulk}), reads:
\begin{eqnarray}
\nonumber
W_1(x,z, \alpha) &=& -{i \over 2} \sqrt{\Delta} F(x,z) e^{i \alpha} {\sqrt{2 \pi} \over V^{1/4} } +  {\cal{O}}(\Delta^{3/2})\nonumber \\
W_2 (x,z,\alpha) &=& {1 \over 2} \sqrt{\Delta} F(x,z) e^{i \alpha} {\sqrt{2 \pi} \over V^{1/4}}+ {\cal{O}}(\Delta^{3/2}) = i W_1 + {\cal{O}}(\Delta^{3/2}),\nonumber \\
W_3 &=&  {\cal{O}}(\Delta^{3/2})\,,\quad W_4 =  {\cal{O}}(\Delta^{3/2})\,,\quad S_n =  {\cal{O}}(\Delta)\,.
\label{orderdeltabackgroundbulk}
\end{eqnarray}
We stress that the ${\cal{O}}(\Delta)$ contribution to $S_n$ is determined by the ${\cal{O}}(\sqrt{\Delta})$ terms shown above, in a manner described in Appendix \ref{constructing}, but explicit expressions will not be needed.
The arbitrary phase $\alpha$ is due to the gauge freedom to rotate around the $\tau^3$ isospin direction, and the function $F(x,z)$ is given by the expression:
 \begin{eqnarray}
\label{efgloriousbulk}
F(x,z) &=& \sqrt{L_2 L_4} \sum\limits_{n_1,n_3= -\infty}^\infty  e^{ -i 2\pi (n_1 {x_1 \over L_1} + n_3 {x_3 \over L_3})}  e^{ - i {z_2 \over L_2}(x_2 - n_1 L_2)  - i {z_4 \over L_4} (x_4- n_3 L_4)} \nonumber \\
&& ~~~~~\times~~ h_0^{12}(x_2 -   (n_1-{z_1 \over 2 \pi}) L_2)  \; h_0^{34}(x_4 -   (n_3-{z_3 \over 2 \pi}) L_4)~, \end{eqnarray}
 with the normalization $\int_{\mathbb T^4} |F|^2 =\prod_{i=1}^4L_i= V$. The functions $h_0^{12,34}$ are the same harmonic oscillator ground state wave functions appearing in eqn.~(\ref{selfdualzeromodes}) and described there.
 
It remains to check that this solution obeys the equations of motion (\ref{classical EOM}), and in particular, that the fermionic source in the bosonic equation of motion vanishes identically.  To this end, we need to find the fermion zero modes, which we do in Section \ref{sec:fermionzero}. For this purpose, we will need to use the field strength, a computation carried out in detail in Appendix \ref{appx:fieldstrength}. We find that to order ${\cal O}(\sqrt{\Delta})$:
 \begin{eqnarray}
 \nonumber
 F_{12}&=&\frac{2\pi}{L_1L_2}\,, \quad  F_{34}=\frac{2\pi}{L_3L_4}\,,\\
 \nonumber
 F_{13} &=& -i \gamma^* e^{i \alpha} \sqrt{2\pi\over 2 L_3L_4} G(x,z) \tau^+ + i \gamma e^{- i \alpha} \sqrt{2\pi \over 2 L_3L_4} G^*(x,z) \tau^-, \\
 F_{14} &=& \gamma^* e^{i \alpha}  \sqrt{2\pi \over 2 L_3L_4} G(x,z) \tau^+  +   \gamma e^{- i \alpha} \sqrt{2\pi \over 2 L_3L_4} G^*(x,z) \tau^- \,,
 \label{noncartanfbulk}
 \end{eqnarray}
where  $\gamma = {i \over 2}  \sqrt{2 \pi \Delta \over \sqrt{V}}$ and the function $G(x,z)$ is  \begin{eqnarray}
\label{gfglorious11}
G(x,z) &=& \sqrt{L_2 L_4} \sum\limits_{n_1,n_3= -\infty}^\infty  e^{ -i 2\pi (n_1 {x_1 \over L_1} + n_3 {x_3 \over L_3})}  e^{ - i {z_2 \over L_2}(x_2 - n_1 L_2)  - i {z_4 \over L_4} (x_4- n_3 L_4)} \nonumber \\
&& ~~~~~\times~~ h_0^{12}(x_2 -   (n_1-{z_1 \over 2 \pi}) L_2)  \; h_1^{34}(x_4 -   (n_3-{z_3 \over 2 \pi}) L_4)\,.
\end{eqnarray}
The function $h_{1}$ is the first excited state of the simple harmonic oscillator, and just like $F(x,z)$, $G$ is dimensionless and similarly normalized $\int_{\mathbb T^4} |G|^2 =\prod_{i=1}^4L_i= V$. 
 To find the field strength to order ${\cal O}(\Delta)$ and establish the self-duality of the solution, one needs to solve for the functions $S_n$, see (\ref{equationfors2}). In Appendix \ref{constructing}, we discuss the relations obeyed by the field strength of $S_n$ (explicit expressions will not be needed in our work). These are important for   establishing the self-duality of the solution, and thus of the fact that its action $S_0$ saturates the BPS bound, $S_0 = {4\pi^2 \over g^2}$.

Finally, we stress that the solutions, whose construction via a small-$\Delta$ expansion was described above, have been subjected to comparison with ``exact'' solutions obtained by numerically minimizing the lattice Yang-Mills theory action with an $n_{12}=n_{34}=1$ twist. A good qualitative (and in some cases quantitative) agreement of the gauge invariants charactering the solution, computed analytically\footnote{This is also done in our Appendix \ref{gaugeinvariants}, where the $z_n$ dependence of the local and non-local gauge invariants characterizing the solution is discussed in detail.} to ${\cal{O}}(\Delta)$, with the numerical approximation of the exact solution was found, for $\Delta$ in the range $0.02 - 0.09$ for various lattice sizes, see \cite{GarciaPerez:2000aiw} for a detailed discussion. 

The existence of a consistent solution in the form  (\ref{expansion}) to all orders in $\Delta$ has been shown in \cite{GarciaPerez:2000aiw}, but explicit calculations beyond the leading order have not yet been performed.  The radius of convergence in $\Delta$ is also not known. In what follows, we assume that the all-order $\Delta$-expansion gives rise to a unique self-dual solution 
of action $S_0 = {4\pi^2 \over g^2}$, at least for sufficiently small $\Delta$. This is supported by the consistency of  (\ref{expansion}) and the agreement with numerical tests \cite{GarciaPerez:2000aiw}. 

\subsection{Fermion zero modes}
\label{sec:fermionzero}

The fermion zero modes can be found by solving the Weyl equations from (\ref{classical EOM}),  $ D_n \bar \sigma_n \lambda=0\,, D_n \sigma_n \bar \lambda=0$ in the background (\ref{generalbkgrdbulk}).  We perform these explicit computations in Appendix \ref{appx:fermionviadirac}. 

The fermion zero modes can also be obtained via the supersymmetry transformations (\ref{susyhermitean1bulk}), with the result agreeing with the direct solution of the Weyl equations.
Consider the effect of the supersymmetry transformation (\ref{susyhermitean1bulk}) in the background of the bosonic solution (\ref{generalbkgrdbulk}), with fermions set to zero, $\bar\lambda = \lambda = 0$. Since our solution  is self-dual, i.e. obeys $\bar\sigma^{mn} F_{mn} = 0$, the SUSY transformation only produces $\lambda_\alpha$ variations:
\begin{eqnarray}
\delta\lambda=-\sigma_{mn} F_{mn}\zeta\,,\quad \delta\bar\lambda=0\,,
\label{susy transform for zero modes}
\end{eqnarray}
which is true to all orders in $\Delta$.
 Computing $\delta\lambda$  we obtain, see Appendix \ref{appx:fermionsusy}, to ${\cal O}(\sqrt{\Delta})$:
  \begin{eqnarray}
 \left(\begin{array}{c}  \delta\lambda_1 \cr  \delta\lambda_2 \end{array} \right) &=&-  2 i F_{12}^{(0)} \left(\begin{array}{c}\zeta_1 \cr -\zeta_2 \end{array} \right)  + 2  F_{13}^{(1)}  \left(\begin{array}{c} \zeta_2 \cr -  \zeta_1 \end{array} \right) -i  2  F_{14}^{(1)} \left(\begin{array}{c}\zeta_2 \cr  \zeta_1 \end{array} \right) \nonumber\\
    &=&\left[ \left(\begin{array}{c}\eta_1 \cr \eta_2 \end{array} \right) {\tau^3 \over 2} + {V^{1 \over 4} \over \pi^{1 \over 2}} \gamma^* e^{i \alpha} G(x,z)  \left(\begin{array}{c} \eta_2 \cr0 \end{array} \right) \tau^+     +{V^{1 \over 4} \over \pi^{1 \over 2}} \gamma e^{- i \alpha}  G^*(x,z) \left(\begin{array}{c} 0 \cr \eta_1 \end{array}  
    \right)\tau^-    \right]\,,\nonumber \\ \label{lambdasusy4}
 \end{eqnarray}
 where in going from the first to second line we used the definitions $\eta_1\equiv 4\pi V^{-1/2}\zeta_1$, $\eta_2 \equiv -4\pi V^{-1/2}\zeta_2$.
 Thus, as expected, we find $2$ fermion zero modes in accordance with the index theorem. One can easily see that $\mbox{tr}\bar\lambda \bar\sigma_n[T^A, \lambda]=0$, and hence, the self-dual instanton (\ref{generalbkgrdbulk}) on the asymmetric $\mathbb T^4$ solves the sourceless equations of motion $D_m F_{mn}=0$ and is a consistent background for SYM. 
 
 \subsection{Bosonic zero modes}
 
 For every fermionic zero mode  
 \begin{eqnarray}\phi_\alpha^{\; (\beta) \; A}=-(\sigma_{mn})_{\alpha}^{(\beta)}F_{mn}^A ,  \beta=1,2 ,  A=1,2,3 , 
\label{zeromodephi}
 \end{eqnarray}there are two bosonic zero modes. Thus, in total, there are four independent bosonic zero modes. The advantage of the discussion that follows is that the bosonic zero 
modes automatically obey the background gauge condition and, furthermore, their construction holds to arbitrary orders in $\Delta$.
 
 The four-vector expressions for the bosonic zero modes are denoted by $Z_n^{(\beta) \; A}$ and $Z_n^{(\beta \; ') \; A}$, where $\beta,\beta' = {1,2}$. These modes are determined as  described in e.g. \cite{Vandoren:2008xg,Dorey:2002ik}: from  each zero mode $\phi_\alpha^{(\beta)}$ of the undotted Dirac equation one builds two four-vector bosonic zero modes, denoted by $Z_n^{(\beta)}$ and $Z_n^{(\beta \; ')}$  (from here on, we suppress the Lie-algebra index $A$ to reduce clutter). Their four-vector components    are then  (see  Appendix \ref{appx:bosonicmodes}, eqn.~(\ref{zeromodesfromdirac})):
\begin{eqnarray}\label{fourvectorzeromodesbulk}
Z_n^{(\beta)} &=& \left\{ \Im \phi_2^{(\beta)}, - \Re \phi_2^{(\beta)},  \Im \phi_1^{(\beta)} , \Re \phi_1^{(\beta)}\right\}, ~
Z_n^{(\gamma \; ')} =  \left\{ \Re \phi_2^{(\gamma)}, \Im \phi_2^{(\gamma)}, \Re \phi_1^{(\gamma)}, - \Im \phi_1^{(\gamma)} \right\}. \nonumber
\end{eqnarray}
Using the expression of $\phi_\alpha^{(\beta)}$ in terms of $F_{mn}$, we can also express $Z_n^{(\beta)}$ and $Z_n^{(\gamma \; ')}$  in terms of $F_{mn}$:
\begin{eqnarray}\label{fourvectorzeromodes12bulk}
 Z_n^{(1)} &=& \left\{ \Im \phi_2^{(1)}, - \Re \phi_2^{(1)},  \Im \phi_1^{(1)} , \Re \phi_1^{(1)}\right\} = \left\{ - 2 F_{14},  2 F_{13},  - 2 F_{12}, 0\right\},\nonumber \\
 Z_n^{(2)} &=& \left\{ \Im \phi_2^{(2)},  -\Re \phi_2^{(2)},  \Im \phi_1^{(2)} , \Re \phi_1^{(2)}\right\}=  \left\{2 F_{12}, 0, -2 F_{14}, 2 F_{13}\right\},\nonumber \\
 Z_n^{(1 \; ')} &=&   \left\{ \Re \phi_2^{(1)}, \Im \phi_2^{(1)}, \Re \phi_1^{(1) },  -\Im \phi_1^{(1)} \right\} = \left\{ -2 F_{13},  -2 F_{14}, 0, 2 F_{12}\right\},\nonumber \\
  Z_n^{(2\; ')} &=&    \left\{ \Re \phi_2^{(2)}, \Im \phi_2^{(2)}, \Re \phi_1^{(2)},  \Im \phi_1^{(2)} \right\} =  \left\{0, 2 F_{12}, 2 F_{13}, 2 F_{14}\right\}\,.
\end{eqnarray}  
It is a simple exercise to check $D_nZ_n^{(\beta)} =0$ and $D_nZ_n^{(\gamma \; ')}=0$, and thus, these zero modes solve the classical equations of motion as expected. 

Further, one can write down $ Z_n^{(\beta)}$ and $ Z_n^{(\gamma')}$ as the derivatives with respect to the collective coordinates $\{z_i\}$ modulo gauge transformations. To this end, we define new zero modes $Y_n^{(i)}$, $i=1,2,3,4$, by relabeling (\ref{fourvectorzeromodes12bulk}) as follows
\begin{eqnarray}
Y_n^{(3) } \equiv Z_n^{(1)}\,,\quad Y_n^{(1)}\equiv -Z_n^{(2)} \,,\quad Y_n^{(4)}\equiv - Z_n^{(1 \; ')} \,,\quad
Y_n^{(2) }\equiv - Z_n^{(2\; ')}\,,
\end{eqnarray}  
and show in Appendix \ref{derivatives} that
\begin{eqnarray}
Y_n^{(k)} =\frac{4\pi L_k}{\sqrt{V}}\frac{\partial A_n}{\partial z_k}+D_n(A^{cl}) \Lambda^{(k)}\,.
\label{new zero modes}
\end{eqnarray}  
There, we also explicitly find the expressions for the background-gauge restoring gauge transformations $\Lambda^{(k)}$ to order ${\sqrt{\Delta}}$ and show that they obey boundary conditions preserving the transition functions. For later convenience,  we also define the inner product  of the zero-mode wave functions (or moduli space metric)
 \begin{equation}\label{bosonicproductbulk}
 U^{kl} = {2 \over g^2} \int_{\mathbb T^4} \tr Y_n^{(k)} Y_n^{(l)}
 \end{equation} 
 which in diagonal basis is simply  $U^{kl}=  \delta_{lk} u_l$. Using (\ref{fourvectorzeromodes12bulk}), see  Appendix \ref{appx:bosonicmodes} for details, we readily obtain
  \begin{equation}\label{ugeneralbulk}
 U^{kl} =\frac{4 \delta_{kl}}{g^2} \int_{\mathbb T^4} \tr F_{mn}\tilde F_{mn}=4\delta_{kl}S_0= \delta^{kl}  \; {16 \pi^2 \over g^2}\,.
 \end{equation}
This result is valid to all orders in $\Delta$, since the action of a fractional instanton does not depend on the size or shape of $\mathbb T^4$.

\section{The path integral: bosonic and fermionic measures}
\label{The path integral: bosonic and fermionic measures}

The contributions from the non-zero modes of fermions, bosons, and ghosts cancel in the path integral, thanks to supersymmetry.  Thus, in our subsequent discussion, we only discuss the contribution from zero modes to the bosonic and fermionic measures.

\subsection{Fermionic measure}
\label{Fermionic measure}

The fermion zero modes measure will be inferred from the one for the non-zero modes. We expand the fermions as eigenfunctions of the second order Hermitian operators
  \begin{eqnarray}\label{secondorder2bulk}
D\bar D &=& D^2 + i F_{mn} \sigma^{mn}~, ~~~~~ ~ - (D \bar D)_\alpha^{~\beta} \lambda_\beta = \omega^2 \lambda_\beta ~\\
\bar D D &=& D^2 + i F_{mn} \bar\sigma^{mn}~ = D^2,  ~~- D^2 \bar\lambda^{\dot\beta} = \omega^2  \bar\lambda^{\dot\beta}~,\nonumber 
\end{eqnarray}
where, in the second line, we used the self-duality of the background. 

To discuss the measure,  we begin by considering the contribution of a single nonzero eigenvalue $\omega$ to the fermion path integral. Let $- (D\bar D)_\alpha^{\;\; \beta} \phi_\beta^i = \omega^2 \phi_\alpha^i$, where $i$ labels the different eigenfunctions, the commuting functions $\phi_\alpha^i$, with the same eigenvalue $\omega$ (we note that there are at least two of them). We expand the nonzero-mode part of the fermion field (for brevity, denoting it with the same letter $\lambda$, $\bar\lambda$) 
 \begin{eqnarray}\label{lambdaexpansionbulk}
 \lambda_\alpha &=& \sum_i \chi^i \; \phi^i_\alpha \\
 \bar\lambda^{\dot\alpha} &=& \sum_i \bar\chi^i  \; {1 \over \omega} \bar D^{\dot\alpha \alpha} \phi_\alpha^i, \nonumber
 \end{eqnarray}
 where we used the fact that the nonzero eigenfunctions of $D\bar{D}$ and $\bar{D} D$ are related as shown and we attach the spinor index to the bosonic solution of the 2nd order equation and not to the Grassmann variable, $\chi^i$ or $\bar\chi^i$ (the fact that there is more than a single solution for every $\omega$ is accounted by the index $i$). 
 We also indicate that the $\lambda$ and $\bar\lambda$ expansions have each their separate Grassmann variables $\chi^i, \bar\chi^i$.
 
Plugging (\ref{lambdaexpansionbulk}) into the fermionic action (\ref{total action}), we obtain after integration by parts and using the fact that $\phi$ is an eigenvector of $D \bar{D}$: 
\begin{eqnarray}\label{fermionproductbulk}
S_F=  {2 \over g^2} \tr (D_n \bar\lambda_{\dot\alpha} \bar\sigma_n^{\dot\alpha \alpha} \lambda_\alpha) &=& \sum_{ij} \bar\chi^j \chi^i \omega \; ({2 \over g^2}  \int_{\mathbb T^4} \tr \phi^{i \alpha} \phi^j_\alpha) \\
 &=& \omega  \sum_{ij} \bar\chi^j \chi^i \; U_F^{ij}~, ~\nonumber
 \end{eqnarray}
where the fermion mode inner product matrix is 
\begin{eqnarray}\label{fermionnormdefinitionbulk}
  U_F^{ij} = {2 \over g^2} \int_{\mathbb T^4} \tr(\phi^{i}_2 \phi^j_1 - \phi^{i}_1 \phi^j_2), ~ U_F^{ij} = - U_F^{ji}.
  \end{eqnarray}
 Then we define the fermion nonzero mode path integral so that it produces $\omega$ (the minimal number of eigenfunctions with the same eigenvalue is two, i.e. $i,j=1,2$, with $U_F$ generically a $2 \times 2$ matrix)
 \begin{equation}\label{nonzeromodemeasurebulk}
\int \prod_{i} d \chi^i d \bar\chi^i \; (\det U_F)^{-1} e^{-S_F}  = \omega~.
\end{equation}
When all nonzero eigenvalues are taken into account, we obtain the square root of the product over all nonzero eigenvalues of $\bar{D} D$ (or $D \bar D$).
 
 The  integrals over the fermion zero modes are defined in the same manner via the same mode normalization matrix, $U_F^{ij}$, defined in (\ref{fermionnormdefinitionbulk}). Recalling that only the undotted spinors $\lambda_\alpha$ have zero modes, we expand\footnote{Here and below, we use $\phi^i_\alpha$ to denote the zero-mode solutions of $D \bar D$,  obeying $(D\bar D)_\alpha^{\;\; \beta} \phi_\beta^i = 0$. The reader should forgive us for using the same letter as in the non-zero mode discussion near (\ref{lambdaexpansion}).}
 \begin{equation}\label{fermionzeromodeexpansionbulk}
 \lambda_\alpha = \sum_i \eta^i \phi^i_\alpha + \text{nonzero modes},
 \end{equation}
 where we use $\eta^i$ to denote the zero-mode Grassmann variable. The fermion zero-mode measure is then taken to be the ``square root'' of (\ref{nonzeromodemeasurebulk}):\footnote{This definition ensures that, upon perturbing with a zero-mode lifting mass term, $\delta S_{m}= {m \over g^2} \tr \lambda^\alpha \lambda_\alpha$, one obtains $m$ for the zero-mode contribution to the path integral.}
  \begin{equation}\label{fermionzeromodemeasurebulk}
d \mu_F =  \prod_i d \eta^i  \; (\det U_F)^{-1/2}~  = \prod_i d \eta^i  \; (\text{Pf} U_F)^{-1}~.
 \end{equation}

 In  Appendix \ref{appx:fermionmodemeasure}, we calculate the Pfaffian, using the zero-mode wave functions (\ref{zeromodephi}), and show that $\text{Pf} U_F= - U_F^{12} = 4 \times {4 \pi^2 \over g^2}$, see eqn.~(\ref{fermionmodematrixlast}). Notice that  
 this  result is valid to all orders in $\Delta$ (thus including the nonabelian part of the zero modes), since $\text{Pf} U_F$ involves integrating the square of the field strength $F_{mn}$ over $\mathbb T^4$, which is proportional to the action, a $\Delta$-independent quantity. 
 
 Furthermore, it follows from (\ref{fermionzeromodeexpansionbulk}) that   $\tr \lambda^\alpha \lambda_\alpha = {1 \over 2} \eta^\alpha \eta^\beta \phi^{(\alpha) A} _\gamma \phi^{(\beta) A}_\delta \epsilon^{\delta\gamma} +$(nonzero modes). Thus, combining with (\ref{fermionzeromodemeasurebulk}), one finally finds:
\begin{eqnarray}
\int d \mu_F \tr\lambda\lambda(x)&=& {g^2 \over 16 \pi^2  } \int d\eta_1d \eta_2 \tr[\lambda^\alpha\lambda_\alpha(x)]  \nonumber \\
&=&
 {g^2 \over 16 \pi^2}  {1 \over 2} \;   ( \phi^{(2) A} _\gamma  \phi^{(1) A}_\delta   -\phi^{(1) A} _\gamma  \phi^{(2) A}_\delta )[x]\epsilon^{\delta\gamma}~.
\label{uf int result}
\end{eqnarray}
We next note that all gauge invariant quantities in the fractional instanton background depend on the combinations of $x_i$ with the dimensionless collective coordinates $z_l$, see eqn.~(\ref{hatz}) in Appendix \ref{constructing}. Thus, to calculate the condensate, in eqn.~(\ref{uf int result}) one should replace
\begin{eqnarray}\label{replace}
 x_1 \rightarrow x_1 - {L_1 z_2 \over 2 \pi}, x_2 \rightarrow x_2 + {L_2 z_1 \over 2 \pi}, x_3 \rightarrow x_3 - {L_3 z_4 \over 2 \pi}, x_4 \rightarrow x_4 + {L_4 z_3 \over 2 \pi}.
\end{eqnarray}
That all gauge invariant quantities depend on (\ref{replace}) follows from the actions of translations in our background and is explained in  Appendix \ref{constructing} (see Footnote \ref{hatzdependence} there).

\subsection{Bosonic measure}
\label{bosonic measure}

We express the bosonic field $A_n(x)$,  to be integrated over in the path integral, as a sum of the classical solution $A_n^{cl}(x,z)$ (\ref{generalbkgrdbulk}), the zero mode fluctuations normalized as in (\ref{Ynk1}) of Appendix \ref{derivatives}, and the nonzero modes of the fluctuation operator denoted by $Z_n^{q}$  (of eigenvalues $\omega_q$) using the same notation as in Appendix \ref{leadingzeromode}, see the discussion after eqn.~(\ref{bosonicaction}):
\begin{eqnarray}\label{bosonicvariable1}
A_n(x) &=& A_n^{cl}(x,z)  + \sum\limits_{k=1}^4 \zeta_k^{(0)}  {\sqrt{V} \over 4 \pi L_k} ~ Y_n^{(k)}(x,z) + \sum_q \zeta_q Z_n^{(q)}\,.
\end{eqnarray}
Using the inner product defined in (\ref{ugeneralbulk}), the measure of the bosonic zero modes takes the form 
\begin{eqnarray}
\label{mub1bulk}
d \mu_B = 
\left( \det {V \over 16 \pi^2 L_k L_l} \; U^{kl}\right)^{1\over 2}\; \prod\limits_{k=1}^4 \left[{d \zeta_k^{(0)} \over \sqrt{2 \pi}} \right]
={V \over g^4}  \prod\limits_{k=1}^4 {d \zeta_k^{(0)} \over \sqrt{2 \pi}}\,. 
\end{eqnarray}
In Appendix \ref{jacobian}, we show how we change the variables $\zeta_k^{(0)}$ to the collective coordinates $\{z_m\}$ by inserting a unity, \'a la Faddeev-Popov gauge-fixing method. The resulting expression is
\begin{eqnarray}
\label{mub2bulk}
d \mu_B  
&=&{V \over g^4}  \prod\limits_{k=1}^4 {d z_k \over \sqrt{2 \pi}} ~.
\end{eqnarray}
%

\section{The Hamiltonian formalism, Wilson loops, and the moduli space}
\label{The Hamiltonian formalism, Wilson loops, and the moduli space}

Up to this point, we have all pieces to compute the gaugino condensate except for the shape and size of the moduli space ${\cal M}$, or in other words, the range of integration over the collective coordinates $\{z_m\}$. Determining ${\cal M}$ will force us to deviate, for now, from the path integral to the Hamiltonian formalism. We refer the reader to \cite{Cox:2021vsa} for a detailed description of the Hilbert space, while here, we only provide a synopsis needed to study the Wilson loops and moduli space.

\subsection{Pure Yang-Mills theory}
\label{sec:pureymhamiltonian}

The moduli space is determined in the absence of fermions, and thus, we start by studying the Hamiltonian formalism of pure Yang-Mills theory with vacuum angle $\theta$ on $\mathbb T^3$ with a unit 't Hooft magnetic flux. 

Consider $L_{1,2,3}$ of $\mathbb T^3$ as space of volume $V_3$ and $L_4$ as Euclidean time, where  $V_3$ is to be taken much smaller than the inverse of $\Lambda^3$, the strong-coupling scale, while $L_4$ can be varied from small to large w.r.t. $\Lambda^{-1}$. Recall that we apply the twists $m_3\equiv n_{12}=1$, this is a ``magnetic flux'' piercing the 1-2 plane, and $k_3\equiv n_{34}=1$. The latter is related to a twist of the partition function by a centre symmetry transformation, to be described below. The physical Hilbert space lives along the constant time slices and, thus,  is in the $m_3=1$ ``magnetic flux'' background. Let $|\psi\rangle$ denote a state in the physical Hilbert space on $\T^3$ with twisted boundary conditions, ${\cal H}_{m_3=1}$. 

We introduce the operator $\hat T_3$ as the generator of the $\mathbb Z^{(1)}_2$ $1$-form center symmetry in the $L_3$ direction. The generator $\hat T_3$ commutes with the Hamiltonian, $[\hat T_3, \hat H]=0$, and thus, they can be diagonalized in the same basis.   It can be shown that the action of $\hat T_3$ on its eigenstates, $|\psi \rangle = |e_3\rangle$, in the physical Hilbert space is $\hat T_3 |e_3 \rangle=e^{i\pi e_3-i\frac{\theta}{2}}|e_3\rangle$, where $e_3\in\{0,1\}$ is the $\mathbb Z_2$ ``electric flux'' of the state. This terminology can be explained as follows.  Since $\hat T_3$ is a center-symmetry generator, it acts on Wilson loops winding around the $L_3$ direction as $\hat T_3 \hat W_3=-\hat W_3 \hat T_3$. Then, one readily finds $\hat T_3 (\hat W_3 |e_3 \rangle)=e^{i\pi (e_3+1)-i\frac{\theta}{2}}(\hat W_3 |e_3 \rangle)$, i.e., the action of $\hat W_3$ on a state increases the $\mathbb Z_2$ flux of this state by one unit. The fact that  $\hat W_3$ can be thought of as creating a winding electric flux tube explains the terminology. We conclude that $\hat T_3$ measures the center flux of a given state in Hilbert space.  The generators of the center symmetries in the other two spatial directions, $x_1$ and $x_2$, $\hat T_{\ell}$, $\ell = 1,2$, act similarly on Wilson loops winding in the corresponding directions. They also commute with the Hamiltonian and thus all eigenstates of $\hat{H}$ are also labeled by $e_{1}\in\{0,1\}$ and $e_{2}\in\{0,1\}$.\footnote{The action of $\hat T_1$ on states of electric flux $e_1$ is $\hat T_1 | e_1 \rangle = e^{i \pi e_1} |e_1\rangle$ (and similar for $\hat T_2$), without the $\theta/2$ factor in the action of $\hat T_3$ (which is due to the $m_3=1$ twist).}

The twisted partition function of pure Yang-Mills theory, the one we have been studying in the absence of fermions, is the one with an insertion of a $\hat T_3$ twist:
\begin{eqnarray} \label{twistedz}
\nonumber
 {\cal Z}_{YM}[n_{12}=1, n_{34}=1]&=& {\text{tr}}_{{\cal{H}}_{m_3 = 1}} \left[ e^{- L_4 \hat H} \hat T_3 \right]\\
 \nonumber
 &=&\sum_{E(e_3),\, e_3=\{0,1\}}\langle E(e_3), e_3| e^{- L_4 E(e_3)} \hat T_3| E(e_3), e_3 \rangle\\
 &=&\sum_{E(e_3),\, e_3=\{0,1\}}\langle E(e_3), e_3| e^{- L_4 E(e_3)}| E(e_3), e_3 \rangle e^{i\pi e_3-i\frac{\theta}{2}}\,,
\end{eqnarray}
where in going from the first to the second line, we summed over a complete set of eigenstates in the physical Hilbert space  ${\cal H}_{m_3=1}$ that diagonalize $\hat H$ and $\hat T_3$ simultaneously (for brevity, omitting the summation over $e_{1,2}$). 
The expectation value of the Wilson loop operator $\hat W_3$ is\footnote{We ignore the normalization of Wilson loops. We will come back to normalization in calculating the gaugino condensate.} $\langle W_3 \rangle={\text{tr}}_{{\cal{H}}_{m_3 = 1}} \left[ e^{- L_4 \hat H} \hat W_3\hat T_3 \right]$. Using the relation $\hat T_3 \hat W_3=- \hat W_3 \hat T_3$ and the fact that $\hat T_3$ (but not $\hat W_3$) commutes with $\hat H$, we immediately find $\langle \hat W_3 \rangle=-\langle \hat W_3 \rangle$, and hence, $\langle\hat W_3 \rangle=0$.  

One can also show that the expectation value of the Wilson loop winding in any other direction vanishes identically.  Since $\hat T_{\ell}$, $\ell = 1,2,3$ and $\hat H$ form a set of commuting operators, they can be diagonalized simultaneously. Let $\{|E(e_1,e_2,e_3), e_1,e_2,e_3\rangle\}$ be a set of orthonormal eigenstates of the set of the $4$ operators $\hat H, \{\hat T_\ell, \ell=1,2,3 \}$, where $e_\ell\in \{0,1\}$ is the $\mathbb Z_2$ electric flux in the $L_\ell$ direction, and we emphasized that, in general, the energy of the state depends on these fluxes. The expectation value of a general Wilson loop $\hat W_p$ wrapping the $L_p$ direction and computed using the $\hat T_3$-twisted partition function (\ref{twistedz}) reads 
\begin{eqnarray}
\langle \hat W_p \rangle&=&\sum_{E(e_1,e_2,e_3),\,e_i=\{0,1\}} e^{-L_4E}\langle E(e_1,e_2,e_3),e_1,e_2, e_3 |\hat W_p\hat T_3|E(e_1,e_2,e_3),e_1,e_2, e_3\rangle \\
&=&\sum_{E(e_1,e_2,e_3),\,e_i=\{0,1\}} e^{-L_4E}\langle E(e_1,e_2,e_3),e_1,e_2, e_3 |\hat W_p|E(e_1,e_2,e_3),e_1,e_2, e_3\rangle e^{i\pi e_3-i\frac{\theta}{2}}\,. \nonumber
\end{eqnarray}
However, the insertion of $\hat W_p$ increases the electric flux by one unit in the $L_p$ direction and produces a sum over different energy eigenstates. Thus, its diagonal matrix element in the $|E(e_1,e_2,e_3),e_1,e_2, e_3\rangle$ state vanishes. We immediately conclude that a Wilson loop that wraps $L_1$, $L_2$, or $L_3$ must vanish on $\mathbb T^4$ with twisted boundary conditions.  The vanishing of $\langle \hat W_4\rangle$ follows by applying a $90$-degree rotation to any spatial Wilson loop, i.e. upon considering a different $\T^4$ direction as time. Thus, we conclude that for any winding Wilson loop $\hat W_p$,
\begin{eqnarray} \label{Wptwisted}
\langle \hat W_p \rangle={\text{tr}}_{{\cal{H}}_{m_3 = 1}} \left[ e^{- L_4 \hat H} \hat W_p\hat T_3 \right] = 0.
\end{eqnarray}
We spent so much time explaining the expected result (\ref{Wptwisted}) because its consistency with semiclassics is one of our main criteria used to determine the moduli space of the fractional instanton.

Thus, we now contrast the general result (\ref{Wptwisted}) with the computation of the expectation value of a Wilson loop in the path integral formalism, in the semiclassical approximation. Consider a Wilson loop $ W(C_{n_1,n_2,n_3,n_4})$, with $C$ beginning  at an arbitrary point $x$ in $\mathbb T^4$ and winding $n_\ell$ times around each direction $L_\ell$:
\begin{eqnarray} \label{wilsonopr}
 W(x,C_{n_1,n_2,n_3,n_4}) = \tr \left( {\cal{P}} e^{i \int_{C_{n_1,n_2,n_3,n_4}} {A}_k (x') dx'_k}\;   \Omega_2^{n_2} (x) \; \Omega_4^{n_4} (x) \right), 
\end{eqnarray}
where we inserted $ \Omega_2^{n_2} (x) \; \Omega_4^{n_4} (x)$ to enforce the gauge invariance of $ W(x,C_{n_1,n_2,n_3,n_4})$. Using the classical self-dual background (\ref{generalbkgrdbulk}),(\ref{orderdeltabackgroundbulk}), we show in eqn.~(\ref{wtoorderdelta}) in the Appendix,  that $W$ to order $\Delta$ is:
\begin{eqnarray}
\nonumber
 &&W(x,C_{n_1,n_2,n_3,n_4}) \\
 &=&  2 \cos\left[{1\over 2}\left(n_1 (z_1 + {2\pi x_2 \over L_2}) + n_2 (z_2 - {2 \pi x_1 \over L_1}) + n_3 (z_3 + {2\pi x_4 \over L_4}) + n_4 (z_4 - {2\pi x_4 \over L_2})\right)   \right]. \nonumber \\
 &&\times \left[1+\Delta {\cal F}(x,z)\right]\,.
  \label{windingwilsonzero1bulk}
 \end{eqnarray}
The ${\cal{O}}(1)$ and ${\cal{O}}(\Delta)$ contributions come from the abelian and nonabelian components of (\ref{generalbkgrdbulk}). The cosine function has $4\pi$ periodicity in $\{z_i\}$, while the ${\cal{O}}(\Delta)$ piece ${\cal F}(x,z)$ is a periodic and even function of $\{z_1 + {2\pi x_2 \over L_2}, z_2 - {2 \pi x_1 \over L_1}, z_3 + {2\pi x_4 \over L_4}, z_4 - {2\pi x_4 \over L_2}\}$ with periodicity $2\pi$ for $\{z_i\}$. 

Now, using the results from the previous section and limiting our discussion to pure Yang-Mills theory and ignoring issues of normalization, the expectation value of a general Wilson loop is%
\begin{eqnarray}
\nonumber \label{wilsonlooplimits}
&&\langle  W(x,C_{n_1,n_2,n_3,n_4}) \rangle=\sum_{\nu}\int [D A_\mu] W(x,C_{n_1,n_2,n_3,n_4})e^{-S_{YM}-i\theta(\nu + {1\over 2})}|_{n_{12}=1, n_{34}=1}\\
\nonumber
&&\sim e^{-S_0 - i {\theta \over 2}} {V \over g^4}  \int_{\cal M} \prod\limits_{k=1}^4 {d z_k \over \sqrt{2 \pi}} \; W(x,C_{n_1,n_2,n_3,n_4})+ e^{-S_0+i{\theta \over 2}} {V \over g^4} \int_{\cal M}  \prod\limits_{k=1}^4 {d z_k \over \sqrt{2 \pi}} \; \tilde W(x,C_{n_1,n_2,n_3,n_4})\,,\\
\end{eqnarray}
where in going from the first to the second line, we ignored the quantum loops and used the bosonic zero-mode measure (\ref{mub2bulk}). We limited the r.h.s. to the contributions of the $\nu=0$ sector, with total topological charge $Q=\frac{1}{2}$, and the $\nu=-1$ sector, with total topological charge $Q=-\frac{1}{2}$. Ignoring higher-$Q$ contributions is justified in the limit when semiclassics holds, i.e. for a small $\T^4$. 

Furthermore, in writing the above expression, we assumed  that the fractional instantons (and antinstantons\footnote{$\tilde W$ in the second term is the Wilson loop (\ref{wilsonopr}), but evaluated in the anti-instanton background, whose explicit form is similar to (\ref{windingwilsonzero1bulk}) and shall not be needed.}) we obtained are the only $|Q| = {1\over 2}$ saddle points in the path integral, contributing to the first and second terms above, respectively.  While we have no proof of uniqueness, the favourable comparison of ``exact'' (i.e. obtained by numerically minimizing the action) fractional instantons on an asymmetric $\T^4$ with the solutions obtained by small-$\Delta$ expansion gives  support in favor of this assumption, at least for small enough $\Delta$. In this regard, we note that ref.~\cite{GarciaPerez:2000aiw} found good agreement for $\Delta$ as large as $0.08$.

Our point now is that the $d^4 z$ integral in (\ref{wilsonlooplimits}) vanishes, and is thus consistent with (\ref{Wptwisted}), for all values of $\theta$ and for all $x$, if and only if the limits of integration are taken $z_i\in [0, 4\pi)$. This leads us to conclude that the moduli space ${\cal M}\sim \mathbb T^4$ with period $4\pi$ in every direction.\footnote{Moreover, it is easy to check that changing the limits of integration to any other values yields a position-dependent Wilson loop. For example, consider $\langle W(x,C_{n_1=1,n_2=0,n_3=0,n_4=0})\rangle$ and take the range of $z_i\in [0, 2\pi)$, which yields $\langle W(x,C_{n_1=1,n_2=0,n_3=0,n_4=0})\rangle\sim -\sin\left(\pi \frac{x^2}{L}\right)+{\cal O}(\Delta)$. This result should not be expected on physical grounds since our background has constant field strength on $\mathbb T^4$ (to leading order in $\Delta$); the expectation value of the Wilson loop should, at most, be a constant (to leading order in $\Delta$).}

A further argument in favour of this identification of the moduli space is that the classical fractional instanton field configurations with $z_n$ differing by $2 \pi$ are distinguished by the gauge invariant winding Wilson loop operators, while those differing by $2 \pi$ are not distinguished. Thus one expects that   $z_n$ and $z_n + 2 \pi$ are not to be identified. In contrast, no local gauge invariant operators can distinguish between $z_n$ and $z_n + 2 \pi$, see Section \ref{gaugeinvariants} in the Appendix for detail.\footnote{\label{sunfootnote}We   note, without discussing the details, that a virtually identical  argument for extending the    limits of $z_n$ integration holds also in the $SU(N)$ case in the background of the $Q=1/N$ solutions found by 't Hooft \cite{tHooft:1981nnx}, recently extended in the framework of the small-$\Delta$ expansion \cite{Gonzalez-Arroyo:2019wpu}.}

\subsection{SYM theory} 

 Before turning to the calculations of the gaugino condensate using the path integral method, we pause here to discuss SYM theory using the Hamiltonian formalism, akin to our discussion of pure Yang-Mills of Section \ref{sec:pureymhamiltonian}. The partition function of SYM with twisted boundary conditions in the 1-2 and 3-4 planes, identical for bosons and fermions, is given by 
\begin{eqnarray}
{\cal Z}_{SYM}[n_{12}=1,n_{34}=1]=\mbox{tr}_{{\cal H}_{m_3=1}}\left[(-1)^Fe^{-L_4\hat  H}\hat T_3\right]\,. \label{twistedZ1}
\end{eqnarray}
The insertion of the fermion number $(-1)^F$ guarantees that both fermions and bosons obey periodic boundary conditions in the time direction.  Let $\hat X$ be the $\mathbb Z_4^{d\chi}$ discrete chiral symmetry generator. Then, it can be shown that the symmetry operators $\hat T_3$ and $\hat X$ obey the algebra:
\begin{eqnarray}
\left[\hat H, \hat T_3\right]=0\,,\quad \left[\hat H, \hat X\right]=0\,, \quad \hat T_3 \hat X=e^{i\pi}\hat X \hat T_3\,,
\label{SYM algebra}
\end{eqnarray}
where the last relation is the result of the mixed anomaly between the $0$-form discrete chiral and $1$-form center symmetries \cite{Cox:2021vsa}.\footnote{The same algebra arises in the 2-D massless charge-$2$ Schwinger model \cite{Anber:2018jdf}, due to a similar anomaly.} Since $\hat H$ commutes with $\hat T_3$, we can as before, label the physical states in ${\cal H}_{m_3=1}$ by $|E(e_3),e_3\rangle$. The algebra (\ref{SYM algebra}) requires that $\hat X|E(e_3),e_3\rangle=|E(e_3), e_3-1\rangle$, and thus, it is easily seen the states $|E(e_3), e_3=1\rangle$ and $|E(e_3), e_3=0\rangle$ are degenerate: $\hat H|E, e_3=0\rangle=E(e_3=0)|E, e_3=0\rangle$, $\hat H|E, e_3=1\rangle=E(e_3=0)|E, e_3=1\rangle$, at any size of the spatial $\T^3$, as a consequence of the anomaly.  

Now, we can readily calculate the partition function (\ref{twistedZ1}) to find
 \begin{eqnarray}
 \nonumber
{\cal Z}_{SYM}[n_{12}=1,n_{34}=1]&=& {\text{tr}}_{{\cal{H}}_{m_3 = 1}}\left[ (-1)^F e^{- L_4\hat H} \hat T_3\right]\\
\nonumber
 &=& \sum\limits_{E,\, e_3 = \{0,1\}} (-1)^F (-1)^{e_3} \langle E, e_3 | e^{- E L_4} | E, e_3 \rangle \\
 &=& \sum\limits_{E,\, e_3 =\{ 0,1\}} (-1)^F (-1)^{e_3}  e^{- E L_4} = 0~. \label{twistedZ}
 \end{eqnarray}
 This equation provides one way to see that the twisted partition function of SYM vanishes, by noting that all states in Hilbert space are doubly degenerate and their contributions cancel each other in the twisted partition function (\ref{twistedZ}). 
 
 Now we insert the gaugino condensate, the $\tr \lambda^2$ operator, where the trace is taken in the color space. The operator is inserted at $x_4 =0$ and some position on $\T^3$:
  \begin{eqnarray}
  \nonumber
 \langle \tr \lambda \lambda \rangle &=&  {\text{tr}}_{{\cal{H}}_{m_3 = 1}}\left[ (-1)^F e^{- L_4 \hat H}  \tr\lambda^2 \; \hat T_3\right] \over {\text{tr}}_{{\cal{H}}_{m_3 = 1}}\left[ (-1)^F e^{- L_4 \hat H} \right] \\
 &=& {\cal N}^{-1}\sum\limits_{E,\, e_3 =\{ 0,1\}} (-1)^F (-1)^{e_3} e^{- E L_4} \langle E, e_3 |  \tr\lambda^2  | E, e_3 \rangle\,. \label{trlambdanormalized}
 \end{eqnarray}
Here, we took the normalization constant  ${\cal N}$ to be the Witten index: ${\cal N}={\text{tr}}_{{\cal{H}}_{m_3 = 1}}\left[ (-1)^F\right]=2$, given by the partition function on $\T^3$ with an $n_{12}$ twist, but  without a $\hat T_3$ twist in the time direction (recall (\ref{the main gaugino condensate normalization}) in the path integral framework).

Next, we know that $\hat X |E, e_3 \rangle = |E, e_3 + 1\rangle$ from the anomaly (\ref{SYM algebra}) and that $\hat X$ acts on the condensate as $\hat X \tr \lambda^2 = - \tr \lambda^2 \hat X$. Thus, because
 \begin{equation}\langle E, 0 |  \hat X^\dagger \hat X \tr\lambda^2  | E, 0 \rangle = - \langle E, 1 |  \tr \lambda^2  | E, 1 \rangle\,, \end{equation}  $\tr \lambda^2$  has opposite expectation values in  the two degenerate states $e_3=0$ and $e_3=1$. Thus, 
  we find
  \begin{eqnarray}
  \nonumber
 \langle \tr \lambda \lambda \rangle  &=&{\cal N}^{-1} \sum\limits_{E,\, e_3 = \{0,1\}} (-1)^F e^{- E L_4} \langle E, 0 |  \tr\lambda^2  | E, 0 \rangle_{V_3, n_{12}}\\
  &=&   \sum\limits_{E} (-1)^F e^{- E L_4} \langle E, 0 |  \tr\lambda^2  | E, 0 \rangle_{V_3, n_{12}}\,,\label{condensatefiniteL}
 \end{eqnarray}
 where the sum is over only the half of the Hilbert space with $e_3 = 0$ and 
 we are reminded that the condensate is being computed in a small $V_3$ theory Hilbert space with boundary conditions twisted by $n_{12}$.

 If $L_4$ is small compared to the energy scales in the problem, there is no reason to assume that any particular values of $E$ dominate. Thus, eqn.~(\ref{condensatefiniteL}) is the Hamiltonian expression of the expectation value we have computed for general $L_4$.
 However, we can try take the $L_4 \rightarrow \infty$ limit to obtain\footnote{Equating the result of our calculation, eqn.~(\ref{condensatefinal}) below, to the  r.h.s. of (\ref{lambdaresultlargetime}) is only possible if we assume that the semi-classical treatment holds as $L_4$ is varied between small and large, and, in particular, there are no new contributions that contribute to the l.h.s. This can be only justified by understanding the convergence radius of the $\Delta$ expansion.}
  \begin{eqnarray}\label{lambdaresultlargetime}
\langle \tr \lambda \lambda \rangle =     \langle 0, 0 |  \tr\lambda^2  | 0, 0 \rangle_{V_3, n_{12}}\,,
 \end{eqnarray}
 where we used that supersymmetry is unbroken and the small-$V_3$ ground state is bosonic, $(-1)^F = 1$.\footnote{We note that a similar equation holds in $SU(N)$ theories for one of the $N$ vacua of the twisted, small-$V_3$ theory. There ${\cal{N}}=N$ and there is $N$-fold degeneracy of all energy states on $\T^3$ with $n_{12}=1$.}
 
 In conclusion, eqn.~(\ref{lambdaresultlargetime}) shows that in the large-$L_4$ limit, the calculation of $\langle \tr \lambda^2 \rangle$ using the $\hat T_3$-twisted parition function and normalized by the Witten index, as in (\ref{trlambdanormalized}), yields the gaugino condensate in one of the two degenerate ground states of the  small-$V_3$ theory with a unit 't Hooft flux $n_{12}$. The state $|0,0 \rangle_{V_3, n_{12}}$ is further expected, as $V_3 \rightarrow \infty$, to become one of the two degenerate   ground states of the $SU(2)$ SYM theory on $\R^4$.

 \section{The gaugino condensate}
 \label{The gaugino condensate}
 
 Finally, we put pieces together to read out the condensate. We use the fact that non-zero modes in  (\ref{the main gaugino condensate}) cancel between the bosons, fermions, and ghosts. Then, we combine the integral of $\tr \lambda^2$ over the zero mode measure for the fermions from eqn.~(\ref{uf int result}) with the boson zero mode measure (\ref{mub2bulk}). We also recall that we normalize by the Witten index ${\cal N}=2$ to obtain: %
 \begin{eqnarray}
 \nonumber
 \langle \tr \lambda\lambda \rangle&=& {M_{PV}^3  e^{-\frac{4\pi^2}{g^2}}  \over {\cal N}}
 \; {V \over g^4} \\
 \nonumber
 &&\times
  \int_{z_i\in [0,4\pi]} \prod\limits_{k=1}^4 {d z_k \over \sqrt{2 \pi}} \; {g^2 \over 16 \pi^2}  {1 \over 2} \;   \epsilon^{\delta\gamma} ( \phi^{(2) A} _\gamma \phi^{(1) A}_\delta    -\phi^{(1) A} _\gamma  \phi^{(2) A}_\delta)[\{x\}]\bigg\vert_{x_1 \rightarrow x_1 - {z_2 L_1 \over 2 \pi},\; \text{see eqn.}\; (\ref{replace})}  
\nonumber  \end{eqnarray}
  \begin{eqnarray} \label{lambdalambda1}
  &=& {\Lambda^3 \over 2} {V \over 4 \pi^2} {1 \over 16 \pi^2} \int\limits_{-2\pi}^{2 \pi} d^4 z \; ( \phi^{(2) A} _2 \phi^{(1) A}_1    -\phi^{(1) A} _2  \phi^{(2) A}_1)[  - {z_2 L_1 \over 2 \pi},  {z_1 L_2 \over 2\pi},     - {z_4 L_3 \over 2\pi},   {z_3 L_4 \over 2 \pi}]. 
 \end{eqnarray}
To obtain the last line, we recalled the substitution of (\ref{replace}), the fact that all local gauge invariants are periodic functions of  $\hat z_k$ (defined in (\ref{replace})) of period $2\pi$, and used  shifts of the $z_n$ variables to set the $x_k$-coordinates (of period $L_k$) to zero. 
Further, Pauli-Villars regularization has been utilized to renormalize the theory, and the scale $M_{PV}$ is the Pauli-Villars mass. The factor $M_{PV}^3$ comes from regularizing the boson and fermion determinants, with each zero-mode  contributing an appropriate factor of $M_{PV}$:  the bosonic determinant gives $M_{PV}^4$ and the fermionic determinant gives $M_{PV}^{-1}$ (we do not describe this in detail, as it is a standard procedure in supersymmetric instanton calculations \cite{Davies:1999uw,Davies:2000nw,Dorey:2002ik,Vandoren:2008xg}).  In going from the first to the second line of (\ref{lambdalambda1}), we used  the fact that $\Lambda^3 = {M_{PV}^3 \over g^2} e^{-\frac{4\pi^2}{g^2}}$ in the scheme often  used for gaugino condensate calculations 
\cite{Dorey:2002ik,Vandoren:2008xg}, notably, in the weakly-coupled $\R^3 \times \S^1$ set up and its comparison with the weak-coupling calculation on $\R^4$ \cite{Davies:2000nw}.
 
Next, we rescale the $z_k$ variables in (\ref{lambdalambda1}) by appropriate factors of $2\pi \over  L_p$,  contributing an overall Jacobian ${(2\pi)^4 \over V}$.
 Taking the liberty to again call the rescaled variables $x_k$, we obtain for the condensate
  \begin{eqnarray}
 \langle \tr \lambda\lambda \rangle&=& {\Lambda^3 \over 2} {1 \over 4}   \prod_k \int\limits_{-L_k}^{L_k} dx_k  \; ( \phi^{(2) A} _2 \phi^{(1) A}_1    -\phi^{(1) A} _2  \phi^{(2) A}_1)[  - x_1,  x_2,   - x_3,   x_4] \nonumber \\
 &=& {\Lambda^3 \over 2} {2^4  \over 4 }    \int\limits_{\T^4} d^4 x   \; ( \phi^{(2) A} _2 \phi^{(1) A}_1    -\phi^{(1) A} _2  \phi^{(2) A}_1)[x] =  {\Lambda^3 \over 2} {2^4  \over 4 } {g^2U_F^{21} }   ~.
 \end{eqnarray}
As indicated, the integrand on the second line is the same as the one appearing in  $U_F^{21}$ from (\ref{fermionnormdefinitionbulk}), safe for the absence of a factor of $1/g^2$. 
Thus,  recalling that $U_F^{21} = {16 \pi^2 \over g^2}$, as per  the discussion after  (\ref{fermionzeromodemeasurebulk}) (see also (\ref{fermionmodematrixlast})), we finally obtain for (\ref{condensatefiniteL}):
   \begin{eqnarray}\label{condensatefinal}
 \langle \tr \lambda\lambda \rangle&=&  32 \pi^2  \Lambda^3. 
 \end{eqnarray}

Further, assuming that the result extends to $L_4 \rightarrow \infty$, as described in the previous section, we  use (\ref{lambdaresultlargetime}) and (\ref{condensatefinal}) to conclude
  \begin{eqnarray}\label{lambdaresultlargetimeV3}
 \langle 0, 0 |  \tr\lambda^2  | 0, 0 \rangle_{V_3, n_{12}} = 2 \times 16 \pi^2  \Lambda^3~.
 \end{eqnarray}
Thus, the condensate in one of the two degenerate vacua of the small-$V_3, n_{12}=1$ theory   has twice the value calculated on $\R^4$ as described in the Introduction.\footnote{In the spirit of Footnote \ref{sunfootnote}, without discussing the details, we remark  that a similar result is obtained, upon taking the large-$L_4$ limit of 't Hooft's $SU(N)$ solutions with $Q={1\over N}$, constructed via the $\Delta$ expansion in \cite{Gonzalez-Arroyo:2019wpu}. One finds, instead of (\ref{lambdaresultlargetimeV3}), that  the condensate in one of the $N$ degenerate vacua on $\T^3$ with an $n_{12}=1$ twist, $\langle 0, 0 |  \tr\lambda^2  | 0, 0 \rangle_{V_3, n_{12}} = N \times 16 \pi^2  \Lambda^3$, i.e. equals $N$ times the $\R^4$ result.}

{\bf {\flushleft{Acknowledgments:}}}  E.P. is grateful to Antonio Gonzalez-Arroyo for many conversations on fractional instantons and the $\Delta$-expansion during the ``Fluxtube-22'' Workshop at KITP during the Winter of 2022, supported in part by the National Science Foundation under Grant No. NSF PHY-1748958. We also thank F. David Wandler for comments on the manuscript.  Both authors would like to thank the Simons Centre for Geometry and Physics,  where this work started. E.P. acknowledges the hospitality of Durham University during work on this project  and M.A.  acknowledges the hospitality of  the University of Toronto, where this work was completed.  M.A. is supported by STFC through grant ST/T000708/1.   E.P. is supported by a Discovery Grant from NSERC. 

\appendix
  
\section{Constructing instantons on the asymmetric $\T^4$ with twists}
\label{Constructing instantons on the asymmetric T4 with twists}

Our $SU(2)$ Lie algebra convention is as follows. We denote the generators in the Cartan as $\tau^3\over 2$ and the non-Cartan as $\tau^{\pm}$, where $\tau^{+} = \left(\begin{array}{cc}0 &1 \cr 0 & 0\end{array}\right) = (\tau^-)^\dagger$. We expand the Hermitean gauge field as
\begin{equation}
\label{LieAlgebraSU2}
A_n = A_n^{3} {\tau^3 \over 2} + A_n^{+} \tau^+ + A_n^{-} \tau^-, ~ A^{-} = (A^{+})^\dagger, ~n=1,2,3,4.
\end{equation}
An identical expression (but without the reality condition) also holds for the adjoint representation fermions $\lambda_\alpha$, $\bar\lambda_{\dot\alpha}$ ($\alpha, \dot\alpha=1,2$), which are independent variables  in Euclidean space, to be discussed in more detail later. 

The fields are smooth and defined on $\R^4$, with coordinates $x_n$, $n=1,2,3,4$. The restriction to a $\T^4$ with periods $L_n,$  and with 't Hooft twists is accomplished by imposing periodicities defined by transition functions $\Omega_n(x)$, $x\in \R^4$. 
As explained in the main text, we choose a gauge where the $\T^4$ transition functions are trivial in the $x_1$ and $x_3$ directions and nontrivial in the $x_2$ and $x_4$ directions
\begin{equation}\label{omega1}
\Omega_2(x) = e^{- i 2 \pi {x_1\over L_1} {\tau^3 \over 2}}, \; \; \Omega_4(x) = e^{- i 2 \pi {x_3 \over L_3} {\tau^3 \over 2}}, ~\text{while} ~ \Omega_1 = \Omega_3 = 1_{2 \times 2}, ~\forall x \in \R^4.
\end{equation}
Thus, there are two nontrivial twists, $n_{12} = n_{34} = 1$, since (\ref{omega1}) obey the cocycle conditions
\begin{equation}
\label{cocycle}
\Omega_i (x+ L_j \hat e_j) \; \Omega_j(x) = e^{i\pi n_{ij}}\;  \Omega_j (x+ L_i\hat e_i)\; \Omega_i  (x),~ i,j = 1,2,3,4, \; \forall x \in \R^4,
\end{equation}
where $\hat{e}_n$ is a unit vector in the $x_n$ direction.
 The  periodicity  condition on the gauge fields in $\R^4$ define the $\T^4$ fields:
\begin{equation}\label{bc}
A(x + \hat{e}_n L_n) = \Omega_n(x) (A(x) - i d) \Omega_n^{-1}(x), ~ n=1,2,3,4, ~
\forall x \in \R^4, \end{equation}
where we denoted $A(x) =\sum\limits_{n=1}^4 A_n(x) d x_n$. The Euclidean path integral is performed over fields obeying (\ref{bc}), with fixed transition functions $\Omega_n(x)$.
Gauge transformations $g(x) \in SU(2), x \in \R^4$ act on the gauge field the usual way\footnote{
Our convention  for gauge transformations, $A(x) \rightarrow g(x) (A(x)-id )g^{-1}(x)$, implies $F_{mn} = \partial_m A_n - \partial_n A_m + i [A_m,A_n]$ and $D_m \phi = \partial_m \phi + i [A_m, \phi]$ for any adjoint $\phi$. Adjoint fermions obey the same periodicity conditions as (\ref{bc}), but without  the inhomogeneous term.} while their action on transition functions is 
\begin{equation}\label{omegaperiodicgauge}
\Omega_n(x) \rightarrow g(x + \hat{e}_n L_n) \Omega_n (x) g^{-1}(x) \; ~, ~n=1,2,3,4, ~\forall x \in \R^4.
\end{equation}
Gauge transformations $g(x)$ which leave the transition functions invariant will be called  ``$\Omega$-periodic.'' With our choice of gauge for the transition functions, $\Omega$-periodic gauge transformations $g(x)$ are periodic in $x_1$ and $x_3$ but not in $x_2$ and $x_4$. The only constant $\Omega$-periodic transformations are the abelian ones,  $g  = e^{i \alpha \tau^3}$.  

For future use, we note that a  fundamental-representation Wilson loop along a unit-winding   loop $C_1$ (winding once along $x_n$) is
\begin{eqnarray}
W(C_1, x) = \tr \left( {\cal{P}} e^{ i\int\limits_{x}^{x+ \hat e_n L_n} A_k(x') d x'_k } \; \Omega_n (x) \right).
\end{eqnarray}
Here, the insertion of $\Omega_n(x)$ ensures invariance under (\ref{omegaperiodicgauge}).

A particular field configuration obeying (\ref{bc}) is the constant field strength Abelian background, the ``fractional instanton'' introduced by 't Hooft, see \cite{tHooft:1981sps,tHooft:1981nnx,vanBaal:1984ar}:\begin{eqnarray}
\label{abelian}
\bar A_n(x,z) = \bar A_n^3(x,z) {\tau^3 \over 2}:~~ \bar A_1^{3} &=& {2 \pi x_2 \over L_1 L_2} + {z_1 \over L_1}, \\
\bar A_2^{3} &=& {z_2 \over L_2}, \nonumber\\
\bar A_3^{3} &=& {2 \pi x_4 \over L_3 L_4} + {z_3 \over L_3}, \nonumber \\
\bar A_4^{3} &=& {z_4 \over L_4}. \nonumber
\end{eqnarray}
Here, $z_n$ are constants whose significance as collective coordinates associated with the instanton will be discussed at length later. 
The field strength of the abelian background (\ref{abelian}) is:  \begin{eqnarray}
\label{Fabelian0}
F_{mn}^{(0)} &=&{\tau^3 \over 2} \left(\begin{array}{cccc}0& - {2 \pi \over L_1 L_2}& 0 &0 \cr
{2 \pi \over L_1 L_2}&0&0&0 \cr 0 &0& 0& -{2 \pi \over L_3 L_4} \cr
0 &0& {2 \pi \over L_3 L_4}&0 \end{array} \right)~.
\end{eqnarray}
The abelian background (\ref{abelian},\ref{Fabelian0}) has the following   properties:
\begin{enumerate}
\item The field strength $F_{mn}^{(0)}$ from (\ref{Fabelian0}) can be used to explicitly verify that 
the abelian background (\ref{abelian})  has topological charge $1/2$. This can be seen by recalling that the topological charge only depends on the transition functions. Its fractional nature is owing to the nonzero twists  $n_{12} = n_{34} =1$. \item
In addition, it also follows from (\ref{Fabelian0}), that for a ``symmetric'' $\T^4$---one where $L_1 L_2 = L_3 L_4$---the background (\ref{abelian}) is self-dual and hence stable, i.e. it has minimal action for the given topological charge. The action of the self-dual abelian solution is   $S_0 = {4 \pi^2 \over g^2}$, half that of the BPST instanton.
\item 
For use below, it is convenient to introduce the variables
\begin{equation} \label{hatz}
(\hat z_1, \hat z_2, \hat z_3, \hat z_4)  \equiv ( z_1 + {2 \pi x_2 \over L_2}, z_2 - {2 \pi x_1 \over L_1}, z_3 + {2 \pi x_4 \over L_4}, z_4 - {2 \pi x_3 \over L_3})
\end{equation}
The  $\hat z$-variables are important since all gauge invariants characterizing the nonabelian instanton background  depend on $\hat z_n$ only. In the gauge we are using, the fact that the background depends on $\hat z_1, \hat z_3$ is already evident in (\ref{abelian}). The appearance of the combinations $\hat z_2, \hat z_4$ follows from the action of translations in $x_1$ and $x_3$: in order to preserve the transition functions, an $x_1, x_3$ translation is accompanied by a  non-$\Omega$-periodic gauge transformation which shifts $z_2$ and $z_4$. This ensures that all gauge invariant quantities only depend on $\hat z_n$.\footnote{\label{hatzdependence}For example, a translation $x_1 \rightarrow x_1 + \epsilon_1$ is accompanied by $g_{1}(\epsilon_1, x_2) = e^{ i 2 \pi {\epsilon_1 x_2 \over L_1 L_2}{\tau^3 \over 2}}$  ensuring that $\Omega_2(x_1)$ is invariant: as per (\ref{omegaperiodicgauge}), $\Omega_2(x_1)$ transforms  into $g_1(x_2 = L_2) \Omega_2(x_1 + \epsilon_1) g_1^{-1}(x_2 = 0) = \Omega_2(x_1)$. At the same time, the $g_1$ action on $\bar A$ shifts $z_2 \rightarrow z_2 - {2\pi \epsilon_1 \over L_1}$, showing that the variable $\hat z_2 = z_2 - {2 \pi x_1 \over L_1}$ is invariant under the combined action of $g_1$ and translation $e^{ \epsilon_1 \partial_1}$. Our nonabelian solution also exhibits this property: considering, for example, its $W_1$ component of eqn.~(\ref {orderdeltabackground}), we observe that it is invariant under the combined action of translation of $x_1$, the shift of $z_2$ given above, and a gauge transformation by $g_1$, which multiplies $W_1$ by a phase. Gauge invariants built from the solution will then only depend on the $\hat z_n$ combinations.}
\end{enumerate}

\subsection{Constructing the self-dual fractional instanton for small $\Delta$} 
\label{constructing}

As explained in the main text, there are issues regarding the abelian solution in the tuned $\T^4$ that have not yet been addressed in full. Notably, they concern the lifting of the extra bosonic zero modes (found in \cite{vanBaal:1984ar}) present in the $L_1 L_2 = L_3 L_4$ limit.  These issues, as well as our desire to probe the more interesting asymmetric $\T^4$ limit\footnote{This is because the asymmetric $\T^4$  connects to various semiclassical limits that have been discussed in the literature, see the main text.} prompt us to introduce a ``detuning'' parameter  $\Delta$, which we define as follows
\begin{equation}\label{deltadef}
\Delta = {L_3 L_4 - L_1 L_2 \over \sqrt{L_1 L_2 L_3 L_4}} =  {L_3 L_4 - L_1 L_2 \over \sqrt{V}}\,.
\end{equation} 
We always take $L_3 L_4 > L_1 L_2$, i.e. $\Delta > 0$.
For small positive $\Delta$,  a nonabelian solution of the self-duality equations has been constructed   as a series expansion in powers of $\sqrt{\Delta}$ \cite{GarciaPerez:2000aiw}, with the leading contribution being the abelian background (\ref{abelian}). 

In this Section, we exhibit this solution, to the leading nontrivial order in $\Delta$, using our notation and carefully including the dependence on $z_n$.
To begin, consider the classical  background 
\begin{eqnarray}
\label{generalbkgrd}
A_n(x,z) = (\bar A_n^{3}(x,z) + S_n(x,z)) {\tau^3 \over 2} + W_n(x,z) \tau^+ + W_n^*(x,z) \tau^-\,, 
\end{eqnarray}
where $S, W$ are the deviations from the abelian background, to be determined in terms of an expansion in $\sqrt{\Delta}$.  Let us momentarily denote them by $a_n \equiv  S_n  {\tau^3 \over 2} + W_n \tau^+ + W_n^* \tau^- $. 
The construction of the  self-dual solution on the asymmetric torus by means of an expansion in $\Delta$ proceeds by imposing a self-duality condition on the field strength of (\ref{generalbkgrd})   and solving the resulting equations in a series expansion in $\sqrt{\Delta}$  \cite{GarciaPerez:2000aiw}.
In order to solve for  $a_n $, it is  subjected to   the background-Lorentz gauge condition:
\begin{eqnarray} 
D^n(\bar A) a_n &=&   \partial^n a_n + i [\bar A^n, a_n] = 0, ~ \text{or in components:}\label{gaugecondition}\\
 ~ \partial_n S^n &=&0, \nonumber \\~ (\partial_n + i \bar A_n^{3}) W^n  &=&  0. \nonumber
\end{eqnarray}
Further, the boundary conditions (\ref{bc}) imply that $S_n$ is periodic in all $x_n$, while $W_n$ is periodic in $x_1$ and $x_3$, but not in $x_2$ and $x_4$. Explicitly the boundary conditions are\begin{eqnarray}
S_n (x + \hat{e}_k L_k)  &=& S_n(x), \forall k,  \label{periodicities}\\
W_n(x+ \hat{e}_1 L_1) &=& W_n(x), \nonumber\\
W_n(x+ \hat{e}_2 L_2) &=& e^{- i{2 \pi   x_1 \over L_1}}W_n(x),\nonumber\\
W_n(x+ \hat{e}_3 L_3) &=& W_n(x),\nonumber\\
W_n(x+ \hat{e}_4 L_4) &=& e^{-i {2 \pi   x_3 \over L_1}}W_n(x).\nonumber
\end{eqnarray}

 To proceed with the construction of the instanton on the twisted asymmetric $\T^4$, following  \cite{GarciaPerez:2000aiw}, we 
   calculate the field strength of (\ref{generalbkgrd}), set its antiselfdual part to zero, and solve the equations imposing self-duality in a series expansion in powers of $\sqrt{\Delta}$.   To write the subsequent equations, it is convenient to use a quaternion notation. Thus, we  introduce the matrices \footnote{The $n=1,2,3$ components of the four-vector matrix $\sigma^n$ defined here  should not be confused with $\vec\sigma = \{\sigma_a, a = 1,2,3\} = (\sigma_1, \sigma_2,\sigma_3)$, the usual Pauli matrices. We use lower-case indices $a=1,2,3$ to denote the usual Pauli matrices $\sigma_a$ and reserve $\tau^a$ for the $SU(2)$ generators.} \begin{eqnarray} \label{wandsigma}
w &=& \sigma^n W_n ~,~ \text{where} ~\sigma^n = ( i\vec\sigma, 1_{2 \times 2}), ~\bar\sigma^n = (\sigma^n)^\dagger~,
\end{eqnarray}
\begin{eqnarray}
s &=& \sigma^n S_n~,\nonumber\\
w_c &=& C w^* C~, ~ \text{where} ~ C = \left(\begin{array}{cc}0 & -i \cr i & 0 \end{array} \right)~.
 \nonumber
\end{eqnarray}
Here $\vec{\sigma}$ are the usual Pauli matrices, acting in spinor space, not to be confused with the $\tau^3, \tau^\pm$ group generators from earlier. 
In terms of the matrices (\ref{wandsigma}), the condition that the field strength of (\ref{generalbkgrd}) be self dual is $F_{mn} \bar\sigma^m  \sigma^n=0$. This, using (\ref{gaugecondition}), becomes
\begin{eqnarray}\label{selfduality}
F_{mn} \bar\sigma^m  \sigma^n=0 \implies \bar\sigma^n \partial_n s &=& {2 \pi  {\Delta} \over \sqrt{V}} i \tau^3- i (w_c^\dagger w_c - w^\dagger w),  \\
\text{and} ~ \bar\sigma^n(\partial_n + i \bar{A}_n^{3}) w &=& - {i \over 2} (s^\dagger w - w_c^\dagger s). \nonumber
\end{eqnarray}
The self-duality equations (\ref{selfduality}) admit a solution as a series expansion in the $\T^4$ detuning parameter ${\Delta}$ (\ref{deltadef}),
\begin{eqnarray} \label{series1}
w &=& \Delta^{1/2} \; w_1 + \Delta^{3/2} \; w_2 + \Delta^{5/2} \; w_3 + \ldots , \\
s &=& \Delta \; s_1 + \Delta^2 \;s_2 + \Delta^3 \; s_3 + \ldots \nonumber~.
\end{eqnarray} 
The difference in powers of $\sqrt{\Delta}$ in the  series for $w$ and $s$ follows by the structure of the self-duality equations (\ref{selfduality}).

Here, we shall study only the leading-order solution. To exhibit it explicitly, we  substitute the expansions in (\ref{series1}) in (\ref{selfduality}) and keeping the leading term in each equation, we find that $s_1, w_1$ obey
\begin{eqnarray}\label{leadingselfduality}
\bar\sigma^n \partial_n s_1 &=& {2 \pi   \over \sqrt{V}} i \tau^3- i (w_{1 c}^\dagger w_{1 c} - w_1^\dagger w_1)\,, \\
\bar\sigma^n(\partial_n + i \bar{A}_n^{3}) w_1 &=& 0\,. \nonumber
\end{eqnarray}
We begin with the equation for $w_1$---the complex quaternion $\sigma^m W_m$, obeying the boundary conditions (\ref{periodicities}). Periodicity in $x_1, x_3$ is obeyed by writing a Fourier series  
\begin{eqnarray} 
\label{wexpansion0}
w_1 = \sum\limits_{n_1,n_3= -\infty}^\infty  e^{ -i 2\pi (n_1 {x_1 \over L_1} + n_3 {x_3 \over L_3})} w_1^{n_1,n_3}(x_2, x_4),
\end{eqnarray}
while the conditions
\begin{eqnarray}\label{periodicities2}
w_1^{n_1,n_3}(x_2 + L_2, x_4) &=& w_1^{n_1+1, n_3}(x_2, x_4)~, \\
w_1^{n_1,n_3}(x_2, x_4 + L_4) &=& w_1^{n_1, n_3+1}(x_2, x_4)~, \nonumber
\end{eqnarray}
 ensure the rest of  (\ref{periodicities}) and, by induction, imply  that all $w^{n_1,n_3}$ can be expressed via a single function,
 \begin{equation}\label{periodicities3}
 w_1^{n_1,n_3}(x_2, x_4) = u_1(x_2 - n_1 L_2, x_4 - n_3 L_4).
 \end{equation}
 Before we continue, we note that (\ref{periodicities3}) together with (\ref{wexpansion0}) implies that the norm of $w_1$ on $\T^4$ can be expressed as an $\R^2$-integral
 \begin{eqnarray}
 \label{w1norm}
 ||w_1|| &\equiv& \int_{T^4} d^4 x \tr w_1^\dagger w_1 = \sum_{n_1,n_3} \int\limits_{0}^{L_2} dx_2 \int\limits_{0}^{L_4}dx_4 \tr (w_1^{n_1,n_3})^\dagger w_1^{n_1,n_3}\nonumber \\
 &=& \int\limits_{-\infty}^{\infty} dx_2 \int\limits_{-\infty}^{\infty}dx_4 \tr u_1^\dagger (x_2,x_4) u_1(x_2,x_4)~. 
 \end{eqnarray} 
 Thus, finiteness of $||w_1||$ implies that the elements of the quaternion $u_1(x_2,x_4)$ should be square-normalizable functions on the $\R^2$-plane spanned by  $x_{2}, x_4$. Any complete basis of such functions on $\R^2$ can be used, but below, we shall argue that an especially convenient basis   is given by appropriately chosen simple harmonic oscillator eigenfunctions.
 
Continuing with our self-duality equations (\ref{leadingselfduality}), now the second equation in (\ref{leadingselfduality}) together with (\ref{periodicities3}) implies that
\begin{eqnarray}\label{w1equation2}
\left(\begin{array}{cc} (\partial_4 + i {z_4 \over L_4}) + {2 \pi \over L_3 L_4}[x_4 - (n_3 - {z_3 \over 2 \pi})L_4] & - (\partial_2 + i {z_2 \over L_2}) +  {2 \pi \over L_1 L_2}[x_2 - (n_1 - {z_1 \over 2 \pi})L_2] \\(\partial_2 + i {z_2 \over L_2}) +  {2 \pi \over L_1 L_2}[x_2 - (n_1 - {z_1 \over 2 \pi})L_2]
&  (\partial_4 + i {z_4 \over L_4}) - {2 \pi \over L_3 L_4}[x_4 - (n_3 - {z_3 \over 2 \pi})L_4]  \end{array} \right) \nonumber \\
\nonumber
 \times u_1(x_2 - n_1 L_2,x_4 - n_3 L_4)
= 0\,.\\
\end{eqnarray}
The holonomies in the partial derivatives can be absorbed by defining a new quaternion $\tilde u_1(x_2 - n_1 L_2, x_4 - n_3 L_4)$ by means of the redefinition
\begin{equation}\label{absorbholonomies}
u_1(x_2 - n_1 L_2, x_4 - n_3 L_4) = e^{ - i {z_2 \over L_2}(x_2 - n_1 L_2)  - i {z_4 \over L_4}(x_4 - n_3 L_4)} \tilde u_1(x_2 - n_1 L_2, x_4 - n_3 L_4).
\end{equation}
Using this,   equation  (\ref{w1equation2}) becomes
\begin{eqnarray}
\nonumber
\left(\begin{array}{cc} \partial_4 + {2 \pi \over L_3 L_4}[x_4 - (n_3 - {z_3 \over 2 \pi})L_4] & - \partial_2 +  {2 \pi \over L_1 L_2}[x_2 - (n_1 - {z_1 \over 2 \pi})L_2] \\\partial_2 +  {2 \pi \over L_1 L_2}[x_2 - (n_1 - {z_1 \over 2 \pi})L_2]
&  \partial_4  - {2 \pi \over L_3 L_4}[x_4 - (n_3 - {z_3 \over 2 \pi})L_4]  \end{array} \right)\\
\times\tilde u_1(x_2 - n_1 L_2,x_4 - n_3 L_4) 
 = 0\,. 
 \label{equationw1}
\end{eqnarray}

As already alluded to, we now recognize that the first order differential operators appearing above are the creation and annihilation operators of simple harmonic oscillators (SHO) of frequencies $2\pi/(L_1L_2)$ and $2\pi/(L_3 L_4)$. Let 
 $h_p^{ij}(x)$ be the normalized $p$-th state of the SHO  of frequency 
 \begin{equation}\label{frequency}
 \omega_{ij} = {2 \pi
 \over L_i L_j}.
 \end{equation} The relations that we need in what follows are the orthonormality relations  \begin{eqnarray}\label{orthogonal}
 \int\limits_{-\infty}^\infty dx h^{ij}_n(x) h^{ij}_m(x) &=& \delta_{nm}~, n,m=0,1,2..., 
 \end{eqnarray}
 as well as the raising and lowering operators
 which obey the usual equations
 \begin{eqnarray} \label{raising1}
{\sqrt{L_i L_j \over 4 \pi} } \; \left[\partial_x + {2 \pi \over L_i L_j} x\right] \; h^{ij}_n(x) &=& \sqrt{  n} \; h^{ij}_{n-1}(x)~~~~\leftrightarrow ~~~A_{ij} |n \rangle_{ij} = \sqrt{n} |n-1 \rangle_{ij},  \\
~{\sqrt{L_i L_j \over 4 \pi} }\; \left[- \partial_x + {2 \pi \over L_i L_j} x\right] \; h^{ij}_n(x) &=& \sqrt{ n+1 } \; h^{ij}_{n+1}(x)~~~~\leftrightarrow ~~~A^\dagger_{ij} |n \rangle_{ij} = \sqrt{n+1} |n+1 \rangle_{ij}\,,\nonumber 
\end{eqnarray}
where we have indicated that the differential operators in the middle and bottom line are the familiar lowering and raising operators $A_{ij}$, $A_{ij}^\dagger$ of SHOs of frequencies $\omega_{ij}$ (\ref{frequency}).
In terms of $A_{12}, A_{12}^\dagger, A_{34}, A_{34}^\dagger$,\footnote{With appropriately shifted center, as in the argument of $\tilde u_1$ below, a fact which we do not make explicit for brevity.} we can rewrite (\ref{equationw1}) as
\begin{eqnarray}
\left(\begin{array}{cc} \sqrt{2 \omega_{34}} A_{34} &   \sqrt{2 \omega_{12}} A_{12}^\dagger \\ \sqrt{2 \omega_{12}} A_{12}
 & -  \sqrt{2 \omega_{34}} A_{34}^\dagger \end{array} \right)\tilde u_1(x_2 - (n_1-{z_1\over 2 \pi}) L_2, x_4 - (n_2 -{ z_3\over 2 \pi}) L_2) = 0\,.
 \label{equationw1last}
\end{eqnarray}
 It is then clear that this equation is satisfied by $\tilde{u}_1$ given by the product of ground state wave functions of the $\omega_{12}$ and $\omega_{34}$ SHOs  times a constant $2 \times 2$ matrix $C_1$:
\begin{equation}\label{tildeu1}
\tilde u_1 = h_0^{12}(x_2 -   (n_1-{z_1 \over 2\pi}) L_2)  \; h_0^{34}(x_4 -   (n_3-{z_3 \over 2 \pi}) L_4) \; C_1,
\end{equation} where $C_1$ obeys, as a consequence of (\ref{equationw1last})
\begin{equation}\label{cmatrix}
\left(\begin{array}{cc} 0 & \sqrt{2 \omega_{12}} \\ 0 & - \sqrt{2 \omega_{34} }\end{array} \right) C_1=0 \implies C_1 = \left(\begin{array}{cc} a & b \\ 0 &0 \end{array} \right)~, 
\end{equation}
with arbitrary complex numbers $a, b$.

{\flushleft{B}}efore continuing, let us make the following important remark:
 { \small 
\begin{itemize}\item
{\flushleft{W}}e stress that   (\ref{tildeu1}, \ref{cmatrix}) is the unique  solution and not simply a consistent one. Because of the $\R^2$-normalizability (\ref{w1norm}), any  solution can be expressed  in terms of  $\omega_{12}$ and $\omega_{34}$ SHO eigenfunctions.  Explicitly, for the $ab$-th element of the quaternion $\tilde{u}_1$, we have $\tilde{u}_{1 \; ab}(x_2, x_4) = \sum\limits_{n,m=0}^{\infty} c_{ab}^{nm} h^{12}_n (x_2) h^{34}_m (x_4)$. Plugging into (\ref{equationw1last}) and  using (\ref{orthogonal}, \ref{raising1}) one finds a set of linear equations for the constants $c_{ab}^{nm}$:
\begin{eqnarray} \label{cequations}
\sqrt{\omega_{34}} \;c_{11}^{p, q+1} \sqrt{q+1} + \sqrt{\omega_{12}} \;c_{21}^{p-1,q} \sqrt{p} &=&0, ~ \text{for}~ p,q=0,1,2,3...,\nonumber \\
\sqrt{\omega_{34}} \;c_{12}^{p, q+1} \sqrt{q+1} + \sqrt{\omega_{12}} \;c_{22}^{p-1,q} \sqrt{p} &=&0, \nonumber \\
\sqrt{\omega_{12}} \;c_{11}^{p+1, q} \sqrt{p+1} - \sqrt{\omega_{34}} \;c_{21}^{p,q-1} \sqrt{q} &=&0,\nonumber \\
\sqrt{\omega_{12}} \;c_{12}^{p+1, q} \sqrt{q+1} - \sqrt{\omega_{34}}\; c_{22}^{p,q-1} \sqrt{q} &=&0.
\end{eqnarray} One begins by noticing that $c_{11}^{00}$ and $c_{12}^{00}$ do not appear above (this is easy to see by inspection recalling the range of $p,q$, which ensure that terms containing the nonexistent coefficients, like $c^{-1,*}_{**}$, are always multiplied by $0$). Then, 
 recursively solving the equations, one   finds that their only solution is  that all $c_{ij}^{nm}$ vanish except for the two that do not appear in  equations (\ref{cequations}), the undetermined  $c_{11}^{00}$ and $c_{12}^{00}$, called $a$ and $b$ in (\ref{tildeu1}).
    
Let us flesh out the argument. We consider  the first and third equations in (\ref{cequations}); the second and fourth can be considered similarly and are left as an exercise. First notice that (\ref{cequations}) imply that $c_{11}^{0,p} = c_{11}^{p,0} = 0$ for $p=1,2,3...$. Further, denote $\delta = \sqrt{\omega_{12}\over \omega_{34}}$ and shift the $p$'s and $q$'s in (\ref{cequations}) such that both equations involve $c_{11}^{p,q}$ we find
 \begin{eqnarray}\label{cshifted}
 c_{11}^{p,q}\sqrt{q} + \delta c_{21}^{p-1,q-1} \sqrt{p} &=&0, \\
 c_{11}^{p,q} \sqrt{p} - {1 \over \delta} c_{21}^{p-1,q-1} \sqrt{q} &=& 0. \nonumber
 \end{eqnarray}
 Next, we 
multiply the top equation by $\sqrt{p}$ and the bottom by $\sqrt{q}$, subtract them, and find
\begin{equation} 
\label{c21}
c_{21}^{p-1,q-1} (p \delta + {q \over \delta}) = 0. \end{equation}
Recalling that the appropriate range for $p,q$ in (\ref{c21}) is $1,2,3...$, this implies that all $c_{21}$'s vanish. Hence by (\ref{cshifted}), all $c_{11}^{p,q}$ with $(p,q) \ne (0,0)$ vanish as well.

 \end{itemize}  }

\bigskip

{\flushleft{W}}e next turn to the first of the self-duality equations in (\ref{leadingselfduality}). Since $s_1$ is periodic, the r.h.s. of the equation has no constant Fourier mode. Thus, consistency implies that the constant mode of the r.h.s vanishes as well; this will be seen to fix the coefficients $a,b$ in terms of the volume of the torus up to an overall phase. To determine them, let us collect everything---in reverse order, eqns.~(\ref{tildeu1}), (\ref{equationw1}), (\ref{absorbholonomies}), (\ref{periodicities3}), and (\ref{wexpansion0})---and exhibit the full solution for $w_1$ we found so far:
\begin{eqnarray} 
\label{wexpansion01}
w_1 = F(x, z) C_1 =  F(x, z)\left(\begin{array}{cc} a & b \\ 0 &0 \end{array} \right)\,,
\end{eqnarray}
where \begin{eqnarray}
\label{efglorious}
F(x,z) &=& \sqrt{L_2 L_4} \sum\limits_{n_1,n_3= -\infty}^\infty  e^{ -i 2\pi (n_1 {x_1 \over L_1} + n_3 {x_3 \over L_3})}  e^{ - i {z_2 \over L_2}(x_2 - n_1 L_2)  - i {z_4 \over L_4} (x_4- n_3 L_4)} \nonumber \\
&& ~~~~~\times~~ h_0^{12}(x_2 -   (n_1-{z_1 \over 2 \pi}) L_2)  \; h_0^{34}(x_4 -   (n_3-{z_3 \over 2 \pi}) L_4)~, \end{eqnarray}
 with the normalization  
 \begin{equation}\label{normalizationF}
 \int_{T^4} |F|^2 = V,
 \end{equation} 
 recalling  (\ref{raising1}). Note that $F$ is dimensionless and $a,b$ have dimension one. 
 In addition,   under the reflection of all $x_n$ and all $z_n$, after relabeling $n_1, n_3 \rightarrow - n_1, - n_3$ and noting that $h_0^{ij}$ are even functions of their arguments, we find
 \begin{equation}
 \label{reflectionef}
 F(-x,-z) = F(x,z)~.
 \end{equation}

With (\ref{wexpansion01}, \ref{efglorious}), we have that the vanishing constant mode of $\partial s$ from equation (\ref{leadingselfduality}) implies  that 
\begin{eqnarray}
\nonumber
{2 \pi  \sqrt{V}} \tau^3 &=& (C_c^\dagger C_c - C^\dagger C) \int\limits_{T^4} d^4 x |F(x,z)|^2~, \text{or}    \\
 {2 \pi   \sqrt{V}} \left(\begin{array}{cc} 1& 0\\0 & -1\end{array} \right) &=& \left(\begin{array}{cc} |b|^2 - |a|^2 & - 2 a^* b \\ - 2 a b^* & |a|^2 - |b|^2\end{array} \right) V\,. 
\end{eqnarray}
We thus conclude that  $a=0$ while $|b|^2 = {2 \pi \over V^{1/2}}$. Thus, our solution has the form
\begin{eqnarray} 
\nonumber
\label{solutiongauge1}
w(x,z) &=& \sigma^n W_n = \sqrt{\Delta} w_1(x,z) =\sqrt{\Delta}  F(x,z) e^{i \alpha} {(2 \pi)^{1/2} \over V^{1/4}}\left(\begin{array}{cc} 0 &1  \\ 0 &0 \end{array} \right) + {\cal{O}}(\Delta)\,,\\
s(x,z) &=& \sigma^n S_n =  {\cal{O}}(\Delta)\,,
\end{eqnarray}
with the phase $\alpha$  due to the gauge freedom to rotate around the $\tau^3$ isospin direction.

Before we continue, let us exhibit the self-dual solution we found, eqn.~(\ref{solutiongauge1}), $w = \Delta^{1/2} w_1$ in terms of the 4-vector gauge potentials,  $W_n, S_n$ of (\ref{generalbkgrd}):
\begin{eqnarray}
\label{orderdeltabackground}
\nonumber
S_n &=&  {\cal{O}}(\Delta)\,,\\
W_1(x,z, \alpha) &=& -{i \over 2} \sqrt{\Delta} F(x,z) e^{i \alpha} {\sqrt{2 \pi} \over V^{1/4} } +  {\cal{O}}(\Delta^{3/2}) \equiv  - \gamma e^{i \alpha} F(x,z),    \text{where}~ \gamma \equiv {i \over 2}  \sqrt{2 \pi \Delta \over \sqrt{V}}, \nonumber \\
W_2 (x,z,\alpha) &=& {1 \over 2} \sqrt{\Delta} F(x,z) e^{i \alpha} {\sqrt{2 \pi} \over V^{1/4}}+ {\cal{O}}(\Delta^{3/2}) = i W_1 + {\cal{O}}(\Delta^{3/2}),\nonumber \\
W_3 &=&  {\cal{O}}(\Delta^{3/2}),  \nonumber \\
W_4 &=&  {\cal{O}}(\Delta^{3/2}).
\end{eqnarray}
We note that the above implies that the only gauge-field zero modes in our background are the one due to translations (corresponding to shifting $z_n$) and constant gauge rotations around $\tau^3$ (corresponding to shifts of $\alpha$). The former are physical and gauge invariant quantities characterizing the solution will be seen to depend on $z_n$, while the latter are gauge artifacts with no physical quantity exhibiting $\alpha$-dependence.

For calculating the action density and gauge-invariant ``electric'' and ``magnetic'' fields to leading nontrivial order, one also needs the expression for $S_1$, which was not given in (\ref{orderdeltabackground}).  Here we simply describe how this can be found. We have, with $w_1$ from (\ref{solutiongauge1})
\begin{eqnarray}
\nonumber
w_1 &=& F \left(\begin{array}{cc} 0 &b\cr 0&0 \end{array}\right)~, ~ b = e^{i \alpha}{ \sqrt{2\pi} \over V^{1 \over 4}}  \implies w_1^\dagger w_1 = |F|^2 \left(\begin{array}{cc} 0 &0\cr 0&|b|^2 \end{array}\right)\,,\\
w_{1 c} &=& F^* \left(\begin{array}{cc} 0 &0\cr - b^*&0 \end{array}\right) \implies w_{1 c}^\dagger w_{1 c} = |F|^2 \left(\begin{array}{cc} |b|^2 &0\cr 0&0 \end{array}\right)\,. 
\end{eqnarray}
Hence, from the top equation in (\ref{leadingselfduality}), we obtain the equation for $s_1$:
\begin{eqnarray}\label{equationfors1}
\bar\sigma^n \partial_n s_1 =i \tau^3  {2 \pi \over \sqrt{V}} \left( 1     - |F(x,z)|^2 \right)~.
\end{eqnarray}
The consistency of this equation (the absence of a constant Fourier mode of the r.h.s.) discussed above is yet again seen to follow from (\ref{normalizationF}). We also note that    equation (\ref{equationfors1}) allows adding a constant to $s_1$; this is a contribution to the constants $z_n$,  already included in our background (\ref{abelian}).
To solve (\ref{equationfors1}), we multiply by $\sigma^m \partial_m$ to obtain 
\begin{eqnarray}\label{equationfors2}
\partial_n \partial^n s_1 = - i \sigma^m \tau^3  {2 \pi \over \sqrt{V}} (F^* \partial_m  F  + F \partial_m F^*) ~.
\end{eqnarray}
This equation is solved by expanding both $s_1$ and the r.h.s. in Fourier modes in $x_1...x_4$ and equating the coefficients. 
This can be explicitly done and a unique $s_1$ can be written down, but we shall not need the explicit form here. As we mentioned in Section \ref{sec:future}, understanding the higher orders in the $\Delta$ expansion is an interesting problem for future studies.

 \subsection{The field strength tensor of the fractional instanton  to order $\Delta$}
 \label{appx:fieldstrength}
 
 To find the gauge invariants of the solution to order $\Delta$, we need the field strength of $s_1$, namely\footnote{As stressed above, we shall not need its explicit form, but only the relations it satisfies, which are important to ensure self-duality.}
\begin{eqnarray}F^{s}_{nm} \equiv ( \partial_n S_{m } - \partial_m S_{n})\,,\end{eqnarray}
  where for brevity we use $S_n$ to denote the coefficient of the ${\cal{O}}(\Delta)$ part of $S_n$. The l.h.s. of (\ref{equationfors1}) has the form
 \begin{eqnarray}
&& \left(\begin{array}{cc} -i \partial_3 + \partial_4 & -i \partial_1 - \partial_2 \cr -i \partial_1 + \partial_2 & i \partial_3 + \partial_4\end{array} \right) \left(\begin{array}{cc}i S_3 + S_4 & iS_1+S_2 \cr iS_1 - S_2 & -i S_3 + S_4 \end{array} \right) \nonumber \\&=&\left(\begin{array}{cc} \partial_n S^n + i (F^s_{12} - F^s_{34}) & - F^s_{13} - F^s_{24} + i(F^s_{23}- F^s_{14}) \cr  F^s_{13}+F^s_{24} + i(F^s_{23}- F^s_{14}) &  \partial_n S^n - i (F^s_{12} - F^s_{34}) \end{array} \right) \,.
\end{eqnarray}
Comparing with the r.h.s. of (\ref{equationfors1}), using $\partial_n S^n=0$, we conclude that
\begin{eqnarray}
\nonumber
\label{strengthofs}
F^s_{12} - F^s_{34} &=& {2\pi \over \sqrt{V}}(1 - |F(x,z)|^2), \\
F^s_{13} + F^s_{24}&=&0,\nonumber \\
F^s_{23} - F^s_{14} &=&0.
\end{eqnarray}
Note that these terms should be multiplied by $\Delta$. Also note that these equations by themselves do not determine the individual $F_{12}^s$, $F_{34}^s$ (etc.) but only their non-self-dual parts. To find the individual $F_{mn}^s$, we need to solve 
(\ref{equationfors2}) to first find $s_1$.

The field strength of the ${\cal{O}}(1)$ abelian background (\ref{abelian}) is $F_{mn}^{(0)}$, already given in (\ref{Fabelian0}). Combining this with the results for $F_{mn}^s$ from (\ref{strengthofs}), we find
  \begin{eqnarray}
  \nonumber
\label{Fabelian1}
F_{mn}^{(0+1(s))} &=&
{\tau^3 \over 2} \left(\begin{array}{cccc}
0& - {2 \pi \over L_1 L_2} +\Delta F_{12}^s&  \Delta F_{13}^s &\Delta F_{14}^s \cr
{2 \pi \over L_1 L_2} - \Delta F_{12}^s&0&\Delta F_{14}^s&- \Delta F_{13} \cr 
- \Delta F_{13}^s & - \Delta F_{14}^s& 0& -{2 \pi \over L_3 L_4} + \Delta F_{34}^s\cr
- \Delta F_{14}^s &\Delta F_{13}^s& {2 \pi \over L_3 L_4}- \Delta F_{34}^s&0 
\end{array} \right)\,,\\
\end{eqnarray}
where $F_{34}^s$ is understood to be expressed via $F_{12}^s$ through the first equation in (\ref{strengthofs}).

To study the complete field strength tensor in the $\tau^3$ direction to order $\Delta$, we must add to (\ref{Fabelian1}) 
the ${\cal{O}}(\Delta)$ contribution to  the $\tau^3$ field strength coming from the commutator $F_{mn}^{(1(w))} = i [W_m \tau^+ + W_m^* \tau^-, W_n \tau^+ + W_n^* \tau^-] = {\tau^3 \over 2} 2 i (W_m W_n^* - W_m^* W_n)$. For the background (\ref{orderdeltabackground}), this is only nonzero for $F_{12}^{(1(w))}$. We find
\begin{eqnarray}\label{worderdeltastrength}
F_{12}^{(1(w))}= -F_{21}^{(1(w))} = {\tau^3 \over 2} {2 \pi   \Delta\over \sqrt{V}}|F(x,z)|^2~, ~\text{all other} \; F_{mn}^{(1(w))}=0. 
\end{eqnarray}

Most of the terms in (\ref{Fabelian1}) obey self-duality because of $F_{13}^{(0+1(s))} = - F_{24}^{(0+1(s))}$ and $F_{23}^{(0+1(s))} =  F_{14}^{(0+1(s))}$ as is evident from (\ref{Fabelian1}), (\ref{strengthofs}).
 The only terms left to consider are the $12$ and $34$ entries. To study their self-duality, we write, using the top equation in (\ref{strengthofs})
\begin{eqnarray}
F_{12}^{(0+1(s))} - F_{34}^{(0+1(s))} &=& {\tau^3\over 2}\left(- 2 \pi {L_3 L_4 - L_1 L_2 \over V} + \Delta (F_{12}^s - F_{34}^s)\right)\nonumber \\
&=& {\tau^3\over 2}\left(- 2 \pi {\Delta \over \sqrt{V}} + 2 \pi {\Delta \over \sqrt{V}}(1 - |F(x,z)|^2\right) \nonumber \\
&=& - {\tau^3 \over 2}  {2 \pi\Delta \over \sqrt{V}} |F(x,z)|^2\,.
\end{eqnarray}
This non-selfdual contribution to $F_{mn}^{(0+1(s))}$ is cancelled by the ${\cal{O}}(\Delta)$ contribution of the $W_n$ shown in (\ref{worderdeltastrength}).
 
Displayed in full glory,  the Cartan part of the field strength to order $\Delta$ is given by the sum of (\ref{Fabelian1}) and (\ref{worderdeltastrength}): 
  \begin{eqnarray}
  \nonumber
  \label{Fabelian2}
&&F_{mn}^{(0+1)}\big\vert_{Cartan} =\\
&&\left(\begin{array}{cccc}
0& - {2 \pi \over L_1 L_2} +\Delta (F_{12}^s + {2 \pi \over \sqrt{V}}|F(x,z)|^2)&  \Delta F_{13}^s &\Delta F_{14}^s \cr
{2 \pi \over L_1 L_2} - \Delta (F_{12}^s + {2 \pi \over \sqrt{V}}|F(x,z)|^2)&0&\Delta F_{14}^s&- \Delta F_{13} \cr 
- \Delta F_{13}^s & - \Delta F_{14}^s& 0& -{2 \pi \over L_3 L_4} + \Delta F_{34}^s\cr
- \Delta F_{14}^s &\Delta F_{13}^s& {2 \pi \over L_3 L_4}- \Delta F_{34}^s&0 
\end{array} \right)~. \nonumber\\
\end{eqnarray}
We stress again that $F_{34}^s$ is understood to be expressed via $F_{12}^s$ through (\ref{strengthofs}), a substitution we have not explicitly done for lack of space. Further, the above expression is understood to be multiplied by $\tau^3/2$ and the undertermined $F_{mn}^s$ are found by solving (\ref{equationfors2}).

The non-Cartan part of the solution, the $W_n$ from (\ref{orderdeltabackground}) also produce a non-Cartan field strength. 
The  ${\cal{O}}(\sqrt{\Delta)}$ field strength due to $W_n$ is $F_{mn}^{(1)}= (\partial_m + i \bar{A}_m^3) W_n \tau^+ +  (\partial_m - i \bar{A}_m^3) W_n^* \tau^- - (m \leftrightarrow n)$ .
 It is self-dual by itself and the ${\cal{O}}(\sqrt{\Delta)}$ field strength of $W_n$ is the only non-Cartan contribution to  order $\Delta$ (the next non-Cartan contribution is of order $\Delta^{3/2}$):
  \begin{eqnarray}\label{Flinearw}
F_{mn}^{(1)} &=&\left(\begin{array}{cccc}0& 0& F_{13}^{(1)} &F_{14}^{(1)}\cr
0 &0&F_{14}^{(1)}&- F_{13}^{(1)} \cr -  F_{13}^{(1)}&- F_{14}^{(1)}& 0& 0 \cr
-F_{14}^{(1)} &F_{13}^{(1)}& 0&0 \end{array} \right) .\end{eqnarray}
The nonzero entries of (\ref{Flinearw}) are
 \begin{eqnarray}\label{noncartanf}
 F_{13}^{(1)} &=& -i \gamma^* e^{i \alpha} \sqrt{\omega_{34} \over 2} G(x,z) \tau^+ + i \gamma e^{- i \alpha} \sqrt{\omega_{34} \over 2} G^*(x,z) \tau^-, \\
 F_{14}^{(1)} &=& \gamma^* e^{i \alpha}  \sqrt{\omega_{34} \over 2} G(x,z) \tau^+  +   \gamma e^{- i \alpha} \sqrt{\omega_{34} \over 2} G^*(x,z) \tau^-  ~,\nonumber 
 \end{eqnarray}
where we shouldn't forget that $\gamma \sim \sqrt{\Delta}$, as $\gamma = {i \over 2}  \sqrt{2 \pi \Delta \over \sqrt{V}}$, as
per (\ref{orderdeltabackground}). The function $G(x,z)$ is  \begin{eqnarray}
\label{gfglorious1}
G(x,z) &=& \sqrt{L_2 L_4} \sum\limits_{n_1,n_3= -\infty}^\infty  e^{ -i 2\pi (n_1 {x_1 \over L_1} + n_3 {x_3 \over L_3})}  e^{ - i {z_2 \over L_2}(x_2 - n_1 L_2)  - i {z_4 \over L_4} (x_4- n_3 L_4)} \nonumber \\
&& ~~~~~\times~~ h_0^{12}(x_2 -   (n_1-{z_1 \over 2 \pi}) L_2)  \; h_1^{34}(x_4 -   (n_3-{z_3 \over 2 \pi}) L_4)~,
\end{eqnarray}
where, just like $F(x,z)$, $G$ is dimensionless.

In conclusion, the full field strength tensor of our solution, to order $\Delta$, is  given by the Cartan part (\ref{Fabelian2}) and the non-Cartan part (\ref{Flinearw}).  Owing to the self-duality, the action of the solution is $S_0 = {4 \pi^2 \over g^2}$, as can also be explicitly inferred from the explicit form of the field strength.
These properties of the solutions are  useful in what follows.

\subsection{The gauge invariants of the fractional instanton background}
\label{gaugeinvariants}

We now consider the 
 $z_n$-dependence of gauge invariants, both local and nonlocal (i.e. winding Wilson loops) characterizing the solution. 
 
 We begin with the $z_n$ dependence of {\it local gauge invariants}, formed of traces of powers the field strength tensor. Using the data given above, we now  compute \begin{equation}\label{efield0}\langle E_i^2\rangle \equiv \int\limits_{0}^{L_3} dx_3 \int\limits_{0}^{L_4}dx_4 \tr F_{i 4}^2(x,z),\end{equation} calling it the ``$i$-th component of the electric field squared,'' averaged over $x_3$ and $x_4$.\footnote{The averaging is done solely in order to shorten the formulae that we display. This was also done   in   \cite{GarciaPerez:2000aiw}, when comparing  the analytic solution to the numerical minimization of the action at given $\Delta$.} Omitting overall constants,  for $i=1$, we find for the leading contribution to (\ref{efield0}), which is ${\cal{O}}(\Delta)$: 
 \begin{eqnarray}\label{efield1}
\langle E_1^2\rangle &\sim &  \sum\limits_{n_1,m_1}
  e^{ -i 2\pi (n_1 - m_1){1 \over L_1}({x_1} - {z_2 L_1 \over 2 \pi})}h_0^{12}(x_2  + {z_1 L_2 \over 2 \pi} -   n_1 L_2)  h_0^{12}(x_2 + {z_1 L_2 \over 2 \pi}-   m_1 L_2)  \nonumber \\
  &\sim& \sum\limits_{n_1,m_1} e^{ -i 2\pi (n_1 - m_1){1 \over L_1}(x_1 - {z_2 L_1 \over 2 \pi}) } e^{- {\pi \over L_1 L_2} \left[(x_2  + {z_1 L_2 \over 2 \pi} -   n_1 L_2)^2 + (x_2 + {z_1 L_2 \over 2 \pi}-   m_1 L_2)^2\right] }\,.
 \end{eqnarray}
The sum over $m,n$ is rapidly converging and can be seen to produce  localized bumps on $\T^4$ when plotted, periodic on $\R^2$ with period $L_1, L_2$ (usually a limit of $-4 \le m,n \le 4$ suffices to make the errors tiny and produces a periodic picture). 
The point of this discussion is to illustrate two facts:\begin{enumerate} \item That all local gauge invariants characterizing our fractional instanton  depend on the combinations $z_2 - {2 \pi x_1  \over L_1}$ and $z_1 + {2 \pi  x_2 \over L_2}$ (this also holds for the dependence on $z_4 - {2 \pi x_3  \over L_2}$ and $z_3 + {2 \pi  x_4 \over L_4}$, not shown above). \item That, furthermore, the local gauge invariants have $2 \pi$ periodicity in these variables.
\end{enumerate}

 Next, we consider the $z_n$ dependence of {\it winding Wilson loops}. To order $\Delta^0$, with the abelian background (\ref{abelian}), this is a rather straightforward task. Consider a Wilson loop beginning at some $x$, going along the $x_1$ direction $n_1$ times (i.e. from $x$ to $x + \hat{e}_1 n_1 L_1$), then $n_2$ times in the $x_2$ direction, $n_3$ times in the $x_3$ direction, and $n_4$ times in the $x_4$ direction, with the final point being   $x+ \sum\limits_{k=1}^4 \hat{e}_k n_k L_k$. The gauge invariant Wilson loop along the loop $C_{n_1,n_2,n_3,n_4}$, beginning at $x$ and consisting of the winding straight segments described above, is given by
 \begin{eqnarray}
 \label{windingwilsonzero}
 W(C_{n_1,n_2,n_3,n_4}) = \tr \left( {\cal{P}} e^{i \int_{C_{n_1,n_2,n_3,n_4}} {A}_k (x') dx'_k}\;   \Omega_2^{n_2} (x) \; \Omega_4^{n_4} (x) \right),  \end{eqnarray}
where  the nontrivial transition functions are inserted to ensure gauge invariance. 

The path-ordering can be disregarded for an abelian background ($\bar{A}$ of (\ref{abelian})) and one can conclude that \begin{eqnarray}
 \label{windingwilsonzero1}
 &&W^{\Delta^0}(C_{n_1,n_2,n_3,n_4}) \\
 &=&  2 \cos\left[{1\over 2}\left(n_1 (z_1 + {2\pi x_2 \over L_2}) + n_2 (z_2 - {2 \pi x_1 \over L_1}) + n_3 (z_3 + {2\pi x_4 \over L_4}) + n_4 (z_4 - {2\pi x_4 \over L_2})\right)   \right]. \nonumber \end{eqnarray}
We now observe that  the ${\cal{O}}(\Delta^0)$ Wilson loops  $W^{\Delta^0}_{n_1,n_2,n_3,n_4}$ are periodic functions of  the variables $\hat z_1 \equiv z_1 + {2\pi x_2 \over L_2}$, $\hat z_2 \equiv z_2 - {2 \pi x_1 \over L_1}$, $\hat z_3 \equiv z_3+ {2 \pi x_4 \over L_4}$, $\hat z_4 \equiv z_4 - {2 \pi x_3 \over L_3}$. However,  in contrast with the local gauge invariants, like the earlier (\ref{efield1}), the periodicity in $\hat z_n$ is $4 \pi$, rather than $2\pi$.
Thus, while   local gauge invariants can not distinguish  values of $\hat z_n$ differing by $2\pi$, nonlocal gauge invariant observables distinguish such values. These values are, therefore, not physically equivalent.

 We are led to conclude that the range of values of $\hat z_n$ which are distinguished by gauge invariant quantities---and are thus physically distinct---is given by  $\T^4$. In our  description, the $\T^4$ has ``circumference'' $4 \pi$, $\hat z_n \equiv \hat z_n + 4 \pi$.
This fact is important in the calculation of the gaugino condensate. 
 
It should be clear that the ${\cal{O}}(\Delta)$ (and higher) contributions to the Wilson loops (\ref{windingwilsonzero}), which require taking the path ordering into account, are proportional to the same overall factor as (\ref{windingwilsonzero1})---since they come upon expanding the path-ordered exponential. Thus, eqn.~(\ref{windingwilsonzero1}), is multiplied by $(1 +{\cal{O}}(\Delta))$. The $ {\cal{O}}(\Delta)$ terms   are $2 \pi$ periodic functions of $\hat z_n$, even with respect to their simultaneous reflection, not affecting our conclusion above. 

As an explicit simple example, consider the $ {\cal{O}}(\Delta)$ contribution to $W(C_{1,0,0,0})$, obtained by expanding the path-ordered exponential and using the explicit form of the solution (\ref{orderdeltabackground}). Omitting the overall constant, we obtain\footnote{We indicated that there is an additional contribution of the Cartan-direction ${\cal{O}}(\Delta)$ component $s_1$, which we have not computed. However, since it is found by solving (\ref{equationfors2}), whose r.h.s. is (schematically) $\sim F^* F$ which has the same  properties as the (also schematically) $|W_1|^2$ contribution shown in (\ref{wtoorderdelta}).}
\begin{eqnarray} \label{wtoorderdelta}
&&W(C_{1,0,0,0})\big\vert_{{\cal{O}}(\Delta)-\text{contribution}} \\
&\sim& \cos {\hat z_1 \over 2} \left( \bigg\vert \sum\limits_{n_3} e^{i n_3 \hat z_4}\; h_0^{34}({\hat z_3 L_4 \over 2 \pi} - n_3  L_4)  \bigg\vert^2 \right. \nonumber \\
&& \left. ~~~~~~~ \times \bigg\vert \sum\limits_{n_1} e^{- i n_1 \hat z_2} \;h_0^{12}({\hat z_1 L_2 \over 2 \pi} - n_1  L_2) \;\int\limits_{0}^1 dt \; e^{i t (\hat z_1 - 2 \pi n_1)} \bigg\vert^2 + {\text{contribution of }} \; s_1 \right). \nonumber
\end{eqnarray}
The expression above illustrates the properties we mentioned earlier: as a function of $\hat z_n$, the ${\cal{O}}(\Delta)$ contribution  multiplying the overall $\cos {\hat z_1 \over 2}$ factor  is $2 \pi$ periodic.

\section{Zero modes of the fractional instanton}
\label{zeromodesection}

\subsection{Leading-order bosonic zero-modes and measure}
\label{leadingzeromode}

Here, we study the leading order bosonic zero modes and construct the ${\cal{O}}(\Delta^0)$ bosonic zero mode measure. In the following Sections, we shall argue that the measure remains the same to all orders in $\Delta$. 

The bosonic zero modes are related to the dependence of the solution on  $z_p$. Their leading-order wave functions are particularly simple
  \begin{equation}\label{zeromodes1}
  Z_n^{0, (p)} = { \partial A_n^{cl.} \over \partial z_p} = 
 { \delta_{np} \over L_p} {\tau_3 \over 2} + ....
 \end{equation} As usual when performing semiclassical instanton calculations, we add a background Lorentz gauge fixing term to the bosonic action\footnote{It should be clear that, to order ${\cal{O}}(\Delta^0)$, the zero modes (\ref{zeromodes1}) already obey the background gauge condition: the leading-order background is in the Cartan algebra and thus commutes with $Z_n^{0,(p)}$ and, furthermore, the leading-order zero modes (\ref{zeromodes1}) are constant.} and  expand to second order in fluctuations $a_m$, $A_m \equiv A_m^{cl} + a_m$, $D_m \equiv D_m(A^{cl}) = \partial_m + i [A_m^{cl}, *]$. We stress that here   $A_m^{cl}$ is the classical solution on the asymmetric  $\T^4$, taken to the desired order in $\Delta$. Thus, we obtain the usual action of the bosonic fluctuations 
  \begin{eqnarray}\label{bosonicaction}
 S_{b,g.f.} &=& \int_{T^4}\left[ {1 \over 2 g^2} \tr F_{mn} F_{mn} + {1 \over g^2} \tr (D_m a_m)^2 \right] \\
 &\simeq& {4 \pi^2 \over g^2} +  {1 \over g^2} \int_{T^4} \tr \left[ a_n O_{nm} a_m\right], ~\text{where} ~~O_{nm} a_m = - D^2 \delta_{nm} a_m -2 i [F_{nm},  a_m]. \nonumber
 \end{eqnarray}
We include the rather well-known details that follow in order to motivate the inner product of modes as well as to follow the factors of 2. Denote by $Z_m^k$ the nonzero-eigenvalue eigenfunctions of $O_{nm}$, i.e. $O_{nm} Z_m^k = \omega_k Z_n^k$ and expand the nonzero-mode part of the fluctuation $a_m$ (for brevity, using the same letter as the full fluctuation which includes the zero modes) as $a_m =\sum_k \zeta_k Z_m^k$. Then
 we have that $$S_{b,g.f.} - {4 \pi^2 \over g^2} =  \sum_{l,k}{\omega_l \zeta_k \zeta_l \over 2} \left( {2 \over g^2} \int_{T^4} \tr Z_m^k Z_m^l \right)  \equiv  \sum_{l,k}{\omega_l \zeta_k \zeta_l \over 2} \; U_{kl},$$
 where the last equation defines the zero-mode norm matrix $U_{kl}$, explicitly spelled out in (\ref{bosonicproduct}) below. We  diagonalize the matrix $U_{kl}$ with eigenvalues $u_k$ and 
   define the measure of the path integral over the nonzero modes as $  \prod_k {d \zeta_k \sqrt{u_k} \over \sqrt{2 \pi}}$, thus normalizing  the path integral to simply produce  the product $\prod_k \omega_k^{-1/2}$. 
 The upshot is that we defined the inner product  (or moduli space metric)
 \begin{equation}\label{bosonicproduct}
 U_{kl} ={2 \over g^2} \int_{T^4} \tr Z_n^{k} Z_n^{l}
 \end{equation} which in a diagonal basis is simply  $U_{kl}=  \delta_{lk} u_l$. 
 
 We use the same inner product  for the zero modes $Z_n^{(0),p}$ of eqn.~(\ref{zeromodes1}), for which we find, neglecting the ${\cal{O}}(\sqrt{\Delta})$ contributions
 \begin{equation}
 \label{zeromodenorm}
 u_k^{(0)} = {V \over g^2 L_k^2}.
 \end{equation} As before, we expand the gauge field as $A_n = A_n^{cl} + \sum_{p=1}^4 \zeta_p^{(0)} Z_n^{(0),p} + (\text{nonzero modes})$, and define the measure over the zero modes  as $\prod_{k=1}^4 d \zeta_k^{(0)} \sqrt{u_k^{(0)} \over 2 \pi}$, the same as for the nonzero modes.  Taken at face value, this integral is undetermined until we find the region of integration over $\zeta_k^{(0)}$.  To change variables  $\zeta_k^{(0)} \rightarrow z_k$, we note that to leading order in $\Delta$ this is quite straightforward, since the dependence of $A^{cl}(z)$ on $z_k$ is linear hence one simply replaces $\zeta_k^{(0)}$ by $z_k$.
  These ${\cal{O}}(\Delta^0)$ considerations allow us to obtain the  bosonic zero-mode measure
 \begin{eqnarray} \label{bosoniczeromodemeasure}
d \mu_B \equiv { d^4 z \over (\sqrt{2 \pi})^4} \prod\limits_{l=1}^4 \sqrt{u_l^{(0)}} = {V \over g^4} {dz_1 dz_2 dz_3 dz_4 \over (\sqrt{2\pi})^4}~.
\end{eqnarray}
As described earlier, Section \ref{gaugeinvariants}, the $z_n$ are integrated over in the range from $0$ to $4 \pi$ modulo an overall reflection. 
 
 We end this introductory discussion of bosonic zero modes with two comments:
\begin{enumerate}
\item
Eqn.~(\ref{bosoniczeromodemeasure}) was obtained by considering only the leading-order solution.  
Our next task is to show that the zero modes of the ${\cal{O}}(\sqrt{\Delta})$ (and higher) solution enjoy the same measure. We need to ensure that the background gauge condition can be satisfied and that the change of variables from $\zeta_k^{(0)}$ to $z_n$ results in the same measure.
\item
An additional question that needs to be discussed is the fact that for $\Delta=0$, the self-dual abelian  solution has extra non-Cartan zero modes,  in addition to the constant modes, as was found long ago \cite{vanBaal:1984ar}. 
In contrast, the asymmetric $\T^4$ self-dual solution has only the zero modes discussed above, as we show below.
\end{enumerate}

\subsection{Fermions and their zero-mode measure}
\label{fermionappx1}

It is well-known (see e.g. \cite{Vandoren:2008xg,Dorey:2002ik}) that there is a relation between the zero modes of the adjoint Dirac operator
and the zero modes of the bosonic fluctuation operator $O_{mn}$ of eqn.~(\ref{bosonicaction}), to be exploited later. To this end, as well as because we are interested in the theory with adjoint fermions, we now consider the adjoint fermions and their path integral in the fractional instanton background.

The Euclidean space Lagrangian density with the fermions included is  \begin{equation}\label{action1}
 {1\over 2 g^2} \tr F_{mn} F_{mn} +{2 \over g^2} \tr (\partial_n \bar\lambda_{\dot\alpha} + i [A_n,\lambda_{\dot\alpha}])\bar\sigma_n^{\dot\alpha \alpha} \lambda_\alpha),\end{equation} where $\lambda$ and $\bar\lambda$ are independent variables.\footnote{\label{sigmamnfootnote}All our notation regarding fermions is as in \cite{Dorey:2002ik}, save for the fact that ref.  \cite{Dorey:2002ik} uses antihermitean gauge fields, necessitating the replacement $A^{\text{that ref.}} = i A^{\text{this paper}}$.  The four-vectors $\sigma^n, \bar\sigma^n$ were already defined in (\ref{wandsigma}). For futher use, notice, in particular, that $\sigma_{mn} = (\sigma^m \bar\sigma^n - \sigma^n \bar\sigma^m)/4$,  and that these matrices are, explicitly, $\sigma_{12} = \sigma_{34} = {i \over 2} \sigma_3$, $\sigma_{13} = -\sigma_{24}= - {i \over 2} \sigma_2$, $\sigma_{14} = \sigma_{23}= {i \over 2} \sigma_1$. As we already noted, we use $\sigma_a$ to denote the usual Pauli matrices,  not to be confused with the components of the four-vector $\sigma^n$ of (\ref{wandsigma}).} For future use, we note that it is invariant under the supersymmetry
\begin{eqnarray}\label{susyhermitean1}
\delta A_n &=& \zeta^\alpha \; \sigma_{n \; \alpha \dot\alpha} \; \bar\lambda^{\dot\alpha} + \bar\zeta_{\dot\alpha} \; \bar\sigma_n^{\dot\alpha \alpha} \; \lambda_\alpha \\
\delta \lambda_\alpha &=& - \sigma_{mn \; \alpha}^{~~~~~\beta} \;\zeta_\beta \; F_{mn} \nonumber \\
\delta \bar\lambda^{\dot\alpha} &=& - \bar\sigma^{~~~~~\dot\alpha}_{mn \;\;~~ \dot\beta} \;\bar\zeta^{\dot\beta} \; F_{mn} \nonumber
\end{eqnarray}
where the $\sigma$'s are the ones of \cite{Dorey:2002ik}. As usual for spinors, $\xi^1 = \xi_2, \xi^2 = - \xi_1$ and likewise for dotted.

    The procedure for the fermions we shall follow is to again start from the nonzero modes. We expand the fermions as eigenfunctions of the second order Hermitean operators
  \begin{eqnarray}\label{secondorder2}
D\bar D &=& D^2 + i F_{mn} \sigma^{mn}~, ~~~~~ ~ - (D \bar D)_\alpha^{~\beta} \lambda_\beta = \omega^2 \lambda_\beta ~\\
\bar D D &=& D^2 + i F_{mn} \bar\sigma^{mn}~ = D^2,  ~~- D^2 \bar\lambda^{\dot\beta} = \omega^2  \bar\lambda^{\dot\beta}~,\nonumber 
\end{eqnarray}
where, in the second line, we used the self-duality of the background (\ref{selfduality}).
To discuss the measure,  begin by considering the contribution  of a single (for brevity) nonzero eigenvalue $\omega$ to the fermion path integral. Let $- (D\bar D)_\alpha^{\;\; \beta} \phi_\beta^i = \omega^2 \phi_\alpha^i$, where $i$ labels the different eigenfunctions, the commuting functions $\phi_\alpha^i$, with the same eigenvalue $\omega$ (we note that there are at least two of them). We expand the nonzero-mode part of the fermion field (for brevity, denoting it with the same letter $\lambda$, $\bar\lambda$) 
 \begin{eqnarray}\label{lambdaexpansion}
 \lambda_\alpha &=& \sum_i \chi^i \; \phi^i_\alpha \\
 \bar\lambda^{\dot\alpha} &=& \sum_i \bar\chi^i  \; {1 \over \omega} \bar D^{\dot\alpha \alpha} \phi_\alpha^i, \nonumber
 \end{eqnarray}
 where we used the fact that the nonzero eigenfunctions of $D\bar{D}$ and $\bar{D} D$ are related as shown and we attach the spinor index to the bosonic solution of the 2nd order equation and not to the Grassmann variable, $\chi^i$ or $\bar\chi^i$ (the fact that there is more than a single solution for every $\omega$ is accounted by the index $i$). 
 We also indicate that the $\lambda$ and $\bar\lambda$ expansions have each their separate Grassmann variables $\chi^i, \bar\chi^i$.
 
Plugging (\ref{lambdaexpansion}) into the fermionic action (\ref{action1}), we obtain after integration of parts and using the fact that $\phi$ is an eigenvector of $D \bar{D}$: 
\begin{eqnarray}\label{fermionproduct}
S_F=  {2 \over g^2} \tr (D_n \bar\lambda_{\dot\alpha} \bar\sigma_n^{\dot\alpha \alpha} \lambda_\alpha) &=& \sum_{ij} \bar\chi^j \chi^i \omega \; ({2 \over g^2} \int \tr \phi^{i \alpha} \phi^j_\alpha) \\
 &=& \omega  \sum_{ij} \bar\chi^j \chi^i \; U_F^{ij}~, ~\nonumber
 \end{eqnarray}
where the fermion mode inner product matrix is 
\begin{eqnarray}\label{fermionnormdefinition}
  U_F^{ij} = {2 \over g^2} \int_{T^4} \tr(\phi^{i}_2 \phi^j_1 - \phi^{i}_1 \phi^j_2), ~ U_F^{ij} = - U_F^{ji}, \end{eqnarray}
 and again we remind ourselves that we are just looking at a single eigenvalue (one can imagine a sum over them).
 Then we define the fermion nonzero mode path integral so that it produces $\omega$ (the minimal number of eigenfunctions with the same eigenvalue is two, i.e. $i,j=1,2$, with $U_F$ generically a $2 \times 2$ matrix)
 \begin{equation}\label{nonzeromodemeasure}
\int \prod_{i} d \chi^i d \bar\chi^i \; (\det U_F)^{-1} e^{-S_F}  = \omega~.
\end{equation}
When all nonzero eigenvalues are taken into account, we obtain the square root of the product over all nonzero eigenvalues of $\bar{D} D$ (or $D \bar D$).
 
 The definition of the integrals over the fermion zero modes are done in the same manner using the same mode normalization matrix, $U_F^{ij}$ defined in (\ref{fermionnormdefinition}). We imagine 
 (as we shall argue to be the case in our background) that only the undotted spinors $\lambda_\alpha$ have zero modes, thus we expand\footnote{Here and below, we use $\phi^i_\alpha$ to denote the zero-mode solutions of $D \bar D$,  obeying $(D\bar D)_\alpha^{\;\; \beta} \phi_\beta^i = 0$. The reader should forgive us for using the same letter as in the non-zero mode discussion near (\ref{lambdaexpansion}).}
 \begin{equation}\label{fermionzeromodeexpansion}
 \lambda_\alpha = \sum_i \eta^i \phi^i_\alpha + \text{nonzero modes},
 \end{equation}
 where we use $\eta^i$ to denote the zero-mode Grassman variable. The fermion zero-mode measure is then taken to be the ``square root'' of (\ref{nonzeromodemeasure}):\footnote{This definition ensures that, upon perturbing with a zero-mode lifting mass term, $\delta S_{m}= {m \over g^2} \tr \lambda^\alpha \lambda_\alpha$, one obtains $m$ for the zero-mode contribution to the path integral.} \begin{equation}\label{fermionzeromodemeasure}
d \mu_F =  \prod_i d \eta^i  \; (\det U_F)^{-1/2}~  = \prod_i d \eta^i  \; (\text{Pf} U_F)^{-1}~.
 \end{equation}

\subsubsection{No zero modes of $D^2 = D \bar D$  on the asymmetric $\T^4$  }
\label{noDsqrtzeromodes}

The reason to include this Section is that in the $\Delta=0$ abelian self-dual background, i.e. in the symmetric-$\T^4$ case,  the $D^2$ $(=  \bar D D)$ operator has zero modes:  $a_n = c_n \tau^3 $, with arbitrary constant $c_n$. These zero modes are also zero modes of $O_{mn}$ of (\ref{bosonicaction}), owing to the abelian nature of the background.
The four Cartan zero modes of the $\Delta=0$ self-dual solution appear in addition to the two non-Cartan (complex) zero modes,  also obeying $O_{mn} a_n = 0$ but with $a_n$ in the $\tau^\pm$ directions, found in \cite{vanBaal:1984ar}. Thus the $\Delta=0$ self-dual abelian background has 8, not 4 bosonic zero modes. 
While it is expected that interactions lift half of these zero modes, this has not been shown in any detail and is a problem for future studies. 

In any case, since we find the $\Delta \ne 0$ detuned $\T^4$  of greater physical interest,  we proceed to show that in our self-dual background (\ref{orderdeltabackground}) the $D^2$ operator has no zero modes. 
Furst, suppose that there exists an adjoint field $\phi$ (generally complex) obeying $D_n D_n \phi=0$ in our background; $\phi$ is, of course, assumed to obey the boundary conditions appropriate to adjoints, i.e. (\ref{bc}) without the nonhomogeneous term. This implies that $\int\limits_{\T^4} \tr \phi^\dagger D_n D_n \phi =0$, or, after integration by parts, noting that the boundary terms vanish if (\ref{bc}) are obeyed,  $ \int\limits_{\T^4}\tr D_n \phi^\dagger D_n \phi = 0$.
This is only possible if $D_n \phi = 0$, or writing explicitly the components of this equation:
\begin{eqnarray}
\label{phizero1}
\partial_n \phi^3 - 2 i A_n^- \phi^+ + 2 i A_n^+ \phi^- &=&0, \nonumber \\
(\partial_n - i A_n^3) \phi^- + i A_n^- \phi^3 &=& 0, \nonumber \\
(\partial_n + i A_n^3)\phi^+ - i A_n^+ \phi^3 &=&0,\end{eqnarray}
where $A_n^{3,\pm}$ are to be substituted by the ${\cal{O}}(\sqrt{\Delta})$ background fields (\ref{orderdeltabackground}). 
Eqns.~(\ref{phizero1}) imply that the $D^2$ zero mode $\phi_0$ of the  abelian background (the $\sim \Delta^0$ term with $A^\pm = 0$) has components $\phi_0^3 = c, \phi_0^\pm = 0$. One can show that there are no other zero mode solutions obeying the right boundary conditions (see \cite{vanBaal:1984ar}, or perform a SHO analysis of (\ref{phizero1}) in the background (\ref{abelian}), similar to the analysis  in Section \ref{constructing}). 

We now want to argue that the ${\cal{O}}(\sqrt{\Delta})$ perturbations around the abelian background of  eqn.~(\ref{orderdeltabackground})  lift the zero eigenvalue of $-D^2$. To this end, we use perturbation theory for the non-negative Hermitean operator $-D^2$ and compute its matrix element in the unperturbed ``eigenstate'' found above, $\phi_0 = c {\tau^3 \over 2}$. For the shift of the eigenvalue, this gives $\int_{\T^4} \tr \phi_0^\dagger (-D^2) \phi_0 = {|c|^2 \over 2}\int_{\T^4}  4 A_n^- A_n^+ $, where only the $n=1,2$ components are nonzero in the order $\sqrt{\Delta}$ solution: $A_1^+=W_1$ and $A_2^+=W_2 = i W_1$.  The integral is positive definite, as it is easily seen to be proportional to $\int_{\T^4} |F|^2$, recall (\ref{normalizationF}), showing that the zero eigenvalue is lifted in the detuned-$\T^4$ self-dual background. This is a welcome feature of this background, compared to the one in the symmetric $\T^4$ studied in \cite{vanBaal:1984ar}.

\subsubsection{The zero modes of $D$ on the asymmetric $\T^4$ via the Dirac equation}
\label{appx:fermionviadirac}

In this Section,  we explicitly study the Dirac equation for the undotted fermions in the $\sim \sqrt{\Delta}$ background and show that they have two zero modes. Since, as shown above, $D^2 = \bar D D$ has no zero modes, the presence of two zero modes of $D \bar D$ is guaranteed by the index theorem in the topological charge $1/2$ background.
Thus, we include this discussion only for completeness. In the later Section, we shall see that these zero modes can be obtained using supersymmetry. 

The Dirac equation for the zero mode is $\bar D^{\dot\alpha \alpha} \lambda_\alpha = 0$ yielding
\begin{eqnarray}
\nonumber
0&=&\bar\sigma^{n \dot\alpha \alpha} \left( \partial_n \lambda_\alpha^3 - 2 i W_n^* \lambda_\alpha^+ + 2 i W_n \lambda_\alpha^-\right)\,, \\
0&=&\bar\sigma^{n \dot\alpha \alpha} \left( \partial_n \lambda_\alpha^- -  i \bar A_n^3 \lambda_\alpha^- +  i W_n^* \lambda_\alpha^3\right)\,,\nonumber \\\ 
0&=&\bar\sigma^{n \dot\alpha \alpha} \left( \partial_n \lambda_\alpha^+ + i \bar A_n^3 \lambda_\alpha^+ - i W_n \lambda_\alpha^3\right)\,.
\end{eqnarray}
In our leading-order background (\ref{orderdeltabackground}), these equations give, for the $\lambda_\alpha^3$ components:
 \begin{eqnarray}\nonumber \label{alpha3eqns1}
(i \partial_1 + \partial_2) \lambda_2^3 + (i \partial_3 - \partial_4) \lambda_1^{3} &=& 4 \gamma^* F(x,z) e^{i \alpha} \lambda_2^-\,,  \\
- (i \partial_1 - \partial_2) \lambda_2^3 + (i \partial_3 + \partial_4) \lambda_1^{3} &=& 4 \gamma F^*(x,z) e^{- i \alpha} \lambda_1^+~. 
\end{eqnarray}
  On the other hand, the non-Cartan 
  $\lambda_\alpha^-$ components satisfy the equations
 \begin{eqnarray}
 \nonumber
  \label{lambdaminuseqn}
   \left[i (\partial_1 - i {z_1 \over L_1})+ (\partial_2 - i {z_2 \over L_2}) + { 2 \pi \over L_1 L_2} x_2 \right] \lambda_2^- + \left[i (\partial_3- i {z_3 \over L_3}) - (\partial_4 - i {z_4 \over L_4}) + {2 \pi \over L_3 L_4} x_4\right] \lambda_1^- && \nonumber \\
  = 0\,, && \nonumber \\
  \nonumber
 &&  \\
  \left[i (\partial_1 - i {z_1 \over L_1}) - (\partial_2 - i {z_2 \over L_2}) +{ 2 \pi \over L_1 L_2} x_2\right] \lambda_1^- - \left[i (\partial_3 - i {z_3 \over L_3}) + (\partial_4 - i {z_4 \over L_4}) + {2 \pi \over L_3 L_4} x_4\right] \lambda_2^- && \nonumber \\
=  2 \gamma F^*(x,z) e^{- i \alpha} \lambda_1^3\,, &&\nonumber   \\
&&
\end{eqnarray}
while the  $\lambda_\alpha^+$ components obey
\begin{eqnarray}\nonumber\label{lambdapluseqn}
   \left[-i (\partial_1 + i {z_1 \over L_1}) - (\partial_2 + i {z_2 \over L_2}) + {2 \pi \over L_1 L_2} x_2\right] \lambda_2^+ + \left[-i (\partial_3 + i {z_3 \over L_3})+  (\partial_4 + i {z_4 \over L_4}) + {2 \pi \over L_3 L_4} x_4\right] \lambda_1^+ && \nonumber  \\
  = 2 \gamma^* F(x,z) e^{i \alpha} \lambda_2^3\,,  && \nonumber \\
  \nonumber
  &&  \\
  \nonumber
     \left[i(\partial_1 + i {z_1 \over L_1}) -  (\partial_2 + i {z_2 \over L_2})  - {2 \pi \over L_1 L_2}x_2\right] \lambda_1^+ + \left[-i (\partial_3 + i {z_3 \over L_3}) - (\partial_4 + i {z_4 \over L_4})  + {2 \pi \over L_3 L_4} x_4\right] \lambda_2^+ && \nonumber \\
 = 0\,. &&\nonumber\\ \end{eqnarray}
 
  We now can follow exactly the same steps as in the study of the self-duality equation for $w_1$, see discussion after (\ref{leadingselfduality}): we introduce $x_1, x_3$ Fourier modes for $\lambda^\pm$ and the functions corresponding to (\ref{periodicities3}); likewise, we can absorb the Wilson lines by a redefinition  similar to (\ref{absorbholonomies}). 
Proceeding thus, we can now solve the undotted fermion zero mode equations  (\ref{alpha3eqns1},\ref{lambdaminuseqn}, \ref{lambdapluseqn}) in an expansion in $\sqrt{\Delta}$.  Keeping in mind that $\gamma \sim \sqrt{\Delta}$, it follows that the solution for $\lambda^3_\alpha$ is of order $\Delta^{0}$  and is simply given by a two-component constant Grassmann spinor $\eta_\alpha^3$. This satisfies the $\lambda^3_\alpha$ equation to leading order, since the r.h.s. of (\ref{alpha3eqns1}) is of order $(\sqrt{\Delta})^2$ as the $\lambda^\pm$ solutions are themselves of order $\sqrt{\Delta}$. Proceeding thus, we find the leading-order fermion zero-modes:
  \begin{eqnarray}
\lambda_\alpha^{3 (0)} &=& \eta_\alpha^3, \nonumber  \\
\lambda_1^{- (0)} &=& 0,  \nonumber  
\end{eqnarray}
\begin{eqnarray}
\lambda_2^{- (0)} &=& - \eta_1^3   \gamma e^{- i \alpha} \sqrt{  L_3 L_4 \over \pi}    \sqrt{L_2  L_4}  \sum\limits_{n_1,n_3}   e^{ i 2 \pi (n_1 {x_1 \over L_1} + n_3 {x_3 \over L_3})} e^{ i ({z_2 \over L_2} (x_2 - n_1 L_2) + {z_4 \over L_4} (x_4 - n_3 L_4))} \nonumber  
\\ 
&& \qquad \times \qquad h_0^{12}(x_2 - n_1 +{ z_1 \over 2 \pi}) h_1^{34}  (x_4 - n_3 +{ z_3 \over 2 \pi}),  \nonumber 
\end{eqnarray}
\begin{eqnarray}\label{lleadingfermion} 
 \lambda_1^{+ (0)} &=& {\eta_2^3 } \gamma^* e^{i \alpha}   \sqrt{  L_3 L_4 \over \pi}    \sqrt{L_2  L_4}   \sum\limits_{n_1,n_3}   e^{- i 2 \pi (n_1 {x_1 \over L_1} + n_3 {x_3 \over L_3})} e^{- i ({z_2 \over L_2} (x_2 - n_1) + {z_4 \over L_4} (x_4 - n_3))} \nonumber   \\
&& \qquad \times \qquad  h_0^{12}(x_2 - n_1 +{ z_1 \over 2 \pi})h_1^{34}  (x_4 - n_3 +{ z_3 \over 2 \pi} ), \nonumber \\
\lambda_2^{+ (0)} &=&0. \nonumber \\
&&
\end{eqnarray}
  That  (\ref{lleadingfermion}) are solutions  to order $\sqrt{\Delta}$ follows by direct substitution and use of (\ref{raising1}). 
   Consider, for example, the  $\lambda_\alpha^-$ components of the zero modes. That $\lambda_2^-$  satisfies the top equation in (\ref{lambdaminuseqn}) follows from the fact that it is acted upon by the lowering operator of the $\omega_{12}$ SHO, while setting $\lambda_1^-$ to zero is necessitated by its being acted upon by the raising operator of the $\omega_{34}$ oscillator as well as by the second equation in (\ref{lambdaminuseqn}). The lowering operator of the $\omega_{34}$ oscillator acting on $\lambda_2^-$ from (\ref{lleadingfermion}) can be easily seen to produce the r.h.s. of the second eqn. in (\ref{lambdaminuseqn}), recalling the definition of $F$ from (\ref{efglorious}).  One similarly verifies that $\lambda^{+(0)}_\alpha$ solve the (\ref{lambdapluseqn}).

  Before we continue, let us write the fermion zero modes (\ref{lleadingfermion}) in a more compact manner, in terms of the function $G(x,z)$ of (\ref{gfglorious1}). The fermion zero modes (\ref{lleadingfermion}), that $\omega_{34} = 2 \pi/(L_3 L_4)$, are now written as:
\begin{eqnarray}
\label{lambdazero1}
 \lambda_\alpha^{(0)}
&=&  \left(\begin{array}{c} \eta_1^3 \cr \eta_2^3 \end{array}\right) {\tau^3\over 2} +   \left(\begin{array}{c}    \eta_2^3\cr 0 \end{array}\right)     \gamma^*   e^{ i \alpha} \sqrt{2 \over \omega_{34}} G(x,z)    \tau^++    \left(\begin{array}{c} 0\cr -  \eta_1^3 \end{array}\right)   \gamma  e^{- i \alpha}\sqrt{2 \over \omega_{34}} G^*(x,z)   \tau^-  \nonumber   \\ &=&  \left(\begin{array}{c} \eta_1^3 \cr \eta_2^3 \end{array}\right) {\tau^3\over 2} +   \left(\begin{array}{c}    \eta_2^3\cr 0 \end{array}\right)     \gamma^*   e^{ i \alpha} {V^{1\over 4 } \over \pi^{1 \over 2}}G(x,z)    \tau^+ +    \left(\begin{array}{c} 0\cr -  \eta_1^3 \end{array}\right)   \gamma  e^{- i \alpha}{V^{1\over 4 } \over \pi^{1 \over 2}} G^*(x,z)   \tau^-    , \end{eqnarray}
with the last equality being true to leading order in $\sqrt{\Delta}$.
  
\subsubsection{Zero modes of $D$ and $\bar D$ on the symmetric $\T^4$} 
\label{symmzeromodes} 

For completeness, let us consider the fermionic zero modes of the abelian self-dual instanton solution of the $\Delta=0$ symmetric $\T^4$. For the $\bar D \lambda=0$ equation, we can use (\ref{alpha3eqns1}, \ref{lambdaminuseqn}, \ref{lambdapluseqn}) with $\gamma$ set to zero, i.e. without r.h.s. Then, it immediately follows that the  zero modes on the symmetric $\T^4$ are 
  \begin{eqnarray}
\lambda_\alpha^{3 (\Delta=0)} &=& \eta_\alpha^3, \nonumber  \\
\lambda_1^{- (\Delta=0)} &=& 0,  \nonumber  
\end{eqnarray}
\begin{eqnarray}
\lambda_2^{- (\Delta=0)} &=& \eta_2^-     \sqrt{L_2  L_4}  \sum\limits_{n_1,n_3}   e^{ i 2 \pi (n_1 {x_1 \over L_1} + n_3 {x_3 \over L_3})} e^{ i ({z_2 \over L_2} (x_2 - n_1 L_2) + {z_4 \over L_4} (x_4 - n_3 L_4))} \nonumber  
\\ 
&& \qquad \times \qquad h_0^{12}(x_2 - n_1 +{ z_1 \over 2 \pi}) h_0^{34}  (x_4 - n_3 +{ z_3 \over 2 \pi}),  \nonumber 
\end{eqnarray}
\begin{eqnarray}\label{symmetrict4fermion} 
 \lambda_1^{+ (\Delta=0)} &=&  \eta_1^+   \sqrt{L_2  L_4}   \sum\limits_{n_1,n_3}   e^{- i 2 \pi (n_1 {x_1 \over L_1} + n_3 {x_3 \over L_3})} e^{- i ({z_2 \over L_2} (x_2 - n_1) + {z_4 \over L_4} (x_4 - n_3))} \nonumber   \\
&& \qquad \times \qquad  h_0^{12}(x_2 - n_1 +{ z_1 \over 2 \pi})h_0^{34}  (x_4 - n_3 +{ z_3 \over 2 \pi} ), \nonumber \\
\lambda_2^{+ (\Delta=0)} &=&0. \nonumber \\
&&
\end{eqnarray}
That these are solutions of (\ref{alpha3eqns1}, \ref{lambdaminuseqn}, \ref{lambdapluseqn})  with $\gamma=0$ follows from observing that the annihilation operators of the $\omega_{12}$ and $\omega_{34}$ SHOs act on $\lambda^+_1$ and $\lambda^-_2$ components only, while the other components are acted upon by creation operators and are thus set to zero.  These four zero modes of  $\bar D^{\dot \alpha \alpha}$ at $\Delta=0$  combine with the two dotted zero  modes, obeying  $  D_{\alpha \dot \alpha} \bar\lambda^{\dot\alpha} =0$, which exist due to the fact that $D^2$ has zero modes at $\Delta=0$ (recall Section \ref{noDsqrtzeromodes}):\footnote{As in our analysis leading to the bosonic solution of eqn.~(\ref{wexpansion01}), one can show that there are no normalizable zero modes of $D$ or $\bar D$ except for (\ref{symmetrict4fermion},\ref{symmetrict4fermionbarred}).}
\begin{eqnarray}
\bar\lambda_{\dot\alpha}^{3 (\Delta=0)} &=& \bar\eta_{\dot\alpha}^3,\nonumber \\
\bar\lambda_{\dot\alpha}^{\pm (\Delta=0)} &=&0, \label{symmetrict4fermionbarred}
\end{eqnarray}

The presence of two dotted (\ref{symmetrict4fermionbarred}) and four undotted (\ref{symmetrict4fermion})  fermion zero modes in the self-dual background on the symmetric $\T^4$ is, of course, in accord with the index theorem. The presence of extra zero modes is the fermionic counterpart of the existence of bosonic zero modes of $D^2$ on the symmetric $\T^4$, as discussed after eqn.~(\ref{phizero1}) of Section \ref{noDsqrtzeromodes}.  As mentioned in the main text, the extra zero modes---bosonic and fermionic---are expected to be lifted once fluctuations around the self dual background on the symmetric $\T^4$ and their interactions  are taken into account, but this has not been yet demonstrated. 

\subsubsection{The zero modes of $D$ on the asymmetric $\T^4$ via supersymmetry}
\label{appx:fermionsusy}

We now check that the asymmetric $\T^4$ undotted fermion zero modes   (\ref{lleadingfermion}) can be obtained via supersymmetry of the ${\cal{O}}(\sqrt{\Delta})$ bosonic background (\ref{abelian}, \ref{orderdeltabackground}). 
Consider the effect of the SUSY transforms (\ref{susyhermitean1}) in the gauge field background (\ref{abelian}, \ref{orderdeltabackground}) with fermions set to zero, $\bar\lambda = \lambda = 0$. Since our solution  is self-dual, i.e. obeys $\bar\sigma^{mn} F_{mn} = 0$, the SUSY transform only produces $\lambda_\alpha$ variations.
 Computing $\delta\lambda$  we obtain\begin{eqnarray}
\label{lambdasusy2}
 \delta\lambda &=&-(\sigma_{mn} F_{mn}^{(0)}  + \sigma_{mn} F_{mn}^{(1)}) \zeta + \ldots
 \end{eqnarray}
 where $F_{mn}^{(0)}$   is the field strength (\ref{Fabelian0}) of the abelian background (\ref{abelian}), 
 and $F_{mn}^{(1)}$ is the order $\sqrt{\Delta}$ contribution from (\ref{Flinearw}, \ref{noncartanf}).
 Thus, combining everything, plugging into (\ref{lambdasusy2}), and recalling Footnote \ref{sigmamnfootnote}, we obtain \begin{eqnarray} \left(\begin{array}{c}  \delta\lambda_1 \cr  \delta\lambda_2 \end{array} \right) &=&-  2 i F_{12}^{(0)} \left(\begin{array}{c}\zeta_1 \cr -\zeta_2 \end{array} \right)  + 2  F_{13}^{(1)}  \left(\begin{array}{c} \zeta_2 \cr -  \zeta_1 \end{array} \right) -i  2  F_{14}^{(1)} \left(\begin{array}{c}\zeta_2 \cr  \zeta_1 \end{array} \right) \nonumber\\
    &=&  i {4 \pi \over \sqrt{V}} \left[ \left(\begin{array}{c}\zeta_1 \cr -\zeta_2 \end{array} \right) {\tau^3 \over 2} + {V^{1 \over 4} \over \pi^{1 \over 2}} \gamma^* e^{i \alpha} G(x,z) \tau^+   \left(\begin{array}{c} -\zeta_2 \cr0 \end{array} \right)    +{V^{1 \over 4} \over \pi^{1 \over 2}} \gamma e^{- i \alpha}  G^*(x,z) \tau^-   \left(\begin{array}{c} 0 \cr -\zeta_1 \end{array} 
    \right)  \right]\nonumber \\ \label{lambdasusy3}
      \end{eqnarray}
      On the last line, we again used the small-$\Delta$ relation $\omega_{34} \simeq 2 \pi V^{-{1 \over 2}}$. 
Comparing (\ref{lambdazero1}) with (\ref{lambdasusy3}), we see that they are identical provided the supersymmetry parameter in (\ref{lambdasusy3}) is identified with the Grassmann coefficient in (\ref{lambdazero1}) as follows
\begin{eqnarray} \label{grassmansusyidentification}
 \eta_1^3 &=& 4 \pi i V^{- {1\over 2}} \zeta_1~, \\
 \eta_2^3 &=& - 4 \pi i V^{- {1\over 2}} \zeta_2~,\nonumber
 \end{eqnarray}
 showing that all is consistent with SUSY.

    \subsection{The moduli space metric, to any order in $ \Delta $}
    \label{orderdeltamoduli}
    
       Here we shall study the bosonic and fermionic moduli space to order $\Delta$. 
       It is well known that for every zero mode solution of the undotted Dirac equation $\phi_\alpha^{(\beta)}$, one can construct two zero mode eigenvalues of the operator $O_{nm}$ which obey the background gauge condition $D_m(A^{cl}) a_m=0$. 
      
      We begin with our notation for the zero modes (\ref{lambdasusy2}): we shall denote the corresponding commuting wave functions by $\phi_\alpha^{\beta \; A}$, where $A=1,2,3$ denotes the $SU(2)$ algebra index. These are the two solutions of the undotted Dirac equation  considered above.
Thus,  with $(\beta)=1,2$ labeling the two zero modes, we have
\begin{eqnarray} \label{phidefinition}
 D^{\dot\alpha \alpha} \phi_\alpha^{(\beta)} &=& 0~,\\
  \phi_\alpha^{(\beta)} &=& - (\sigma^{mn})_\alpha^{(\beta)} F_{mn}~, ~ \text{with} ~ \phi_\alpha^{(\beta)}=\sum\limits_{A=1}^3 \phi_\alpha^{\; (\beta) \; A} T^A, \\
 \phi_\alpha^{\; (\beta) \; A} &=& - (\sigma^{mn} F_{mn}^A)_\alpha^{(\beta)} = - 2 i (\sigma_3)_\alpha^{\; (\beta)} F_{12}^A + 2 i (\sigma_2)_\alpha^{\; (\beta)} F_{13}^A  - 2 i (\sigma_1)_\alpha^{\;  (\beta)} F_{14}^A \\
 &=& - 2 i \;
(\sigma_a)_\alpha^{\; (\beta)}   {V}_a^A  ~, \text{where ${V}_a^A$ has components} \; V_3^A = F_{12}^A, V_2^A = - F_{13}^A, V_1^A = F_{14}^A.\nonumber
\end{eqnarray}
In the second equality, we used self-duality of the background and the explicit form of $\sigma_{mn}$ given in Footnote \ref{sigmamnfootnote}.

    \subsubsection{Fermion zero-mode measure to arbitrary order in $\Delta$}
\label{appx:fermionmodemeasure}

In terms of the zero modes (\ref{phidefinition}) the 
fermion zero-mode expansion (\ref{fermionzeromodeexpansion}), relabeling the index $i$ (used there to label the different zero modes), $i$  $ \rightarrow \beta$, we have\begin{eqnarray}\label{lambdaexpansion2}\nonumber
\lambda_\alpha^A &=& \eta^\beta \phi_\alpha^{(\beta) \; A}, ~{\text{sum understood over}} ~ \beta = 1,2,\\
U_F^{\beta \gamma}&=& {1 \over g^2} \int_{T^4} (\phi^{(\beta) \; A}_2 \phi^{(\gamma) \; A}_1 - \phi^{(\beta) \; A}_1 \phi^{(\gamma) \; A}_2)~,
\end{eqnarray}
where again $A=1,2,3$ is the Lie algebra index, summed over the definition of $U_F$.
The fermion zero mode norm matrix $U_F^{\beta\gamma}$ from (\ref{fermionnormdefinition}) is antisymmetric and we have from (\ref{lambdaexpansion2}) and (\ref{phidefinition})
\begin{eqnarray} \label{fermionnormmatrix2}
U_F^{12} &=& {1 \over g^2} \int_{T^4} V_a^A V_b^A \;(-4) [ (\sigma_a)_2^{\; 1}  (\sigma_b)_1^{\; 2} - (\sigma_a)_1^{\; 1} (\tau^b)_2^{ \; 2}] \nonumber \\
&=& {(-4) \over g^2}  \int_{T^4} V_a^A V_b^A \;  X^{ab}\,.
\end{eqnarray}
Consider now the matrix $X^{ab} = (\sigma_a)_2^{\; 1}  (\sigma_b)_1^{\; 2} - (\sigma_a)_1^{\; 1} (\sigma_b)_2^{ \; 2}$ implicitly defined above.
We have, from the explicit form of the Pauli matrices:
\begin{equation}
\label{xmatrix}
||X^{ab} || = \left(\begin{array}{ccc} 
1&-i & 0 \cr 
i & 1& 0 \cr
0&0 &1 
\end{array}\right) ,
\end{equation}
thus 
\begin{eqnarray}\label{fermionmodematrixlast}
U_F^{12} &=& {-4 \over g^2} \int_{T^4} V_a^A V_a^A  = - {4 \over g^2} \int_{T^4}  ((F_{12}^A)^2 + (F_{14}^A)^2 + (F_{13}^A)^2)=  -{1 \over g^2} \int_{T^4} (F_{mn}^A)^2  \nonumber\\
&=& - 4 \; { 4\pi^2 \over g^2} ~~~~~(\text{since} \; {1 \over 4 g^2} \int_{T^4} (F_{mn}^A)^2 = {4 \pi^2 \over g^2}).
\end{eqnarray}
The Pfaffian of the fermion mode matrix is thus $\text{Pf} U_F= - U_F^{12} = 4 \times {4 \pi^2 \over g^2}$ and the fermion zero-mode measure, defined via the Pfaffian in (\ref{fermionzeromodemeasure}) is
 \begin{equation}\label{fermionzeromodemeasure2}
d \mu_F = d \eta^1 d \eta^2  \; (\text{Pf} U_F)^{-1}~ = 4 \times {4 \pi^2 \over g^2} d \eta^1 d \eta^2~.
 \end{equation}
To obtain a nonzero result for the fermion zero mode integral, we  insert the gaugino bilinear, $\tr \lambda^\alpha \lambda_\alpha = {1 \over 2} \eta^\alpha \eta^\beta \phi^{(\alpha) A} _\gamma \phi^{(\beta) A}_\delta \epsilon^{\delta\gamma} +$(nonzero modes), and obtain
\begin{eqnarray}
\label{fermionmeasure3}
\int d\eta^1 d \eta^2 \; ({\text{Pf}} U_F)^{-1} \; \tr \lambda^{\alpha } \lambda_\alpha &=&{g^2 \over 16 \pi^2} \int d\eta^1 d \eta^2  \; {1 \over 2} \;  \eta^\alpha \eta^\beta \phi^{(\alpha) A} _\gamma \phi^{(\beta) A}_\delta \epsilon^{\delta\gamma} \nonumber \\
&=& {g^2 \over 16 \pi^2}  {1 \over 2} \;   ( \phi^{(2) A} _\gamma \phi^{(1) A}_\delta \epsilon^{\delta\gamma}  -\phi^{(1) A} _\gamma \phi^{(2) A}_\delta \epsilon^{\delta\gamma})\,,
\end{eqnarray}
where we used (\ref{fermionmodematrixlast}) and the explicit form of the Pfaffian.

We stress that the fermion zero mode measure (\ref{fermionzeromodemeasure}) as well as (\ref{fermionmeasure3}) hold to arbitrary order in $\Delta$. Furthermore, the measure and the result (\ref{fermionmeasure3}) are independent on $\Delta$.

\subsubsection{Bosonic zero-modes and moduli-space metric to any order in $\Delta$}
\label{appx:bosonicmodes}

Now to the wave functions of the bosonic zero modes obtained from the fermionic modes and automatically obeying the gauge condition.
 For every fermionic zero mode  $\phi_\alpha^{\; (\beta)}$, $\beta=1,2$, there are two bosonic zero modes. Thus, in total there are four independent bosonic zero modes. The advantage of the discussion that follows is that the bosonic zero 
modes thus obtained automatically obey the gauge condition and, furthermore, that their construction holds to arbitrary orders in $\Delta$.
 
 The four-vector expressions for the bosonic zero modes thus obtained are denoted by $Z_n^{(\beta) \; A}$ and $Z_n^{(\beta \; ') \; A}$, where $\beta = {1,2}$, $A=1,2,3$. These modes are determined as follows (see e.g. \cite{Vandoren:2008xg,Dorey:2002ik}). First one forms the quaternions made out of the zero-mode solutions of the undotted Dirac equation, explicitly:
\begin{eqnarray}
\nonumber
\label{zeromodesfromdirac}
\sigma^n Z_n^{(\beta) \; A} = \left(\begin{array}{cc} Z_4^{(\beta)} + i Z_3^{(\beta)} & Z_2^{(\beta)} + i Z_1^{(\beta)} \cr  - Z_2^{(\beta) } + i Z_1^{(\beta)} & Z_4^{(\beta)}- i Z_3^{(\beta)} \end{array} \right)^A &=&  \left(\begin{array}{cc} \phi_1^{(\beta)} & -  \phi_2^{(\beta) \; *}  \cr  \phi_2^{(\beta)} &  \phi_1^{(\beta)\; *}\end{array} \right)^A, \\
\sigma^nZ_n^{(\beta \;') \; A} = \left(\begin{array}{cc} Z_4^{(\beta \;')} + i Z_3^{(\beta \;')} & Z_2^{(\beta \;')} + i Z_1^{(\beta \;')} \cr  - Z_2^{(\beta \;')} + i Z_1^{(\beta \;')} & Z_4^{(\beta \;')}- i Z_3^{(\beta \;')} \end{array} \right)^A &=&  \left(\begin{array}{cc} i \phi_1^{(\beta)} & i  \phi_2^{(\beta) \; *}  \cr i \phi_2^{(\beta)} & -i  \phi_1^{(\beta) \; *} \end{array} \right)^A~.
\end{eqnarray}
Thus, each zero mode $\phi_\alpha^{(\beta)}$ of the undotted Dirac equation can be used to build two four-vector bosonic zero modes, denoted by $Z_n^{(\beta)}$ and $Z_n^{(\beta \; ')}$.
Their four-vector components    are then inferred from (\ref{zeromodesfromdirac}):
\begin{eqnarray}\nonumber\label{fourvectorzeromodes}
Z_n^{(\beta)\; A} &=& \left\{ \Im \phi_2^{(\beta) \;A}, - \Re \phi_2^{(\beta) \;A},  \Im \phi_1^{(\beta) \;A} , \Re \phi_1^{(\beta) \;A}\right\},\\
Z_n^{(\gamma \; ') \;A} &=&  \left\{ \Re \phi_2^{(\gamma) \;A}, \Im \phi_2^{(\gamma) \;A}, \Re \phi_1^{(\gamma) \;A}, - \Im \phi_1^{(\gamma) \;A} \right\},
\end{eqnarray}

Now, knowing the four-vector components of the zero modes, we can use (\ref{zeromodesfromdirac}) and (\ref{phidefinition}) to find $\det U_{kl}$ and argue for its $\Delta$-independence. We need to compute the $4\times 4$ matrix of different overlaps (\ref{bosonicproduct})
\begin{eqnarray}
\nonumber
U^{\beta, \gamma} &=& {2 \over g^2} \int_{T^4} \tr Z_n^{(\beta)} Z_n^{(\gamma)} = {1 \over g^2} \int_{T^4} Z_n^{(\beta)\; A} Z_n^{(\gamma) \; A} \\
\nonumber
U^{\beta, \gamma+2} = U^{\gamma+2, \beta}&=&{2 \over g^2} \int_{T^4} \tr Z_n^{(\beta)} Z_n^{(\gamma\; ')}= {1 \over g^2} \int_{T^4} Z_n^{(\beta)\; A} Z_n^{(\gamma  \; ') \; A},\\
U^{\beta+2, \gamma+2} &=& {2 \over g^2} \int_{T^4} \tr Z_n^{(\beta\; ')} Z_n^{(\gamma\; ')}= {1 \over g^2} \int_{T^4} Z_n^{(\beta\; ')\; A} Z_n^{(\gamma  \; ') \; A},
\end{eqnarray}
 where the trace is now in the Lie-algebra generator space.
The four-vector inner products are\begin{eqnarray}\nonumber
Z_n^{(\beta)} Z_n^{(\gamma)}=  Z_n^{(\beta  \; ')} Z_n^{(\gamma \; ')} &=&  \Im \phi_2^{(\beta)} \Im \phi_2^{(\gamma)} +  \Re \phi_2^{(\beta)}  \Re \phi_2^{(\gamma)} +  \Im \phi_1^{(\beta)} \Im \phi_1^{(\gamma)} + \Re \phi_1^{(\beta) } \Re \phi_1^{(\gamma)}\nonumber\\\nonumber
&=& \sum\limits_{\alpha = 1}^2{1 \over 2} ( \phi_\alpha^{(\beta)} \phi_{\alpha}^{(\gamma) \; *} + \phi_\alpha^{(\beta)\; *} \phi_{\alpha}^{(\gamma) })\,, \\
Z_n^{(\beta)} Z_n^{(\gamma \; ')} &=& \Im \phi_2^{(\beta)} \Re \phi_2^{(\gamma)} -  \Re \phi_2^{(\beta)}  \Im \phi_2^{(\gamma)} +  \Im \phi_1^{(\beta)} \Re \phi_1^{(\gamma)}- \Re \phi_1^{(\beta) } \Im \phi_1^{(\gamma)} \nonumber \\
&=& \sum\limits_{\alpha = 1}^2 {i \over 2}( \phi_\alpha^{(\beta)} \phi_{\alpha}^{(\gamma) \; *} - \phi_\alpha^{(\beta)\; *} \phi_{\alpha}^{(\gamma) })~,
\end{eqnarray}
where momentarily we omitted the group index $A$ (to be restored below). 
Thus, using (\ref{phidefinition}), with the shorthand $\sigma.F^A \equiv \sigma^{mn} F_{mn}^A$, we find 
\begin{eqnarray}\nonumber\label{uelements}
U^{\beta, \gamma} = U^{\beta+2, \gamma+2}&=&{1\over 2 g^2} \int_{T^4} \sum\limits_{\alpha = 1}^2 ( \phi_\alpha^{(\beta) \; A} \phi_{\alpha}^{(\gamma) \; * \; A} + \phi_\alpha^{(\beta)\; * \; A} \phi_{\alpha}^{(\gamma) \; A })  \\\nonumber
&=& {1\over 2 g^2} \int_{T^4} \sum\limits_{\alpha = 1}^2 ( (\sigma.F^A)_\alpha^{\; (\beta)}(\sigma.F^A)_\alpha^{\; (\gamma) \; *}+ (\sigma.F^A)_\alpha^{\; (\beta) \; *}(\sigma.F^A)_\alpha^{\; (\gamma) } )\,,\nonumber \\
\nonumber
U^{\beta, \gamma+2} = U^{\gamma+2, \beta}&=& {i\over 2  g^2}  \int_{T^4} \sum\limits_{\alpha = 1}^2 ( \phi_\alpha^{(\beta) \; A} \phi_{\alpha}^{(\gamma) \; * \; A} - \phi_\alpha^{(\beta)\; * \;A} \phi_{\alpha}^{(\gamma) \; A}) \\
\nonumber
&=&  {i\over 2 g^2}  \int_{T^4} \sum\limits_{\alpha = 1}^2  ( (\sigma.F^A)_\alpha^{\; (\beta)}(\sigma.F^A)_\alpha^{\; (\gamma) \; *}- (\sigma.F^A)_\alpha^{\; (\beta) \; *}(\sigma.F^A)_\alpha^{\; (\gamma) } ).\\
\end{eqnarray}
Let us now study the matrix 
\begin{equation}\label{amatrix}
A_{\beta\gamma} = {1\over 2 g^2} \int_{T^4} \sum\limits_{\alpha = 1}^2 (\sigma.F^A)^{ \; (\beta)}_\alpha (\sigma.F^A)^{  \; (\gamma) \; *}_\alpha. \end{equation}
Using the notation of eqn.~(\ref{phidefinition}), we rewrite it as
\begin{eqnarray}\label{amatrix2}
A_{\beta\gamma} &=& {1 \over 2 g^2} \int_{T^4} V^A_a V^A_b \; 4\; (\sigma_a)_{\alpha}^{~~(\beta)} ((\sigma_b)^*)_\alpha^{~~(\gamma}) = {2 \over  g^2} \int_{T^4} V^A_a V^A_b    \; (\sigma_2 \sigma_{a}  \sigma_b \sigma_2)_{(\beta)}^{~~(\gamma)} \nonumber \\
&=& \delta_\beta^\gamma\; {2 \over g^2} \int_{T^4} V^A_a V^A_a = {1 \over 2 g^2} \int_{T^4} (F_{mn}^A)^2\,.
\end{eqnarray}
Thus, the $4 \times 4$ matrix $U$ with matrix elements (\ref{uelements}), expressed through $A$ of (\ref{amatrix}) is of the form
 \begin{eqnarray}\nonumber
 U &=& \left( \begin{array}{cccc} 
 A_{11}+ A_{11}^* &A_{12}+ A_{12}^* & i (A_{11} - A_{11}^*) &i (A_{12} - A_{12}^*) \cr
  A_{12}+ A_{12}^* &A_{22}+ A_{22}^*&i (A_{12} - A_{12}^*) &i (A_{22} - A_{22}^*)\cr
 i (A_{11} - A_{11}^*) &i (A_{12} - A_{12}^*)   & A_{11}+ A_{11}^* &A_{12}+ A_{12}^* \cr
 i (A_{12} - A_{12}^*) &i (A_{22} - A_{22}^*) &   A_{12}+ A_{12}^* &A_{22}+ A_{22}^*
 \end{array}\right) \\\nonumber
 &=& \left( \begin{array}{cccc} 
 2 A_{11}  &0 &0 &0 \cr
 0&2 A_{22} &0 &0 \cr
0&0  & 2 A_{11} &0 \cr
0 &0 & 0 & 2 A_{22} 
 \end{array}\right) \\\nonumber
 &=& {\text{diag}}(1,1,1,1) \; {1 \over  g^2} \int_{T^4} (F_{mn}^A)^2\\
 &=& {\text{diag}}(1,1,1,1) \; {4} \; {4\pi^2 \over g^2} ~~~~ ~~(\text{since} \; {1 \over 4 g^2} \int_{T^4} (F_{mn}^A)^2 = {4 \pi^2 \over g^2})~.
 \end{eqnarray}
 where we used the diagonal form of $A$ form (\ref{amatrix2}).
 
 Thus, we have for the norm matrix (\ref{bosonicproduct}) of the bosonic zero modes (to any arbitrary  order in $\Delta$ to which the solution has been find): 
  \begin{equation}\label{ugeneral}
 U^{kl} = \delta^{kl} \; 4 \; {4 \pi^2 \over g^2}~, ~k,l=1,2,3,4.
 \end{equation}
The point, so far, is to argue that the inner-product of the bosonic zero modes obeying the gauge condition---or the moduli space metric (\ref{ugeneral})---are $\Delta$-independent.\footnote{The reader may notice that while we use the same letter, the moduli space metric $ U^{kl}$ of (\ref{ugeneral}) does not equal the one  constructed earlier from the leading-order derivatives of the classical solution w.r.t. $z_n$, the metric $ U^{kl}$ of eqns.~(\ref{bosonicproduct}, \ref{zeromodenorm}). In fact, we have $g^8 \det U^{kl}_{eq. (\ref{ugeneral})} = (4\pi)^8$ while $g^8 \det U^{kl}_{eq. (\ref{zeromodenorm})} = V^2$. This is accounted for by the difference in normalization of the respective zero modes; see Section \ref{derivatives} for explicit expressions. This difference contributes an extra factor of $({4 \pi \over \sqrt{V}})^8 \prod_k (L_k^2) = {(4 \pi)^8 \over V^2}$ to the determinant of  eq.~(\ref{zeromodenorm}). }

The careful reader may remark that the fact that the bosonic zero mode metric is proportional to the classical action, as in (\ref{ugeneral}), is well-known, hence the $\Delta$-independence follows from the fact that the action of the self-dual solution is $\Delta$-independent. 
However, we presented the steps outlined above, especially the explicit relation between the undotted Dirac equation zero modes and the bosonic zero modes of eqn.~(\ref{fourvectorzeromodes}), in order to use them to verify that the derivatives of the ${\cal{O}}(\sqrt{\Delta})$ classical solution we found with respect to $z_n$ are equal to the 
modes (\ref{fourvectorzeromodes}) obeying the background gauge condition, after an appropriate gauge transformation. This somewhat more involved procedure, compared to the similar task for the BPST instanton, is what we   discuss next.
 
 \subsubsection{The derivatives of the ${\cal{O}}(\sqrt{\Delta})$  solution and  the gauge condition}
\label{derivatives}

To this end, let us compute the zero modes (\ref{fourvectorzeromodes})   in the $\Delta$ expansion and compare to the derivatives of the classical solution, $\partial A_n^{cl}/\partial {z_k}$. The zero modes (\ref{fourvectorzeromodes}) obeying the gauge condition\footnote{As these can be constructed from $F_{mn}^{cl.}$ to any order in $\Delta$, we shall call these ``exact'' zero modes.} are determined by the matrices (\ref{phidefinition})
\begin{eqnarray}\nonumber
||\Re \phi_\alpha^{(\beta) \; A} || &=& \left(\begin{array}{cc} 0& 2 F_{13}^A \cr  -2 F_{13}^A &0 \end{array}\right)\,,\\
||\Im \phi_\alpha^{(\beta) \; A} ||&=&\left(\begin{array}{cc} - 2 F_{12}^A&-2 F_{14}^A\cr -2 F_{14}^A& 2 F_{12}^A \end{array}\right)\,.
\end{eqnarray}
This yields for  $Z_n^{(\beta)}, Z_n^{(\beta) '}$ of (\ref{fourvectorzeromodes}),
\begin{eqnarray}\label{fourvectorzeromodes12}
Y_n^{(3) \; A} \equiv Z_n^{(1)\; A} &=& \left\{ \Im \phi_2^{(1) \;A}, - \Re \phi_2^{(1) \;A},  \Im \phi_1^{(1) \;A} , \Re \phi_1^{(1) \;A}\right\} = \left\{ - 2 F_{14}^A,  2 F_{13}^A,  - 2 F_{12}^A, 0\right\},\nonumber \\
Y_n^{(1) \; A}\equiv -Z_n^{(2)\; A} &=& \left\{- \Im \phi_2^{(2) \;A},  \Re \phi_2^{(2) \;A},-  \Im \phi_1^{(2) \;A} , -\Re \phi_1^{(2) \;A}\right\}=  \left\{- 2 F_{12}^A, 0, 2 F_{14}^A, -2 F_{13}^A\right\},\nonumber \\
Y_n^{(4) \; A}\equiv - Z_n^{(1 \; ') \;A} &=&   \left\{ -\Re \phi_2^{(1) \;A}, -\Im \phi_2^{(1) \;A}, -\Re \phi_1^{(1) \;A},  \Im \phi_1^{(1) \;A} \right\} = \left\{ 2 F_{13}^A,  2 F_{14}^A, 0, -2 F_{12}^A\right\},\nonumber \\
Y_n^{(2) \; A}\equiv - Z_n^{(2\; ') \;A} &=&    \left\{- \Re \phi_2^{(2) \;A}, -\Im \phi_2^{(2) \;A}, -\Re \phi_1^{(2) \;A},  \Im \phi_1^{(2) \;A} \right\} =  \left\{0, - 2 F_{12}^A, -2 F_{13}^A, -2 F_{14}^A\right\}.\nonumber
\\\end{eqnarray}  
We relabelled the four zero modes $Z_n^{(1)}$, $Z_n^{(2)}$, $Z_n^{(1) '}$, $Z_n^{(2) '}$ by $Y_n^{(k)}$, where $k$ indicates that in our $\Delta$-expanded solution these correspond to derivatives of the classical background with respect to $z_k$ (as we shall see shortly). The functions $Y_n^{(k)}$ calculated in the leading $\sim \sqrt{\Delta}$ solution are explicitly presented below.

 Let us now also write our $\sim \sqrt{\Delta}$ self dual classical solution (\ref{orderdeltabackground}): 
 \begin{eqnarray}\label{ancl}
 A_n^{cl}(x,z) &=& ({2 \pi n_{nm} x^m \over L_n L_m} + {z_n \over L_n}) {\tau^3 \over 2}  + {\tau^+}(\delta_{n1} + i \delta_{n2})   W  + {\tau^-}( \delta_{n1}  - i \delta_{n2}) W^*, \nonumber\\
 W &=&-{i \sqrt{\pi \Delta} \over \sqrt{2 V^{1/2}}} e^{i \alpha} F(x,z),
 \end{eqnarray}
 where for brevity we introduced the (not anti-symmetric!) tensor $n_{nm}$ where only $n_{12}=n_{34}=1$ are nonzero.
 Consider now  the derivative of the classical solution w.r.t. $z_p/L_p$
 \begin{eqnarray}
L_p { \partial A_n^{cl} \over 
 \partial z_p} = \delta_{np} {\tau^3 \over 2}  + {\tau^+}(\delta_{n1}+ i \delta_{n2}) {L_p \partial W \over 
 \partial z_p} + {\tau^-}( \delta_{n1} - i \delta_{n2})  {L_p \partial W^* \over 
 \partial z_p}). 
 \end{eqnarray}
 The Cartan term is the one we considered before, see Section \ref{zeromodesection}.
To find the zero mode wave functions, an explicit computation using the SHO $h_0$ and $h_1$ properties from (\ref{raising1})  shows that
\begin{eqnarray}\nonumber\label{derivativesofF}
 L_1 {\partial F \over \partial z_1}&=& - \sqrt{L_1 L_2 \over 4 \pi} \tilde{G}\,, \\
L_2 {\partial F \over \partial z_2}&=& i {L_2 z_1 \over 2 \pi} F - i \sqrt{L_1 L_2 \over 4 \pi} \tilde{G}\,, \nonumber \\
L_3 {\partial F \over \partial z_3}&=& - \sqrt{L_3 L_4 \over 4 \pi} {G} \,,\nonumber \\
L_4 {\partial F \over \partial z_4}&=& i {L_4 z_3 \over 2 \pi} F -i \sqrt{L_3 L_4 \over 4 \pi} {G} \,.
\end{eqnarray}
Here, we defined, in addition to $F(x,z)$ (\ref{efglorious}) and $G(x,z)$ (\ref{gfglorious1}), a new function $\tilde G(x,z)$, similar to $G(x,z)$ but  where the 1st-excited state is in the $\omega_{12}$ SHO instead:
 \begin{eqnarray}
\label{efglorious3}
\tilde G(x,z) &=& \sqrt{L_2 L_4} \sum\limits_{n^1,n^3= -\infty}^\infty  e^{ -i 2\pi (n_1 {x_1 \over L_1} + n_3 {x_3 \over L_3})}  e^{ - i {z_2 \over L_2}(x_2 - n_1 L_2)  - i {z_4 \over L_4} (x_4- n_3 L_4)} \nonumber \\
&& ~~~~~\times~~ h_1^{12}(x_2 -   (n_1-{z_1 \over 2 \pi}) L_2)  \; h_0^{34}(x_4 -   (n_3-{z_3 \over 2 \pi}) L_4)~. 
\end{eqnarray}
Using these expressions, we find the derivatives of the classical solution w.r.t. $z_p$:
\begin{eqnarray}\label{aclassicalderivatives}
L_1 { \partial A_n^{cl} \over 
 \partial z_1}
 &=&\delta_{n1} {\tau^3 \over 2}  + \left[{\tau^+}(\delta_{n1} + i \delta_{n2}){i \sqrt{\Delta} \over 2 \sqrt{2}} e^{i \alpha}\tilde{G} + h.c.\right]\,, \nonumber \\
 L_2 { \partial A_n^{cl}  \over 
 \partial z_2} &=&\delta_{n2} {\tau^3 \over 2} + \left[{\tau^+}(\delta_{n1} + i \delta_{n2})(-{i \sqrt{\pi \Delta} \over \sqrt{2 V^{1/2}}} e^{i \alpha})( i {L_2 z_1 \over 2 \pi} F - i \sqrt{L_1 L_2 \over 4 \pi} \tilde{G}) + h.c. \right]\,, \nonumber \\
 L_3 { \partial A_n^{cl}  \over \partial z_3} 
 &=&\delta_{n3} {\tau^3 \over 2}  + \left[{\tau^+}(\delta_{n1} + i \delta_{n2}){i \sqrt{\Delta} \over 2 \sqrt{2}} e^{i \alpha}{G} + h.c.\right]\,, \nonumber \\
 L_4 { \partial A_n^{cl}  \over 
 \partial z_4} &=&\delta_{n4} {\tau^3 \over 2} + \left[{\tau^+}(\delta_{n1} + i \delta_{n2})(-{i \sqrt{\pi \Delta} \over \sqrt{2 V^{1/2}}} e^{i \alpha})( i {L_4 z_3 \over 2 \pi} F - i \sqrt{L_3 L_4 \over 4 \pi} {G}) + h.c. \right]\,. \nonumber\\
 \end{eqnarray}
 
 The expressions for the exact zero modes $Y_n^{(k)}$ from (\ref{fourvectorzeromodes12}) can also be computed using our knowledge of the field strength to order $\sqrt{\Delta}$, $F_{mn}^{(0)}, F_{mn}^{(1)}$ of eqns.~(\ref{Fabelian0}) and (\ref{noncartanf}). Next, we use this information and (\ref{aclassicalderivatives}) to determine the gauge transformations needed to bring the $L_k { \partial A_n^{cl} \over 
 \partial z_k}$ zero modes into the background-Lorentz gauge. We shall see that on the $\T^4$ with twists, this is slightly different from the usual BPST instanton. To find these  gauge transformations, we now consider the derivatives of the classical solution w.r.t. each $z_k$ in turn.   
 
{\flushleft\bf{Zero mode $\mathbf{ \partial\over \partial z_1}$ vs $Y^{(1)}$}}:
 Here, we compare $Y_n^{(1)}$ to the derivative w.r.t. $z_1$. From the above we find that the four-vectors of the exact zero mode and the derivative of the classical solution are
 \begin{eqnarray}
 Y_n^{(1)} &=& \left\{ {4 \pi \over \sqrt{V}} {\tau^3 \over 2}, 0,  \tau^+  (-i {\sqrt{2\Delta} \pi \over \sqrt{ V}} e^{i \alpha} G(x,z)) + h.c., \tau^+  {\sqrt{2\Delta} \pi \over \sqrt{ V}} e^{i \alpha} G(x,z) + h.c.\right\}\,, \nonumber \\
 L_1 { \partial A_n^{cl} \over 
 \partial z_1} &=& \left\{ {\tau^3 \over 2}+(\tau^+ {i \sqrt{\Delta} \over 2 \sqrt{2}} e^{i \alpha}\tilde{G}+ h.c.), - (\tau^+ {\sqrt{\Delta}  \over 2 \sqrt{2}} e^{i \alpha}\tilde{G} + h.c.) ,0,0 \right\}. 
 \end{eqnarray}
Their difference is
 \begin{eqnarray}\nonumber\label{yn1}
 Y_n^{(1)} - {4 \pi \over \sqrt{V}}  L_1 { \partial A_n^{cl} \over 
 \partial z_1}&=& \tau^+{\pi \sqrt{2 \Delta} e^{i \alpha} \over V^{1/2}}\left( \begin{array}{c}
 {- i} \tilde{G}(x,z)\cr
  \tilde{G}(x,z)\cr
 -i G(x,z)\cr
  G(x,z)  \end{array} \right) + h.c.\\
  &=& D_n(A^{cl.}) \Lambda^{(1)} = \partial_n \Lambda^{(1)} + i [A_n^{cl.}\big\vert_{\Delta=0}, \Lambda^{(1)} ]. 
    \end{eqnarray}
After some algebra, we find that the gauge transformation making the derivative of the classical solution obey the gauge condition is   ${\cal{O}}(\sqrt{\Delta})$: 
    \begin{eqnarray}\label{lambda1final}
\Lambda^{(1)}(x,z) = - {\sqrt{2 \Delta\pi} \over  V^{1 \over 4} } e^{i \alpha} F(x,z) \tau^+ + h.c.\,.
\end{eqnarray}

{\flushleft\bf{Zero mode $\mathbf{ \partial\over \partial z_2}$  vs $Y^{(2)}$}}: Here we have the four-vectors
 \begin{eqnarray}
 Y_n^{(2)} &=& \left\{0, {4 \pi \over \sqrt{V}} {\tau^3 \over 2},  \tau^+  {\sqrt{2\Delta} \pi \over \sqrt{ V}} e^{i \alpha} G(x,z) + h.c., \tau^+  {i \sqrt{2\Delta} \pi \over \sqrt{ V}} e^{i \alpha} G(x,z) + h.c.\right\} \,,\nonumber \\
 L_2 { \partial A_n^{cl} \over 
 \partial z_2} &=& \left\{\tau^+ {\sqrt{\Delta \pi} e^{i \alpha} \over \sqrt{2} V^{1\over 4}} \left[  {L_2 z_1 \over 2 \pi}  F(x,z) - \sqrt{L_1 L_2\over 4 \pi} \tilde{G}(x,z)\right] + h.c., \right. \nonumber \\
 &&  \left. {\tau^3 \over 2} + \tau^+ i  {\sqrt{\Delta \pi} e^{i \alpha} \over \sqrt{2} V^{1\over 4}} \left[  {L_2 z_1 \over 2 \pi}  F(x,z) - \sqrt{L_1 L_2 \over 4 \pi} \tilde{G}(x,z) \right] + h.c., 0, 0 \right\}\,.
 \end{eqnarray}
 We again consider the difference
  \begin{eqnarray}
  \label{yn2}Y_n^{(2)} - {4 \pi \over \sqrt{V}}  L_2 { \partial A_n^{cl} \over 
 \partial z_2}&=& \tau^+ {\pi \sqrt{2 \Delta} e^{i \alpha} \over \sqrt{V}}\left( \begin{array}{c}
- {2 \sqrt{\pi} \over V^{1 \over 4}} \left[{L_2 z_1 \over 2 \pi} F(x,z) - \sqrt{L_1 L_2 \over 4 \pi} \tilde{G}(x,z)\right]\cr
 -i{2 \sqrt{\pi} \over V^{1 \over 4}} \left[{L_2 z_1 \over 2 \pi} F(x,z) - \sqrt{L_1 L_2 \over 4 \pi} \tilde{G}(x,z)\right]\cr
 G(x,z)\cr
i  G(x,z)  \end{array} \right) + h.c. \nonumber \\
  &=& D_n(A^{cl.}) \Lambda^{(2)} = \partial_n \Lambda^{(2)} + i [A_n^{cl.}\big\vert_{\Delta=0} + A_n^{cl.}\big\vert_{{\cal{O}}(\sqrt{\Delta})}, \Lambda^{(2)} ].
    \end{eqnarray}
Here, in contrast with (\ref{lambda1final}), we find that the gauge transformation making the zero mode obey the background condition also has a Cartan-subalgebra piece $\sim \Delta^{0}$, in addition to an ${\cal{O}}(\sqrt{\Delta})$ piece which is proportional to (\ref{lambda1final}):  
\begin{equation}\label{lambda2new}
\Lambda^{(2)}(x,z) =   {2 L_2 z_1 \over \sqrt{V}} {\tau^3 \over 2} + (i \Lambda^{(1) + }(x,z) \tau^+ + h.c.)~.
\end{equation} 
Naturally, only terms of order $\sqrt{\Delta}$ are to be kept on the r.h.s. of (\ref{yn2}).

{\flushleft\bf{Zero mode $\mathbf{ \partial\over \partial z_3}$ vs $Y^{(3)}$}}: Remarkably, here we find 
\begin{equation}\label{yn3}
Y_n^{(3)} - {4 \pi \over \sqrt{V}}  L_3 { \partial A_n^{cl} \over 
 \partial z_3}=0 \implies \Lambda^{(3)} = 0,
 \end{equation}
 hence no compensating gauge transform is needed. This follows from the four-vector expressions for the zero modes
\begin{equation}Y_n^{(3)} = \left\{i {\sqrt{2 \Delta} \pi \over \sqrt{V}} e^{i \alpha} G \tau^+ + h.c.,  -{\sqrt{2 \Delta} \pi \over \sqrt{V}} e^{i \alpha} G \tau^+ + h.c., {4 \pi \over \sqrt{V}} {\tau^3 \over 2}, 0\right\}\,,
\end{equation}
and
\begin{eqnarray}
L_3 { \partial A_n^{cl} \over \partial z_3} =\left\{i {\sqrt{\Delta} \over 2 \sqrt{2}} e^{i \alpha} G \tau^+ + h.c., -   {\sqrt{\Delta} \over 2 \sqrt{2}} e^{i \alpha} G \tau^+ + h.c., {\tau^3 \over 2}, 0 \right\}~.
\end{eqnarray}
Notice that $x_3$ is the direction where $A^{cl}$ depends on $x_4 + z_3 L_4/(2 \pi)$ and where the classical $\sqrt{\Delta}$ solution vanishes.

 {\flushleft\bf{Zero mode $\mathbf{ \partial\over \partial z_4}$  vs $Y^{(4)}$}}:
 Here, the four-vectors of the exact zero mode and the derivative of the classical solution are
\begin{eqnarray}Y_n^{(4)} = \left\{-{\sqrt{2 \Delta} \pi \over \sqrt{V}} e^{i \alpha} G \tau^+ + h.c.,  -i{\sqrt{2 \Delta} \pi \over \sqrt{V}} e^{i \alpha} G \tau^+ + h.c.,0, {4 \pi \over \sqrt{V}} {\tau^3 \over 2}\right\}\,,
\end{eqnarray}
and
\begin{eqnarray}
L_4 { \partial A_n^{cl} \over \partial z_4} &=&\left\{{ \sqrt{\pi \Delta} \over \sqrt{2 V^{1/2}}} e^{i \alpha} ( {L_4 z_3 \over 2 \pi} F -  \sqrt{L_3 L_4 \over 4 \pi} {G}) \tau^+ + h.c., \right. \nonumber \\
&&
\left. i {\sqrt{\pi \Delta} \over \sqrt{2 V^{1/2}}} e^{i \alpha}( {L_4 z_3 \over 2 \pi} F -  \sqrt{L_3 L_4 \over 4 \pi} {G}) \tau^+ + h.c.,  0, {\tau^3 \over 2} \right\}~.
\end{eqnarray}
Their difference is 
 \begin{eqnarray}\label{yn4}
 Y_n^{(4)} - {4 \pi \over \sqrt{V}}  L_4 { \partial A_n^{cl} \over  \partial z_4} 
 &=&- \tau^+  {\sqrt{2 \pi \Delta} \over V^{1/2} } e^{i \alpha} {L_4 z_3 \over V^{1\over 4}} 
 \left( 
 \begin{array}{c}
   F(x,z) \cr
 i F(x,z) \cr
 0 \cr
 0 
 \end{array} 
 \right) + h.c. \nonumber \\
  &=& D_n(A^{cl.}) \Lambda^{(4)} = \partial_n \Lambda^{(4)} + i [A_n^{cl.}\big\vert_{\Delta=0} + A_n^{cl.}\big\vert_{{\cal{O}}(\sqrt{\Delta})}, \Lambda^{(4)} ]. \end{eqnarray}
Here, we have that \begin{equation}\label{lambda4new}\Lambda^{(4)}(x,z) =  {2 L_4 z_3 \over \sqrt{V}} {\tau^3 \over 2}
\end{equation} only has a $\Delta=0$ part. The  claim that $\Lambda^{(4)}$ obeys (\ref{yn4})  is straightforwardly verified

{\flushleft\bf{Summary:}} in each case we have verified that to order $\sqrt{\Delta}$,  
\begin{equation}\label{Ynk}
{4 \pi\over \sqrt{V}}    L_k { \partial A_n^{cl} \over  \partial z_k} + D_n(A^{cl}) \Lambda^{(k)} = Y_n^{(k)}
\end{equation}
obeys the background gauge condition, where $\Lambda^{(k)}$, for $k=1,2,3,4$, are given in (\ref{lambda1final}, \ref{lambda2new}, \ref{yn3}, \ref{lambda4new}), respectively.  As follows from these explicit expressions, the gauge transformation $\Lambda^{(k)}$ which makes the derivative of the classical solution obey the background-gauge condition is $\Omega$-periodic (recall the definition after (\ref{omegaperiodicgauge})), i.e. obeys the same periodicity conditions (\ref{bc}) as the classical solution. In fact, the $x$-dependent part of $\Lambda^{(k)}$ is expressed through the ${\cal{O}}(\sqrt{\Delta})$ component of the classical solution $A_k^{cl}$, similar to the BPST case. The $\Omega$-periodicity of $\Lambda^{(k)}$   is important in what follows.

We shall next argue that the measure in terms of the $z_n$ variables remains the one we found earlier  by studying the leading-order zero modes, eqn.~(\ref{bosoniczeromodemeasure}). For use below, we also rewrite the zero modes (\ref{Ynk}) as
\begin{equation}\label{Ynk1}
    { \partial A_n^{cl} \over  \partial z_k} + D_n(A^{cl}) \tilde\Lambda^{(k)} = {  \sqrt{V} \over 4 \pi L_k} Y_n^{(k)}~.
\end{equation}
The motivation for the rescaling evident in (\ref{Ynk1}) is that the zero modes are now directly  proportional to the derivatives of $A_n^{cl}$ with respect to $z_k$. The compensating gauge transformation $\tilde\Lambda^{(k)} = { \sqrt{V} \over 4 \pi L_k }  \Lambda^{(k)}$ is trivially related to $\Lambda^{(k)}$ appearing in (\ref{Ynk}) and appearing in (\ref{lambda1final}, \ref{lambda2new}, \ref{yn3}, \ref{lambda4new}).

 \subsubsection{The Jacobian and the all-order bosonic measure}
\label{jacobian}

We begin by shifting the bosonic field $A_n(x)$, obeying (\ref{bc}),  to be integrated over in the path integral  by the classical solution $A_n^{cl.}(x,z)$ of (\ref{ancl}).  We can choose to expand  the fluctuation in terms of a complete set of eigenfunctions of the hermitean operator $O_{mn}(x,z)$---the zero modes  (\ref{Ynk1}) and the nonzero modes  $Z_n^{q}$:\footnote{Of eigenvalues $\omega_q$, using the same notation as after eqn.~(\ref{bosonicaction}): $O_{mn} Z_n^{(q)} = \omega_q Z_n^{(q)}$.} \begin{eqnarray}\label{bosonicvariable}
A_n(x) - A_n^{cl.}(x,z) &=&  
  \sum\limits_{k=1}^4 \zeta_k^{(0)}  {\sqrt{V} \over 4 \pi L_k} ~ Y_n^{(k)}(x,z) + \sum_q \zeta_q \; Z_n^{(q)}(x,z).
\end{eqnarray}
The coefficients $\zeta_k^{(0)}$ and $\zeta_q$ are the projections of the  gauge field fluctuation $A_n(x) - A_n^{cl.}(x,z) $ onto the $Y^{(k)}(x,z)$ and $Z_q(x,z)$ directions in field space, thus they have implicit $z_k$-dependence. In what follows, we trade the integration over $\zeta_k^{(0)}$ for integration over the   $z_k$.  In our subsequent discussion we shall not explicitly write the contribution of the nonzero modes to the measure of the path integral, due to their cancellation with the contribution of the nonzero modes of the fermions and ghosts.

With the expansion (\ref{bosonicvariable}), the measure of the bosonic zero   modes now takes the form, using (\ref{ugeneral})\footnote{Recalling that (\ref{ugeneral}) calculated $U_{kl} = {2 \over g^2} \int_{\T^4} \tr Y_n^{(k)} Y_n^{(l)} = {16 \pi^2 \over g^2} \delta_{kl}$.} and taking into account the normalization of the zero modes
\begin{eqnarray}
\label{mub1}
d \mu_B &=& 
\left( \det {V \over 16 \pi^2 L_k L_l} \; U_{kl}\big\vert_{eq. (\ref{ugeneral})} \right)^{1\over 2}\; \prod\limits_{k=1}^4 \left[{d \zeta_k^{(0)} \over \sqrt{2 \pi}} \right] = {V \over g^4}  \; \prod\limits_{k=1}^4 \;{d \zeta_k^{(0)} \over \sqrt{2 \pi}} ~.
\end{eqnarray}
In order to integrate over fields orthogonal to the $ Y_n^{(k)}(x,z)$ zero modes, we consider their inner product with (\ref{bosonicvariable})
\begin{eqnarray}\label{fk}
f_k &=&  {2 \over g^2} \tr \int_{\T^4} (A_n(x) - A_n^{cl.}(x,z))  \;Y_n^{(k)}(x,z).
\end{eqnarray}
We then insert unity in the path integral, in the form
\begin{eqnarray}\label{unity}
1 = \prod_k  d z_k \prod_p \delta(f_p) \; |\det {\partial f_k \over \partial z_l}|~,
\end{eqnarray}
where $\partial f_k \over \partial z_l$ is evaluated at the value of $z_n$ making delta function vanish.\footnote{ Evaluating $\int_{\T^4} A^{cl.}_n Y^{(k)}_n$ using the explicit expressions of Section \ref{derivatives} and plugging into   (\ref{fk}), we find that to leading-order  in $\Delta$ the zero of $f_k$ occurs for $z_k = z_k^*$, where 
$z_k^* = - \pi (\delta_{k3}+ \delta_{k4}) +  L_k \tilde A_k^{3} + {\cal{O}}(\sqrt\Delta)$. Here, $\tilde A_k^3$ is the constant $\T^4$-mode of the Cartan component of $A_n$. }
Then, using (\ref{bosonicvariable}), the orthogonality of the zero and nonzero modes, we have that (\ref{ugeneral}), that 
\begin{eqnarray}
 f_p ={\sqrt{V} \over 4 \pi L_p} {16 \pi^2 \over g^2}  \zeta_p^{(0)} ~, 
\end{eqnarray}
so that  the $\delta$-function from (\ref{unity}) sets $\zeta_k^{(0)}=0$ after integrating over $\zeta_k^{(0)}$ with the measure (\ref{mub1}) (to avoid confusion, recall  that $\zeta_k^{(0)}$ has implicit $z$-dependence).

On the other hand, evaluating the derivative of (\ref{fk}) , we find
\begin{eqnarray}
{\partial f_k \over \partial z_l} &=&{2 \over g^2} \int_{\T^4} \tr\left( {\partial (A_n - A_n^{cl}) \over \partial z_l} \; Y_n^{(k)} + (A_n - A_n^{cl})  {\partial Y_n^{(k)} \over \partial z_l}\right) \nonumber \\
&=& {2 \over g^2} \int_{\T^4} \tr\left( -{\partial A_n^{cl.} \over \partial z_l} \; Y_n^{(k)} + (A_n - A_n^{cl})  {\partial Y_n^{(k)} \over \partial z_l}\right).\nonumber
\end{eqnarray}
  Next, we use (\ref{Ynk1}) to replace ${\partial A_n^{cl.} \over \partial z_l}$ by $Y_n^{(k)}$. Using the fact that $\tilde\Lambda^{(k)}$ is $\Omega$-periodic  allows us to integrate by parts on $\T^4$ without a boundary term, i.e. set $\int_{\T^4} \tr D_n \tilde\Lambda^{(l)}  \;Y_n^{(k)} = 0$, since $Y_n^{(k)}$ obeys the background gauge condition.
We thus find
\begin{eqnarray}
{\partial f_k \over \partial z_l}  &=&{2 \over g^2} \int_{\T^4} \tr\left( - {\sqrt{V} \over 4 \pi L_l} Y_n^{(l)} \; Y_n^{(k)} + (A_n - A_n^{cl})  {\partial Y_n^{(k)} \over \partial z_l}\right) \nonumber \\
&=& -   {\sqrt{V} \over 4 \pi L_l}{16 \pi^2 \over g^2} \delta_{kl} + {2 \over g^2} \int_{\T^4} \tr (A_n - A_n^{cl})  {\partial Y_n^{(k)} \over \partial z_l}~. 
\end{eqnarray} Thus, the unity insertion (\ref{unity}) becomes
\begin{eqnarray}
\label{unity1}
1 = \prod_k  d z_k \; \delta( \zeta_k^{(0)} {\sqrt{V} \over 4 \pi L_k} {16 \pi^2 \over g^2} )\; \bigg\vert\det \left( {\sqrt{V} \over 4 \pi L_l}{16 \pi^2 \over g^2} \delta_{kl}- {2 \over g^2} \int_{\T^4} \tr \zeta_q Z_n^{(q)} {\partial Y_n^{(k)} \over \partial z_l}\right)\bigg\vert, 
\end{eqnarray}
where we used  (\ref{bosonicvariable}) and took the liberty to set $\zeta_k^{(0)}=0$ in the determinant, due to integrating the delta function with the measure (\ref{mub1}). The nonzero mode part of the fluctuations can be ignored to leading order. Thus,
collecting everything, we find that the Jacobian factors from the delta function and the determinant in (\ref{unity1}) cancel out and the bosonic measure (\ref{mub1}) becomes, after inserting (\ref{unity1}) and integrating over $\zeta_k^{(0)}$,
\begin{eqnarray}
\label{mub2}
d \mu_B  
&=&{V \over g^4}  \prod\limits_{k=1}^4 {d z_k \over \sqrt{2 \pi}} ~.
\end{eqnarray}
The bosonic zero mode measure is thus equal to the leading-order measure constructed earlier in (\ref{bosoniczeromodemeasure}).

   \bibliography{NewDraft.bib}
 
  \bibliographystyle{JHEP}

\end{document}